\documentclass[11pt,a4paper]{amsart}
\usepackage{amssymb}
\usepackage{amsfonts}
\usepackage{hyperref}

\usepackage{latexsym,amsthm,amsmath,amscd}
\usepackage{graphicx,graphics}
\usepackage[all]{xy}
\usepackage{bbm}
\usepackage{bm}

\usepackage{xypic}
\usepackage{mathrsfs}

\newcommand{\frg}{\mathfrak{g}}

\newcommand{\frh}{\mathfrak{h}}
\newcommand{\frk}{\mathfrak{k}}
\newcommand{\frsl}{\mathfrak{sl}}
\newcommand{\frsu}{\mathfrak{su}}
\newcommand{\frS}{\mathfrak{S}}
\newcommand{\frSym}{\mathfrak{Sym}}

\newcommand{\scrc}{\mathscr{C}}

\newcommand{\scrr}{\mathscr{R}}
\newcommand{\scrd}{\mathscr{D}}
\newcommand{\scrs}{\mathscr{S}}
\newcommand{\scrm}{\mathscr{M}}
\newcommand{\scrh}{\mathscr{H}}
\newcommand{\scrt}{\mathscr{T}}
\newcommand{\scrf}{\mathscr{F}}
\newcommand{\scry}{\mathscr{Y}}
\newcommand{\Mod}{\mathscr{M}od}
\newcommand{\Rep}{\mathscr{R}ep}
\newcommand{\Vect}{\mathscr{V}ect}
\newcommand{\Ind}{\mathscr{I}nd}

\newcommand{\inthom}{{\scrh om}}

\newcommand{\modM}{{\tt M}}
\newcommand{\modN}{{\tt N}}
\newcommand{\modA}{{\tt A}}
\newcommand{\modK}{{\tt K}}
\newcommand{\modV}{{\tt V}}

\newcommand{\Idd}{\mathbbm{1}}   			
\newcommand{\id}{{\rm id}}

\newcommand{\ca}{\mathcal{A}}

\newcommand{\ch}{\mathcal{H}}
\newcommand{\cl}{\mathcal{L}}
\newcommand{\cn}{\mathcal{N}}
\newcommand{\cp}{\mathcal{P}}
\newcommand{\cb}{\mathcal{B}}

\newcommand{\cf}{\mathcal{F}}
\newcommand{\cg}{\mathcal{G}}

\newcommand{\cz}{\mathcal{Z}}
\newcommand{\co}{\mathcal{O}}
\newcommand{\cw}{\mathcal{W}}

\newcommand{\cu}{\mathcal{U}}        

\newcommand{\Ch}{\,{\sf ch}}

\renewcommand{\Im}{\ensuremath{\mathfrak{Im}}}

\DeclareMathOperator{\vol}{vol}       
\DeclareMathOperator{\End}{\sf End}
\DeclareMathOperator{\Ob}{\sf Ob}

\DeclareMathOperator{\Exp}{Exp}
\DeclareMathOperator{\Dim}{\sf dim}
\DeclareMathOperator{\Hom}{\sf Hom}
\DeclareMathOperator{\HS}{HS}

\def\ii{{\,{\rm i}\,}}
\def\={\ =\ }
\def\dd{{\rm d}}

\newcommand{\Tr}[1]{\:{\rm Tr}\,#1}
\def\e{{\,\rm e}\,}
\newcommand{\mbf}[1]{{\boldsymbol {#1} }}
\newcommand{\IZ}{\mathbb{Z}}
\newcommand{\IC}{\mathbb{C}}
\newcommand{\IP}{\mathbb{P}}

\newcommand{\IM}{\mathbb{M}}
\newcommand{\IR}{\mathbb{R}}

\theoremstyle{plain}

\numberwithin{equation}{section}

\newcommand{\beqa}{\begin{eqnarray}}
\newcommand{\eeqa}{\end{eqnarray}}
\newcommand{\beq}{\begin{equation}}
\newcommand{\eeq}{\end{equation}}

\def\appendix#1{\addtocounter{section}{1}\setcounter{equation}{0}
\renewcommand{\thesection}{\Alph{section}}
\section*{Appendix \thesection. #1}
\protect\indent \parbox[t]{11.715cm}
}

\newcommand{\Tube}{     \includegraphics[height=3.5ex]{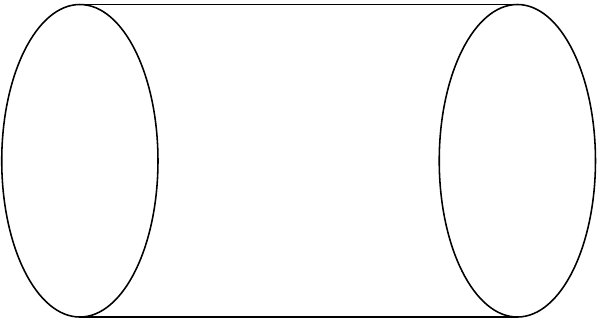}}
\newcommand{\Disk}{  \includegraphics[height=3.5ex]{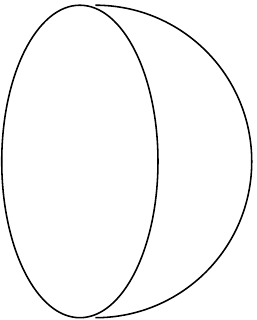}}
\newcommand{\Pants}{\includegraphics[height=6ex]{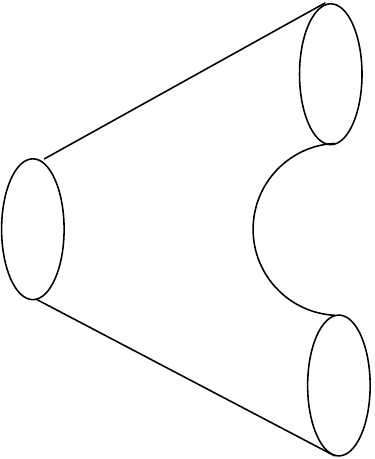}}
\newcommand{\Pantsd}{\includegraphics[height=6ex]{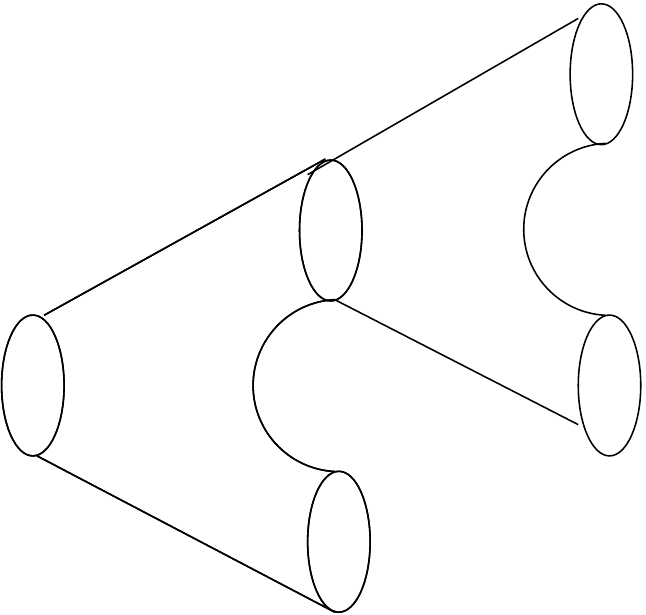}}
\newcommand{\Pantsu}{\includegraphics[height=6ex]{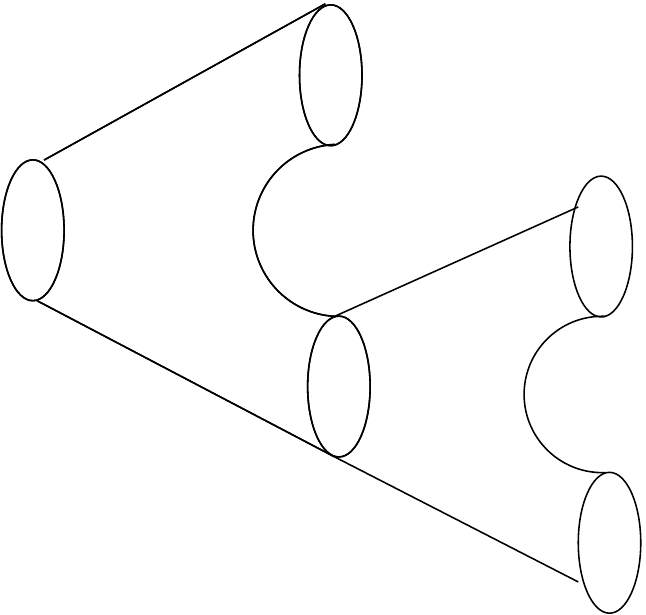}}
\newcommand{\DiskPants}{\includegraphics[height=6ex]{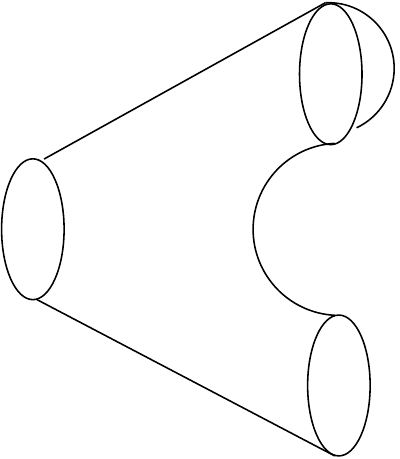}}
\newcommand{\Diskd}{  \includegraphics[height=3.5ex]{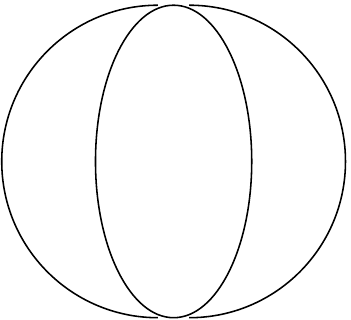}}

\newcommand{\Diskref}{  \includegraphics[height=3.5ex]{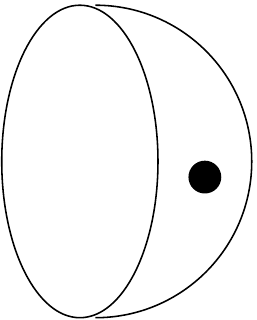}}
\newcommand{\Tuberef}{     \includegraphics[height=3.5ex]{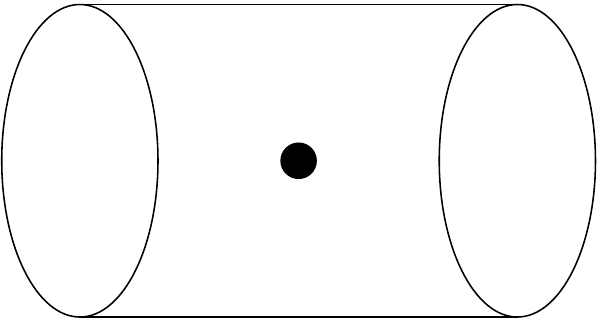}}
\newcommand{\Diskdref}{  \includegraphics[height=3.5ex]{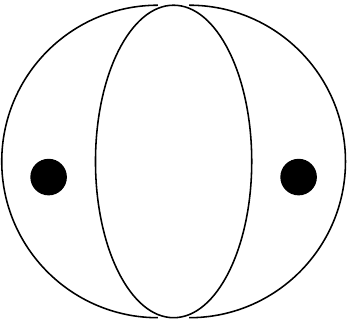}}

\begin{document}

\title[$q$-deformations of two-dimensional Yang-Mills theory]{$\mbf q$-deformations of two-dimensional Yang-Mills theory:
  \\[5pt] Classification, categorification and refinement}
\date{May 2013 \hfill \ EMPG--13--03 \ }
\author{Richard J. Szabo}
\address{\flushleft Department of Mathematics\\ Heriot-Watt University\\ Colin
Maclaurin Building, Riccarton, Edinburgh EH14 4AS, UK\\ Maxwell Institute
for Mathematical Sciences, Edinburgh, UK\\ The Tait Institute, Edinburgh, UK}
\email{R.J.Szabo@hw.ac.uk}
\urladdr{}
\thanks{}
\author{Miguel Tierz}
\address{\flushleft Departamento de An\'{a}lisis Matem\'{a}tico, Facultad de
Ciencias Matem\'{a}ticas \\
Universidad Complutense de Madrid \\ Plaza de Ciencias 3, Ciudad Universitaria,
28040 Madrid, Spain}
\email{tierz@mat.ucm.es}
\urladdr{}
\curraddr{ }
\subjclass{}
\keywords{}

\begin{abstract}
We characterise the quantum group gauge symmetries underlying
$q$-deformations of two-dimensional Yang-Mills theory by studying
their relationships with the matrix models 
that appear in Chern-Simons theory and six-dimensional $\cn=2$ gauge
theories, together with their refinements and supersymmetric extensions. We develop uniqueness results for
quantum deformations and refinements of gauge theories in two dimensions, and describe
several potential analytic and geometric realisations of them. We
reconstruct standard $q$-deformed Yang-Mills amplitudes via gluing rules in the representation category of the quantum group associated
to the gauge group, whose numerical invariants are the usual
characters in the Grothendieck group of the category. We apply this
formalism to compute refinements of $q$-deformed amplitudes in terms
of generalised characters, and
relate them to refined Chern-Simons matrix models and generalized
unitary matrix integrals in the quantum $\beta$-ensemble which compute
refined topological string amplitudes. We also describe
applications of our results to gauge theories in five and seven
dimensions, and to the
dual superconformal field theories in four dimensions which descend
from the $\cn=(2,0)$ six-dimensional superconformal theory.
\end{abstract}

\maketitle

\tableofcontents


\section{Introduction and summary}

Yang-Mills theory in two dimensions has been vigorously studied over the years
because of the analytical tractability of the quantum gauge theory
(see~\cite{review} for a review), originally pointed out by Migdal
\cite{Migdal75} and further developed in \cite%
{Rusakov,Witten:1991we}; an exact lattice gauge theory formalism leads to
the heat kernel expansion for the partition function and correlators on any compact,
connected and oriented Riemann surface $%
\Sigma_h$ of genus $h$. 
In this paper we are concerned with the $q$-deformation of two-dimensional
Yang-Mills theory, studied originally in \cite{Buffenoir:1994fh,Klimcik:1999kg} and
further developed in~\cite{Aganagic:2004js,Beasley:2005vf,Caporaso:2005ta,Blau:2006gh,Griguolo:2006kp,Thompson:2010iy,Ohta:2012ev}. In~\cite{Aganagic:2004js}
it was shown that the partition function of topologically twisted $\mathcal{N%
}= 4$ Yang-Mills theory with gauge group $U(N)$ on the ruled Riemann surface $%
\mathcal{O(-}p\mathcal{)}\rightarrow \Sigma _{h}$ reduces to that of
$q$-deformed $U(N)$ Yang-Mills theory on the base Riemann surface $%
\Sigma _{h}$; this relationship was further clarified and extended
in~\cite{Griguolo:2006kp} to show how the $q$-deformed gauge theory
captures the counting of instantons on Hirzebruch-Jung spaces. From this result it is anticipated that $q$-deformed
Yang-Mills theory provides a
non-perturbative completion of topological string theory on the rank two
Calabi-Yau fibration $\mathcal{O}(p+2h-2)\oplus\mathcal{O}(-p)$
over~$\Sigma_h$. These studies have been reinvigorated in the past few
years with the discovery that for $p=1$ the two-dimensional gauge theory also computes the
partition function of a strongly-coupled $\cn=2$ gauge theory on
$S^1\times S^3$~\cite{Gadde:2011ik,Tachikawa:2012wi,Fukuda:2012jr};
this duality is conjecturally realised within the putative
six-dimensional $\cn=(2,0)$ superconformal theory on $S^3\times S^1\times\Sigma_h$ in
which the four-dimensional gauge theories are specified by the
Riemann surface $\Sigma_h$~\cite{Gaiotto:2009we}. A refinement of this
two-dimensional gauge theory first appeared in its topological
BF-theory form in~\cite{Gadde:2011uv} as the dual to a
four-dimensional $\cn=2$ gauge theory on $S^3\times
S^1$ with two superconformal fugacities; in~\cite{Aganagic:2012si} the
full (non-topological) refined
gauge theory was derived from counting BPS states in refined topological
string theory on the local Calabi-Yau threefold
$\mathcal{O}(p+2h-2)\oplus\mathcal{O}(-p) \to\Sigma_h$ with a
non-selfdual graviphoton background. 

This paper is devoted to an in-depth investigation of quantum
deformations and refinements of two-dimensional Yang-Mills theory,
and their deep and rich connections with a multitude of gauge theories in higher
dimensions. We lay emphasis on understanding the precise quantum group
structure of the gauge symmetries underlying these models, and a
complete characterization of quantum deformations and refinements of gauge theories in two
dimensions. We clarify the relationships between the discrete matrix models that appear in $%
q$-deformed Yang-Mills theory and the continuous matrix models of
Chern-Simons gauge theory using the moment problem, and extend these
equivalences to refined settings. We also relate certain topological versions
of two-dimensional
Yang-Mills amplitudes to gauge theory partition functions in all
dimensions ranging from three to seven by reformulating them in terms of (refined) unitary matrix integrals; in particular, we use this
connection to reproduce (refined) topological string amplitudes from five-dimensional $\cn=1$ gauge theory
partition functions.

As we review in \S\ref{se:2DYMgen}, the partition function
of $q$-deformed Yang-Mills theory on $\Sigma _{h}$ is a simple
variation of the Migdal formula (given explicitly in (\ref{HK}) below), which involves quantum dimensions
rather than ordinary dimensions of representations of the gauge group. This $q$-deformed
gauge theory
bears the same relation to $q$-deformed representation theory as
ordinary Yang-Mills theory does to ordinary representation
theory. It can be regarded as an analytic continuation of
Chern-Simons gauge theory on a Seifert fibration of degree $p$ over the
Riemann surface $\Sigma _{h}$. For genus $h=0$, the Seifert manifold is the
three-sphere
$S^{3}$ for $p=1$ (regarded as the Hopf fibration $%
S^3\to \Sigma _0 =S^{2}$) and the 
lens space $L(p,1)=S^3/\IZ_p$ for $p>1$.
In this particular $q$-deformation the quadratic Casimir eigenvalues are ordinary
integers because they arise from a sum over torus
bundles on $\Sigma_h$ just as in the ordinary case, which do not
undergo any deformation themselves; this is not very natural from the point of view of quantum
group theory. Another somewhat non-canonical feature is the dependence of the
deformation parameter $q$ on the Yang-Mills coupling constant. Alternatively, the approaches of~\cite{Buffenoir:1994fh,Klimcik:1999kg}
deform the gauge symmetry to a quantum group, and thereby avoid some
of these pitfalls. The role of quantum group gauge symmetries
in the complete solution of $q$-deformed two-dimensional Yang-Mills
theory is also stressed in~\cite{deHRT}. Insofar as the $q$-deformed
gauge theory is an analytic continuation of Chern-Simons theory, which
has a well understood relation to quantum groups, it is natural to
explore the relation between the field theories based on an
explicit quantum deformation of the gauge group and the field theories
with undeformed gauge group where the quantum group symmetries seem to
emerge implicitly.

As a first step towards understanding this relationship, in
\S\ref{se:contmm} we relate the discrete matrix models that underlie
$q$-deformed Yang-Mills amplitudes to the continuous Chern-Simons
matrix models; matrix model techniques and relations permeate this
paper and are a driving force in much of our analysis. In particular,
we establish a new relationship with the Stieltjes-Wigert matrix model
for all $p\in\IZ_{\geq0}$ and a new dual formulation of $U(N)$
Chern-Simons theory as a $U(\infty)$ matrix model; this unitary matrix
model is related explicitly to the BF-theory limit of two-dimensional
Yang-Mills theory and also to the $U(\infty)$ matrix models describing
$\cn=2$ gauge theories in six dimensions that we consider in \S\ref{se:Class}.

As a next step, we develop
uniqueness results for quantum deformations of two-dimensional
Yang-Mills theory in \S\ref{se:Class}.
We formulate the $q$-deformation of the heat kernel expansion by quantum dimensions together
with a $q$-deformed Casimir invariant, and show that it is equivalent
to ordinary Yang-Mills theory. We further show that $q$-deformation of the
Boltzmann weight in Migdal's partition function is qualitatively
equivalent to deformation by quantum dimensions. As a consequence, both Klim\v{c}\'{\i}k's partition function for $q$-deformed Yang-Mills theory~\cite{Klimcik:1999kg}
and the partition function for crystal melting with external
potentials are equivalent to that of
ordinary generalized two-dimensional Yang-Mills theory. We extend
these correspondences to $\cn=2$ gauge theories in six dimensions by
rewriting their partition functions as unitary matrix integrals and
applying the strong Szeg\H{o} limit theorem for Toeplitz determinants
to evaluate them in closed form.
Our findings are somewhat in line with the arguments of Brzezi\'nski and Majid~\cite{brzmajid2} that gauge
theories with quantum group gauge symmetries should be defined on
\emph{quantum} spaces in order to get something that is different from
ordinary Yang-Mills theory. We elaborate on this
observation and consider the
diagonalisation technique of~\cite{Aganagic:2004js,Blau:2006gh}
in the context of Yang-Mills theory on the standard Podle\'s quantum sphere
$S_q^2$~\cite{Po87}; this approach effectively abelianizes the non-abelian gauge
theory so that its partition function (and correlation functions) can
be calculated explicitly and straightforwardly by summing the
resulting abelian field theory over all isomorphism classes of torus
bundles. One advantage of this approach is that it computes quantum fluctuations straightforwardly from
Gaussian path integrals of free abelian fields (and hence is essentially rigorous),
and the non-abelian nature of the original field theory is reflected
in the determinants which arise as Jacobians in the diagonalization
procedure. When applied to torus bundles over $S_q^2$, this calculation gives a putative derivation of the heat
kernel expansion involving $q$-deformed Casimir eigenvalues and
moreover explains why quantum group gauge symmetries appear as
ordinary gauge symmetries in this framework.

In \S\ref{se:Cat} we reformulate the standard two-dimensional
topological field theory construction of $q$-deformed
Yang-Mills amplitudes~\cite{Aganagic:2004js} in terms of a functor whose target is the
semisimple ribbon category of representations of the quantum universal
enveloping algebra associated to the gauge group $U(N)$. This
construction ``categorifies'' the usual building blocks of Yang-Mills
amplitudes in terms of $U(N)$ characters, which now appear as
numerical invariants in the corresponding Grothendieck group of the
ribbon category. A related modular tensor category is used
in~\cite{Iqbal:2011kq} to
reformulate the three-dimensional topological field theory
construction of (refined) Chern-Simons theory, which is usually based
on the finite category of integrable representations of $U(N)$. Our
ribbon category can be regarded as a certain completion of this
category (which we describe explicitly) involving direct sums of
infinitely many simple objects, the irreducible representations of the
quantum group. In this way we are able to build Yang-Mills
amplitudes as numerical invariants of this category in a way in which
the quantum group gauge symmetry is manifest simply by construction,
and which moreover exhibits interesting relationships amongst
different correlators. For completeness and convenience of exposition,
we review all category theory concepts and results that we use in this
paper.

One of the main advantages of our categorical reformulation is that it
straightforwardly allows for various generalizations, e.g. to gauge
groups other than $U(N)$. In \S\ref{se:Ref} we use this general treatment to
construct refinements of $q$-deformed Yang-Mills amplitudes; they are
analytic continuations of the correlators in the refined Chern-Simons
theory originally defined in~\cite{Aganagic:2011sg}. Now the
numerical invariants arising from integration of morphisms in the
ribbon category are given by generalised characters. We extend the
discrete/continuous matrix model equivalence to show that the refined
$q$-deformed Yang-Mills partition function for all $p\in\IZ_{\geq0}$ is also equivalent to the
refined Chern-Simons matrix model derived in~\cite{Aganagic:2011sg}
from a refined counting of BPS states of spinning M2-branes in certain M-theory compactifications, which is a quantum
$\beta$-deformation of the usual Stieltjes-Wigert matrix ensemble. By
applying the strong Szeg\H{o} limit theorem to evaluate generalized unitary matrix
models in the quantum $\beta$-ensemble, we
relate various refinements of the two-dimensional gauge theories,
together with their supersymmetric extensions, to supersymmetric gauge
theories in five, six and seven dimensions; we also describe some
uniqueness results for refinements of two-dimensional gauge
theories. In particular, building on constructions from the unrefined
case~\cite{Szabo:2010sd}, we expect that our refined $U(\infty)$
matrix models could be relevant to Macdonald refined stochastic
processes~\cite{Borodin}. We conclude by applying
some of the formalism of generalised characters to discuss possible
physical derivations of refined $q$-deformed Yang-Mills amplitudes
directly from the perspective of two-dimensional quantum field theory.

Three appendices at the end of the paper include some of the more
technical details which are used in the main text: Appendix~A
summarises pertinent aspects of the quantum universal enveloping
algebra of the Lie algebra of the gauge group $U(N)$ that we use,
Appendix~B describes techniques for evaluating Toeplitz determinants
that are used in our analysis of unitary matrix models throughout the
paper, and Appendix~C contains more intricate details concerning our category
theory constructions.

\section*{Acknowledgments}

We thank Mark Adler, Jacob Christiansen, David Evans, Jim Howie, Kurt
Johansson, Anatoly Konechny,
Giovanni Landi,
Mark Lawson, Sanjaye Ramgoolam, Ingo Runkel, Christoph Schweigert, Simon Willerton and
Masahito Yamazaki for
helpful discussions and correspondence. MT thanks J{\o}rgen Andersen
for the warm hospitality during his visit to the Centre for Quantum
Geometry of Moduli Spaces, Aarhus University, Denmark, where part of
this work was carried out. The work of RJS was partially supported by the Consolidated Grant
ST/J000310/1 from the UK Science and Technology Facilities Council,
and by Grant RPG-404 from the Leverhulme Trust. The work of MT is
supported by a Juan de la Cierva Fellowship.

\section{Two-dimensional Yang-Mills theory and its $q$-deformations}\label{se:2DYMgen}

In this section we review some computational aspects and exact analytical
expressions for the quantum partition function of two-dimensional
Yang-Mills theory. We then survey some of the various $q$-deformed
versions which have appeared in the literature.

\medskip

\subsection{Ordinary gauge theory}\label{se:ordgt}~\\[5pt]
The action of Yang-Mills theory with compact gauge group $G$ on an
arbitrary connected and oriented Riemann surface $\Sigma$ with unit area form $\dd\mu$ is given by%
\begin{equation}
S_{\mathrm{YM}}[A] =\frac1{4g_s}\, \big(F_A,F_A\big):=-\frac{1}{2g_s}\,\int_{\Sigma }\,\mathrm{d}\mu ~\Tr%
\big(F_A^{2} \big) \ , \label{continuum}
\end{equation}%
where the positive parameter $g_s$ plays the role of the coupling
constant, $F_A \in\Omega^2(\Sigma,\frg)$ is the curvature of a gauge
connection $A\in\Omega^1(\Sigma,\frg)$ on a trivial
principal $G$-bundle over $\Sigma$, and $\Tr$ is an invariant
quadratic form on the Lie algebra $\frg$ of $G$. Note that the
definition (\ref{continuum}) does not
depend on any choice of metric on the surface, but only on a choice of
measure $\dd\mu$ on $\Sigma$ which represents the generator of $H^2(\Sigma,\IZ)=\IZ$.
In this paper we are mainly interested in the case of a unitary gauge group
$G={U}(N)$; then $\Tr$ refers to the trace in the fundamental
representation of $G$. We denote by $\ca= \Omega^1(\Sigma,\frg)$ the affine
space of gauge connections and by $\cg=\Omega^0(\Sigma,G)$ the group
of gauge transformations.

The quantum gauge theory is defined by the path integral
\beq
Z_{\rm YM}(g_s;\Sigma):=\frac1{\vol(\cg)}\, \Big(\, \frac1{2\pi\, g_s}\,
\Big)^{\dim\cg/2} \ \int_\ca\,\scrd\mu[A] \ \exp\big(- S_{\rm
  YM}[A]\big) \ , \label{YMpartfndef}
\eeq
where $\scrd\mu[A]$ is the translation-invariant Riemannian measure induced by the metric $(-,-)$ on $\ca$. We
similarly define an invariant metric on $\cg$ (by the same formula), which formally
determines the volume of $\cg$.

We can also write this partition
function as
\beq
Z_{\rm YM}(g_s;\Sigma) =
\frac1{\vol(\cg)}\, \int_\ca\,
\scrd\mu[A] \ \int_{\Omega^0(\Sigma,\frg)}\, \scrd\mu[\phi] \
\exp\big(- S_{\rm BF}[\phi,A] \big) \ , \label{YMpartfnBF}
\eeq
where 
\beq
S_{\rm BF}[\phi,A] =\ii\, \langle F_A,\phi\rangle+ \frac {g_s}2 \, \big(\phi\,,\,\phi\big) =
- \int_{\Sigma}\, \Tr \Big( \ii \phi\, F_A-\frac {g_s}2 \, 
\phi^2 \ \dd\mu \Big) \label{YMpartfnBFaction}
\eeq
is the first order form of the Yang-Mills action
functional; in the weak coupling limit $g_s=0$ this is the action for
topological BF-theory on the Riemann
surface $\Sigma$ and in this case (\ref{YMpartfnBF}) computes the
symplectic volume of the moduli space of flat $G$-connections on $\Sigma$. The Euclidean measure
$\scrd\mu[\phi]$ on the Lie algebra of $\cg$ is determined by the same
invariant form that we use to define the volume $\vol(\cg)$. We may regard the curvature $F_A\in\Omega^2(\Sigma,\frg)$ as an element of the dual of the Lie algebra of
$\cg$; then $\langle-,-\rangle$ denotes the pairing between the Lie
algebra of $\cg$ and its dual. The equality between
(\ref{YMpartfndef}) and (\ref{YMpartfnBF})
follows from the functional Gaussian integration over the scalar field
$\phi\in\Omega^0(\Sigma,\frg) $.

The Yang-Mills partition function can be defined and evaluated
rigorously via a combinatorial formalism obtained through lattice regularization~\cite{Witten:1991we,review}. This leads to
the combinatorial heat kernel expansion for the partition function on a compact
Riemann surface $%
\Sigma _{h}$ of genus $h$ which is given by Migdal's formula
\begin{equation}
\mathcal{Z}_{\mathrm{M}}(g_s; \Sigma _{h}) =\Big(\,
\frac{\vol(G)}{(2\pi)^{\dim G}}\, \Big)^{2h-2} \ \sum_{\lambda }\,\left(
\dim \lambda \right) ^{2-2h}\,\exp \Big(-\frac {g_s}2\,C_{2}(\lambda )\Big)\ ,
\label{HK}
\end{equation}%
where the sum runs over all isomorphism classes of irreducible unitary
representations of the gauge group $G$, $\dim\lambda$ is the dimension of the
representation $\lambda$, and $C_2(\lambda)$ is the quadratic Casimir
invariant of $\lambda$ associated with the invariant quadratic form
$\Tr$ on the Lie algebra
$\frg$. Here $\vol(G)$ is the volume of $G$ determined
by $\Tr$; we will usually
drop this volume factor (as well as other normalization
constants) in the following. The formula (\ref{HK}) can also be
derived in an operator formalism from canonical quantization of the continuum gauge theory
defined by the action (\ref{YMpartfnBFaction})~\cite{Witten:1992xu}. For $G={U}(N)$ the representations can be
identified with $N$-component partitions $\lambda=(\lambda_1,\dots,\lambda_N)$
(equivalently Young diagrams), which are the highest weights. The dimension and
quadratic Casimir eigenvalue of the corresponding representation are
then given by the
explicit formulas
\beq
\dim\lambda=\prod_{i<j}\, 
\frac{\lambda_i-\lambda_j+j-i}{j-i} \qquad \mbox{and} \qquad
C_2(\lambda) =\langle\lambda,\lambda+2\rho\rangle
= \sum_{i=1}^N \, \big(\lambda_i^2+(N+1-2i)\, \lambda_i\big) \ , \nonumber
\eeq
where $\langle\lambda,\mu\rangle:=\sum_i\, \lambda_i\,\mu_i$ is the invariant
bilinear pairing induced by $\Tr$ and $$\rho_i=\frac{N-2i+1}2$$ are
the components of the Weyl vector $\rho$ of $G$ (the half-sum of positive roots).

In the following we will also exploit the relationship with the specialization of the Schur polynomials $s_\lambda(x)$ in $N$
variables $x=(x_1,\dots,x_N)$, which form an orthonormal basis on the
ring of symmetric polynomials with respect to the Hall inner
product~\cite{Macdonald}; they can be defined as the character of the $\mathfrak{g}$-module
$\lambda$ evaluated on a matrix $X$ whose eigenvalues are
$x_1,\dots,x_N$, i.e. $s_\lambda(x)=\Tr_\lambda(X)$. Then the Weyl
dimension formula for the irreducible representation of $G$ with
highest weight $\lambda$ can be expressed as
\beq
\dim \lambda =s_{\lambda }(1,\dots,1) \ . \nonumber
\eeq

Rather than working directly with (\ref{HK}), it will sometimes be more practical to
change summation variables $\mu _{i}=\lambda _{i}+N-i$ for
$i=1,\dots,N$ and use the
associated discrete Gaussian matrix model~\cite{Gross:1994mr}
\begin{equation}
\mathcal{Z}_{\mathrm{M}}(g_s;\Sigma_h) =\sum_{\mu\in 
\mathbb{Z}^N
} \ \prod\limits_{i<j} \, \left( \mu _{i}-\mu _{j}\right) ^{2-2h} \, \exp\Big(-%
\frac {g_s}2\, \sum_{i=1}^N\, \mu_i^2\Big) \ ,
\label{discrmat}\end{equation}%
which agrees with (\ref{HK}) up to a standard area-dependent
renormalization $\mu_i\to \mu_i-\frac{N-1}2$.
The dimensions in (\ref{HK}) lead to a
Vandermonde determinant and the Casimir
invariants give
the Gaussian potential of the matrix model, but in terms of the discrete
eigenvalues $\mu= (\mu_1,\dots,\mu_N)\in\IZ^N$.

The expansion (\ref{discrmat}) can
also be derived directly from (\ref{YMpartfnBF}) by
the technique of
diagonalization~\cite{blau1}. As we will make reference to it later
on, let us briefly review this calculation. The crux of the technique is the
classical Weyl integral formula for Lie algebras: Let $f(\phi)$ be
a conjugation invariant function on the Lie algebra $\frg$,
$$
f\big(g^{-1}\, \phi\, g\big)= f(\phi) \qquad \mbox{for} \quad g\in G \
,
$$
and assume that it is integrable with respect to the normalized invariant Haar
measure $\dd\phi$ on $\frg$. The Lie algebra element $\phi\in\frg$ can
be conjugated by an element $U\in G$ into the Cartan subalgebra
$\frh\subset\frg$; then $f(\phi)$ depends only on the eigenvalues
$(\phi_1,\dots,\phi_N)\in\IR^N$. Integrating over $U$ gives a factor
of the order of the Weyl group $W=\frS_N$ of $G=U(N)$, which is the residual group of conjugation
symmetries of $\frh$ acting by permutations of the eigenvalues. Then the
Weyl integral formula reads as
\beq
\int_{\frg}\, \dd\phi \ f(\phi)=\frac1{N!}\, 
\int_{\IR^N} \ \prod_{i=1}^N \, \dd\phi_i \ \prod_{j<k}\, (\phi_j-\phi_k)^2 \
f(\phi_1,\dots,\phi_N) \ ,
\label{Weylintalg}\eeq
where the Vandermonde determinant factor arises as the Jacobian for
the change of variables induced by conjugation into the Cartan
subalgebra.

Let us now apply the diagonalization formula (\ref{Weylintalg}) to the BF-form of the partition
function (\ref{YMpartfnBF}). For this, we observe that the functional 
\beq
F[\phi]:= \frac1{{\rm vol}(\cg)}\, \int_\ca\,
\scrd\mu[A] \ \exp\big(- S_{\rm BF}[\phi,A] \big)
\label{Fphi}\eeq
for fixed $\phi\in \Omega^0(\Sigma,\frg)$ is conjugation invariant. We can thus use a local gauge transformation
$U$ to impose
the torus gauge condition where the scalar field
$\phi\in\Omega^0(\Sigma_h,\frg)$ takes values in the Cartan subalgebra
$\frh$ of the Lie algebra $\frg=\frh\oplus \frk$. However, while the
diagonalised field
$\phi=(\phi_1,\dots,\phi_N)\in\Omega^0(\Sigma_h, \frh)$ is
globally-defined and smooth, there
are obstructions to finding smooth gauge functions $U\in\cg$
globally; these obstructions are parametrized by the set of all isomorphism classes of non-trivial
torus bundles $\cl_n\to\Sigma_h$ with structure group the maximal torus
$T=U(1)^N$, which arise as restrictions of the original trivial principal
$G$-bundle over $\Sigma_h$~\cite{Blau:1994rk}. The $\frh$-component $A^\frh$ of the gauge connection
$A=(A^\frh,A^\frk) \in\Omega^1(\Sigma_h,\frg)$ is a gauge field on a
principal torus bundle
$\cl_n\to\Sigma_h$, while the $\frk$-components $A^\frk$ are sections
of the associated vector bundle $\cl_n\times_T\frk$. The path integral
version of the Weyl integral formula (\ref{Weylintalg}) should then
include a sum over
contributions from connections on all isomorphism classes of
$T$-bundles; we denote the affine space of abelian gauge connections on a principal
$T$-bundle $\cl_n\to\Sigma_h$ by $\ca_n$. The maximal torus bundles
$\cl_n\to\Sigma_h$ are parametrized by their first Chern classes
$c_1(\cl_n)=n=(n_1,\dots,n_N)\in \IZ^N$; by Chern-Weil
theory one has
$$
\frac1{2\pi}\, \int_{\Sigma_h} \, \dd A^\frh_i=n_i
$$
for $A^\frh=(A^\frh_1,\dots,A^\frh_N) \in\ca_n$.
Then the Weyl integral formula for the partition function
(\ref{YMpartfnBF}) gives
\beqa
Z_{\rm YM}(g_s;\Sigma_h) &=& \frac1{{\rm vol}(\cg)} \ \sum_{n\in\IZ^N} \ \int_{\ca_n} \,
\scrd\mu[A^\frh] \ \int_{\Omega^1(\Sigma_h,\cl_n\times_T\frk)} \,
  \scrd\mu[A^\frk] \nonumber\\ && \qquad \times \
  \int_{\Omega^0(\Sigma_h,\IR^N)} \ \prod_{i=1}^N \,
  \scrd\mu[\phi_i] \ \Big[\,\prod_{j<k}\, (\phi_j-\phi_k)^2\, \Big] \
  \exp\big(- S_{\rm BF}[\phi,A^\frh,A^\frk] \big) \ , \nonumber
\eeqa
where
\beq
S_{\rm BF}[\phi,A^\frh,A^\frk]=\sum_{i=1}^N \ \int_{\Sigma_h} \, \Big(-\ii \phi_i\,
\dd A^\frh_i+\frac {g_s}2\, \phi_i^2\, \dd \mu\Big) +\sum_{\alpha\in{\rm Ad}(G)}\
\int_{\Sigma_h}\, \langle\alpha,\phi\rangle\, A^\frk_\alpha\wedge
A_{-\alpha}^\frk
\label{BFgaugefixed}\eeq
and $\alpha\in{\rm Ad}(G)$ are the roots of the Lie algebra $\frg$. Integrating over $A^\frk_\alpha$
gives an inverse functional determinant induced by a 
one-form in $\Omega^1(\Sigma_h)$. As demonstrated
in~\cite{blau1}, by using the Hodge decomposition of forms on
$\Sigma_h$ one finds that only harmonic
forms contribute and yield Vandermonde determinant factors $\prod_{i<j}\, (\phi_i-\phi_j)^{-2h}
$,
as there are $2h$ linearly
independent harmonic one-forms on the Riemann surface
$\Sigma_h$. Altogether this reduces the path integral
(\ref{YMpartfnBF}) to that of an abelian gauge theory based on the
maximal torus $T=U(1)^N$. Any gauge field $A^\frh\in\ca_n$ can be
decomposed as $A^\frh=a^\frh+\tilde
A{}^\frh$, where the monopole connections $a^\frh$ obey $\dd a^\frh_i=
2\pi \,n_i\, \dd\mu$ and
integrating over $\tilde
A{}^\frh_i\in\Omega^1(\Sigma_h)$ gives
delta-function constraints
$$
\dd\phi_i=0
$$
which imply that the scalar fields $\phi_i$ are constant on $\Sigma_h$ for each
$i=1,\dots,N$. The complete path integral is now reduced to a sum of
finite-dimensional integrals
$$
Z_{\rm YM}(g_s;\Sigma_h) = \sum_{n\in\IZ^N} \ 
\int_{\IR^N} \ \prod_{i=1}^N \, \dd\phi_i\ \e^{2\pi\ii n_i\, \phi_i-\frac{g_s}2\,
  \phi_i^2} \ \prod_{j<k}\, (\phi_j-\phi_k)^{2-2h} \ .
$$
The sum over torus
bundles $n\in\IZ^N$
gives periodic delta-functions via the Poisson resummation formula
$$
\sum_{n_i=-\infty}^\infty\, \e^{2\pi\ii n_i\,
  \phi_i}=\sum_{\mu_i=-\infty}^\infty\, \delta(\phi_i-\mu_i) \ ,
$$
and we arrive finally at the discrete Gaussian matrix model (\ref{discrmat}).

\medskip

\subsection{$q$-deformed gauge theories}\label{se:qdefgt}~\\[5pt]
An interesting variant of the model described above is the ``$q$-deformation'' of
two-dimensional Yang-Mills theory. Formally, it arises when one
considers the partition function (\ref{YMpartfnBF}) but now with the
Gaussian integral taken over the domain
$\phi\in\Omega^0(\Sigma,G)$. The partition function
of $q$-deformed Yang-Mills theory on $\Sigma _{h}$ can
again be computed by
diagonalization~\cite{Aganagic:2004js,Blau:2006gh} and it results in a simple
variation of (\ref{HK}) given by the expansion%
\begin{equation}
\mathcal{Z}_{\mathrm{M}}^{(p) }(q; \Sigma _{h})
=\sum_{\lambda }\,\left( \dim _{q}\lambda \right) ^{2-2h}\,\exp \Big(%
-\frac{p\, g_s}2 \,C_{2}(\lambda )\Big)\ , \label{qYM1}
\end{equation}%
but now involving the \emph{quantum} dimensions
\beq
\dim_q\lambda = s_\lambda\big(q^{\rho} \big)=q^{|\lambda|/2} \
s_{\lambda }\big(1,q,\dots,q^{N-1} \big)=\prod_{i<j}\, 
\frac{[\lambda_i-\lambda_j+j-i]_q}{[j-i]_q}
\nonumber
\eeq
of the representations of $G$ given by the Weyl character formula,
where $p\in\IZ_{>0}$, the deformation
parameter is $q:=\e^{-g_s}$ and we have defined $q^\rho:= (q^{\rho_1},\dots,q^{\rho_N})$. Here $|\lambda|:=\sum_i\,\lambda_i$ is the number of boxes in the
Young diagram corresponding to the partition $\lambda$, and
the symmetric $q$-number
\begin{equation}
[x]_q := \frac{q^{x/2} - q^{-x/2}}{q^{1/2} - q^{-1/2}}
\label{eq:q-integer}
\end{equation}
is defined for $q \neq 1$ and any $x \in \IR$; note that
\beq
[n]_q=\sum_{i=1}^n\, q^{\frac12\, (n-2i+1)} \qquad \mbox{for} \quad n\in \IZ_{>0} \
. \nonumber
\eeq
One can analytically continue this
expression to arbitrary
values $q\in\IC$. When $q=\e^{2\pi \ii/(k+N)}$ is a
root of unity, the series (\ref{qYM1}) acquires an affine Weyl symmetry and
should be truncated to the Weyl alcove consisting of integrable
representations of $G$ at level $k$; the resulting sum is then the
partition function of Chern-Simons gauge theory at level $k\in\IZ$ on
a circle bundle of degree $p$ over the Riemann surface $\Sigma_h$. For $p=1$ it is related to the $G_k$-WZW
model on $\Sigma_h$. For $p=0$ the gauge theory is a $q$-deformation of
two-dimensional BF-theory and the partition function
reproduces the Verlinde formula for the dimension of the Hilbert space of
$G_k$-WZW conformal blocks on $\Sigma_h$; this $q$-deformed BF-theory is the
gauged $G/G$ WZW model, or equivalently Chern-Simons theory on the
trivial circle bundle $\Sigma_h\times S^1$, and it may be regarded as a non-linear
deformation of ordinary Yang-Mills theory. For generic $p$ the
partition function (\ref{qYM1}) computes certain intersection indices
on the moduli space of Yang-Mills connections on $\Sigma_h$, including
those which are not flat. The connections between these
two-dimensional and three-dimensional field theories are also analysed in~\cite{Naculich:2007nc}.

The discrete matrix model corresponding to (\ref{qYM1}) involves a
$q$-deformed Vandermonde determinant and an ordinary Gaussian
potential; it is given by~\cite%
{Aganagic:2004js}%
\begin{equation}
\mathcal{Z}_{\mathrm{M}}^{(p) }(q; \Sigma _{h})
= \sum_{\mu\in 
\mathbb{Z}^N
} \ \prod\limits_{i<j}\, \left[ \mu_{i}-\mu_{j}\right] _{q}^{2-2h}\,
\exp\Big(- \frac{p\, g_s}2 \, \sum_{i=1}^N\, \mu_i^2\Big) \ . \label{qmat}
\end{equation}%
The ordinary Yang-Mills partition functions $\cz_{\rm M}(\tilde
g_s;\Sigma_h)$ from (\ref{HK}) and
(\ref{discrmat}) are respectively recovered from (\ref{qYM1}) and
(\ref{qmat}) in the double scaling limit $p\to\infty$, $g_s\to0$ (so
that $q\to1$) with
the renormalized coupling constant $\tilde g_s:=p\, g_s$ fixed.

In the formulation of~\cite{Aganagic:2004js,Blau:2006gh}, the $q$-deformation arises from the change of integration measure for the
compact scalar field $\phi$, or alternatively, from the point of view
of diagonalization, from a restricted sum over torus
bundles to those which are of torsion class $\IZ_p$; it can be understood in terms of the
difference between the Weyl integral formulas for Lie \emph{groups}
and for Lie \emph{algebras}. For this, let $f(U)$ be a conjugation invariant
function on the Lie group $G$, 
$$
f\big(g^{-1}\, U\, g\big)=f(U) \qquad \mbox{for} \quad g\in G\ ,
$$
and assume that it is integrable with respect to the normalized
invariant Haar
measure $\dd U$ on $G$.
We can again conjugate $U\in G=U(N)$ into the maximal torus
$T=U(1)^N$; then $f(U)$ depends only on the eigenvalues
$(u_1,\dots,u_N)\in (S^1)^N$. We parametrize these eigenvalues as
$u_i=\e^{\ii\phi_i}$ with $\phi_i\in[0,2\pi)$ for $i=1,\dots,N$. Then the
Weyl integral formula relates the measure on $G/{\rm Ad}(G)=T/W$
coming from the Haar measure on $G$ with the Haar measure on $T$, and
it modifies (\ref{Weylintalg}) to
\beq
\int_G\, \dd U\ f(U) = \frac1{N!}\, \int_{[0,2\pi)^N} \ \prod_{i=1}^N \,
\frac{\dd\phi_i}{2\pi}\ \prod_{j<k}\, 4\sin^2\Big(\,
\frac{\phi_j-\phi_k}2\, \Big) \ f(\phi_1,\dots,\phi_N)
\label{Weylintgroup}\eeq
where
$f(\phi_1,\dots,\phi_N):=f(\e^{\ii\phi_1},\dots,\e^{\ii\phi_N})$; here
the determinant of the conjugation map on $G$ is the Weyl determinant. We
now follow the same steps that led to the formula (\ref{discrmat}) from
(\ref{Weylintalg}). The path integral is given by
\beqa
Z^{(p)}_{\rm YM}(q;\Sigma_h) &=& \frac1{{\rm vol}(\cg)} \ \sum_{n\in\IZ^N} \ \int_{\ca_n} \,
\scrd\mu[A^\frh] \ \int_{\Omega^1(\Sigma_h,\cl_n\times_T\frk)} \,
  \scrd\mu[A^\frk] \nonumber\\ && \qquad\qquad \qquad \times \ 
  \int_{\Omega^0(\Sigma_h,(\IR/2\pi\, \IZ)^N)} \ \prod_{i=1}^N \,
  \scrd\mu[\phi_i] \ \bigg[\,\prod_{i<j}\, 4\sin^2\Big(\,
\frac{\phi_j-\phi_k}2\, \Big)\, \bigg] \nonumber \\ && \qquad\qquad\qquad\qquad
\qquad\qquad \qquad \qquad \qquad \times \
  \exp\big(- S^{(p)}_{\rm BF}[\phi,A^\frh,A^\frk] \big) \ , \nonumber
\eeqa
where the action (\ref{BFgaugefixed}) is now modified to
$$
S^{(p)}_{\rm BF}[\phi,A^\frh,A^\frk]=\frac1{g_s} \, \sum_{i=1}^N \ \int_{\Sigma_h} \, \Big(-\ii \phi_i\,
\dd A^\frh_i+\frac p2\, \phi_i^2\, \dd \mu\Big) +\sum_{\alpha\in{\rm Ad}(G)}\
\int_{\Sigma_h}\, \big(1-\e^{\ii\langle\alpha,\phi\rangle} \big) \, A^\frk_\alpha\wedge
A_{-\alpha}^\frk
$$
and we have rescaled $g_s\to p\, g_s$ and $\phi\to\phi/g_s$ in the original
BF-theory action (\ref{YMpartfnBFaction}). The sum over
torus bundles in this way yields the $q$-deformed discrete Gaussian matrix model (\ref{qmat}).

A similar formula for the
partition function is derived in the lattice formulation of~\cite{Buffenoir:1994fh}, wherein the
gauge algebra is taken to be the quantum universal enveloping algebra
$\cu_q(\frg)$ (see Appendix~A); invariance of the
lattice gauge fields under the coaction of $\cu_q(\frg)$ implies that
the algebra of gauge fields coincides with the
noncommutative exchange algebra of gauge fields which arises in
Hamiltonian quantization of lattice Chern-Simons theory~\cite{alekseev}. By
using a suitable $q$-analog of the usual lattice Yang-Mills-Haar
measure, the partition function for the resulting
$q$-deformation of Yang-Mills theory is given by a similar series over
$q$-weights.

In~\cite{Klimcik:1999kg}, Klim\v{c}\'{\i}k considers a general class of
``Poisson-Lie Yang-Mills theories'' which are obtained by an
isotropic gauging of the WZW model on a double
$D(G)$ of the gauge group $G$; these field theories are gauge invariant with
respect to two mutually commuting actions of $\cg$, and their zero
coupling limits are Poisson sigma-models on $\Sigma$. When
$D(G)=T^*G\cong G\rtimes\frg^*$ is the cotangent bundle of $G$, the partially gauge-fixed action is
that of the standard Yang-Mills theory (\ref{YMpartfnBFaction}) and
the associated Poisson sigma-model is BF-theory on $\Sigma$. When
$D(G)=G\times G$, the associated Poisson sigma-model is the
$G/G$ WZW model. Our main interest here is the Lu-Weinstein-Soibelman
Drinfel'd double $D(G)=G^{\IC}$, with the
complexification of $G$ regarded as a real
group. In this
case there is an extra parameter $\hbar^{-1}$ which multiplies the
invariant non-degenerate form $\Im\,\Tr$ on the Lie algebra of
$D(G)$, and we set $q:=\e^{-4\pi\, \hbar}$ (now independently of the
coupling $g_s$); this defines a one-parameter deformation of the
standard Yang-Mills theory, which is recovered in the limit
$q\to1$. The associated Poisson sigma-models now include the standard
ones whose perturbation series compute the Kontsevich formality maps
for global
deformation quantization~\cite{cattaneo}. Klim\v{c}\'{\i}k finds a suitable
extension of the diagonalization technique to
accomodate the double $D(G)=G^\IC$, which follows from a generalization
of the Weyl integral formula (\ref{Weylintgroup}) based on an alternative form of the
Cartan decomposition of elements of $G^\IC$. Proceeding as
before to write the path integral as a sum over contributions from all
torus bundles on $\Sigma_h$, one derives the combinatorial
expansion~\cite[eq.~(128)]{Klimcik:1999kg}
\beq
\mathcal{Z}_{\mathrm{K}}(g_s,q; \Sigma _{h}) =
\sum_{\lambda}\, (\dim_q\lambda)^{2-2h} \, \exp\Big(- \frac {g_s}2 \,
\sum_{i=1}^N \, \big(
[\lambda_i+\rho_i]_q^2-[\rho_i]_q^2 \big)\Big) \ .
\label{klim}\eeq
When $q=1$ this is just the heat kernel expansion (\ref{HK}) of ordinary
Yang-Mills theory. For $g_s=0$ and $q$ a root of unity, the
truncation of this series to integrable highest weights gives the
Verlinde formula.

Brzezi\'nski and Majid~\cite{brzmajid1} also consider a version of
$q$-deformed Yang-Mills theory based on a general construction of
quantum group-valued connections on quantum principal bundles with
Hopf algebra fibre. However, its quantization is not clear, as it is not obvious how to
define the path integral on a space of connections taking values in
$\cu_q(\frg)$ such that the ordinary Yang-Mills partition function is
recovered in the classical limit $q\to1$. We shall return to the issue
of properly implementing the quantum group gauge symmetry in \S\ref{se:Hilbertref}.

\section{Continuous matrix models\label{se:contmm}}

In this section we will explicitly relate the discrete matrix models
(\ref{qmat}) for
$q$-deformed Yang-Mills theory on the sphere $\Sigma_0=S^2$ to the
continuous Stieltjes-Wigert matrix models for Chern-Simons gauge
theory on the lens space $L(p,1)=S^3/\IZ_p$ for all $p\in\IZ_{>0}$.

The Stieltjes-Wigert matrix model is characterized by certain mathematical
features which are not present in classical random matrix ensembles \cite{RM}%
. In particular, it is defined by a weight function on $\IR_{>0}$ of log-normal type%
\begin{equation}
\omega _{\mathrm{SW}}( x;s) := \frac s{\sqrt\pi} \, \e^{-s \log
^{2}x} \qquad \mbox{with} \quad s\in\IR_{>0} \ , \label{SW}
\end{equation}%
which leads to an indeterminate moment problem \cite{moment}. This implies
that the Stieltjes-Wigert orthogonal polynomials that solve the Chern-Simons matrix
model \cite{Tierz} are not dense in the Hilbert space ${L}^{2}\big(\IR_{>0}\,,\,\omega_{\mathrm{SW}}
( x;s)\, \dd x
\big)$~\cite{Tierz,deHaro:2005rz}; one consequence of this feature is the exact discretization of the matrix
model~\cite%
{deHaro:2005rz}. The Stieltjes-Wigert polynomials depend on the
$q$-parameter $q=\e^{-1/2s^2}$.

Recall from (\ref{discrmat}) that the partition function of ordinary two-dimensional Yang-Mills
theory on $S^{2}$ can be expressed as a discrete Gaussian matrix model
\begin{equation}
\mathcal{Z}_{\mathrm{M}}(g_s;S^{2})=\sum_{u\in \IZ^N} \, \exp\Big(-\frac
{g_s}2\,
  \sum_{i=1}^N\, u_{i}^{2} \Big) \ \prod_{j<k}\, (u_{j}-u_{k})^{2} \ .
\nonumber
\end{equation}%
Unlike the continuous case, which is a Gaussian unitary
ensemble that is solved with
Hermite polynomials~\cite{RM}, the discrete Gaussian weight does not have a
closed system of orthogonal polynomials associated to it~\cite{Gross:1994mr}.
In fact, the large $N$ phase transition of the gauge theory is related to the
discrepancy between the discrete and continuous matrix models~\cite%
{Gross:1994mr}.

In marked contrast, the orthogonal polynomials for the discrete matrix model of
Chern-Simons theory are, just as in the continuous case, the
Stieltjes-Wigert polynomials \cite{Tierz}; this is related
to the fact that the orthogonality measure for the Stieltjes-Wigert
polynomials is not uniquely determined, and also to the absence of a
large $N$ phase transition in the $q$-deformed gauge theory in this case~\cite{Caporaso:2005ta}. We demonstrate this in
detail below, showing explicitly the equivalence between the
discrete and continuous versions of the Chern-Simons matrix
model. This calculation fills in
the details of the derivation in~\cite[eqs.~(25)--(26)]{deHaro:2005rz} and gives explicitly the corresponding
normalization constants. It specifically relates $q$-deformed Yang-Mills
theory on $S^{2}$ for $p=1$, which is characterized by a discrete matrix model, with
Chern-Simons theory on $S^{3}$, which was originally described by a continuous matrix
model \cite{Marino:2002fk}. The moment problem will also be employed
below to study the analogous relationship in the
case $p\neq 1$; we show that in this instance the discrete matrix
model is related to certain correlators in the Stieltjes-Wigert
ensemble. We further comment on some aspects of the equivalent representation of
the Stieltjes-Wigert matrix model as a unitary matrix model, as this
correspondence will be exploited in our later considerations; in
contrast to the weight functions defined on $\IR$, weight functions on
the unit circle $S^1$ are always moment-determined~\cite{RM}.

\medskip

\subsection{$L(1,1) $ matrix model}\label{se:p=1}~\\[5pt]
We begin with the discrete matrix model (\ref{qmat}) for the partition function
(\ref{qYM1}) on $\Sigma_0=S^2$ for $p=1$ and rewrite it as a
continuous matrix model. We have
\begin{align}
\mathcal{Z}_{\mathrm{M}}^{(1) }(q; S^2)& =
\sum_{u \in\IZ^N}\, \exp\Big(-%
\frac {g_s}2\, \sum_{i=1}^N \, u_{i}^{2}\Big) \ \prod_{j<k}\, 4\, \sinh
^{2}\Big(\, \frac {g_s}2\, (u_{j}-u_{k}) \, \Big) \notag \\[4pt]
& =\sum_{u\in\IZ^N}\, \exp\Big(-\frac {g_s}2\, \sum_{i=1}^N \, u_{i}^{2}
\Big) \,
\exp\Big( (N-1)\, g_s\, \sum_{i=1}^N \, u_{i} \Big) \ \prod_{j<k}\, \left(
\e^{-g_s\, u_{j}}-\e^{-g_s\, u_{k}}\right) ^{2}  \notag \\[4pt]
& =\sum_{u \in\IZ^N}\, q^{\frac12\, \sum_{i}\, u_{i}^{2}}\, (\sigma \, q)^{\sum_{i}\,
  u_{i}} \ \prod_{j<k}\, \left(
q^{u_{j}}-q^{u_{k}}\right) ^{2} \notag
\end{align}
where we have introduced $\sigma:= \e^{N\, g_s}=q^{-N}$ with $q:=
\e^{-g_s}$ as usual. This gives
\begin{align}
\mathcal{Z}_{\mathrm{M}}^{(1) }(q; S^2) & =\sum_{u\in\IZ^N}
\ \prod_{i=1}^{N}\,\sigma^{u_{i}}\,q^{\frac12\, u_{i}^{2}+u_{i}}
\ \prod_{j<k}\,
\left( q^{u_{j}}-q^{u_{k}}\right) ^{2}  \notag \\[4pt]
& =\sigma^{N\, (1-N)}\, \int_{\IR_{>0}^N} \ \prod_{i=1}^N \, \mathrm{d}x_{i} \ \sum_{n_i=-\infty }^{\infty
}\, \sigma^{n_i}\,q^{\frac12\, n_i^{2}+n_i}\, \delta (x_i-\sigma \, q^{n_i}) \ \prod_{j<k}\, \left(
x_{j}-x_{k}\right) ^{2}  \notag \\[4pt]
& =\sigma^{N\,(1-N)}\, q^{N/2} \, M(q,\sigma)^{N}\, 
\int_{\IR_{>0}^N } \ \prod_{i=1}^N \, 
\mathrm{d}x_{i} \ w_{d}(x_{i};q,\sigma) \ \prod_{j<k}\, \left(
  x_{j}-x_{k}\right) ^{2} \ ,
\label{ZM1comp}\end{align}
where the normalization $M(q,\sigma)$ has a triple product form%
\begin{equation}
M(q,\sigma):= \big(-\sigma \, q^{3/2}\,;\,q\big)_{\infty
}\, (-\sigma^{-1}\, q^{-1/2}\,;\, q \big)_{\infty }\, \big(q\,;\, q\big)_{\infty } \ ,
\end{equation}%
and for $a,q\in\IC$ with $|q|<1$ and $k\in\IZ_{>0}\cup\{\infty\}$ we
use the standard hypergeometric notation for the
$q$-shifted factorial
\beq
(a;q)_k := \prod_{n=0}^{k-1} \, \big(1-a\, q^n\big) \qquad \mbox{and}
\qquad (a;q)_0:=1 \ .
\nonumber
\eeq
In (\ref{ZM1comp}) we have introduced the two-parameter family of
discrete measures on $\IR_{>0}$ given by
\begin{equation}
w_{d}(x;q,\sigma):= {\frac{1}{{\sqrt q}\,M(q,\sigma)}} \ \sum_{n=-\infty }^{\infty
}\,\sigma^{n}\,q^{\frac12\, n^{2}+n}\, \delta (x-\sigma\, q^{n}) \ .
\label{discretemeas}\end{equation}
The family $\big(
w_{d}(x;q,\sigma)\big) _{\sigma>0}$ is completely determined by the values
$\sigma \in
\left( q,1\right] $, and translation invariance of the sums leads to
the discrete scaling symmetry~\cite%
{Berg}%
\begin{equation}
M(q,q\, \sigma)=\frac{M(q,\sigma)}{\sigma \, \sqrt{q}} \qquad \mbox{and} \qquad
w_{d}(x;q,q\, \sigma)=w_{d}(x;q,\sigma) \ .
\label{wdtransl}\end{equation}%

The indeterminacy
of the moment problem implies that the discrete measure
(\ref{discretemeas}) is equivalent to the
continuous distribution (\ref{SW}%
)~\cite{Chi,Chi2,Christ}. Note that even though there are infinitely many discrete measures $w_{d}(x;q,\sigma)
$ which are equivalent to $\omega _{\mathrm{SW}}( x;\sigma) $, in
(\ref{ZM1comp}) we take  $\sigma=q^{-N}$, which by (\ref{wdtransl}) is
equivalent to the choice $\sigma=1$. The equivalence between $w_{d}(x;q,1)$ and $\omega _{\mathrm{SW}}\big(
x;\frac1{\sqrt{2g_s}}\big) $ enables us to write%
\begin{equation}
\mathcal{Z}_{\mathrm{M}}^{(1) }(q; S^2)=\sigma^{N\, (1-N)}\,
q^{N/2}\, M(q,\sigma)^{N}\, \int_{\IR_{>0}^N} \ \prod_{i=1}^N \,
\mathrm{d}x_{i} \ \omega
_{\mathrm{SW}}\big(x_{i}\,;\, \mbox{$\frac1{\sqrt{2g_s}}$} \big) \
\prod_{j<k}\, \left(
x_{j}-x_{k}\right) ^{2} \ ,
\label{SWCSS3}\end{equation}%
which up to overall normalization is the partition function $Z_N(q)$ for Chern-Simons
gauge theory on the three-sphere $S^3$~\cite{Tierz}.

In this way one arrives at a rather simple relation between the discrete and continuous
Stieltjes-Wigert ensembles given by%
\beqa
Z_N(q) &:=& \Big(\, \frac{g_s}{2\pi }\, \Big) ^{-N/2}\ 
\int_{\IR^N } \ \prod_{i=1}^N\, 
{\frac{\mbox{d}u_{i}}{2\pi }} \ \e^{-u_{i}^{2}/2g_s} \ \prod_{j<k}\,
4\, \sinh^2 \Big(\,\frac{u_{j}-u_{k}}2\,\Big) \label{SWdisccont} \\[4pt]
&=& \Big( \,
{\frac{q^{-\frac12\, (1-2N+3N^{2})}}{\big(-q^{\frac32-N}\,;q \,\big)_{\infty
}\, \big(-q^{N-\frac12}\,;\,q \big)_{\infty }\, \big(q\,;\,q\big)_{\infty }}}\,
\Big) ^{N}\nonumber \\ && \times\
\sum_{n \in\IZ^N }\, \exp\Big( -\frac {g_s}2\, \sum_{i=1}^N \,
  n_{i}^{2} \Big) \ 
\prod_{j<k}\, 4\, \sinh^2 \Big(\, \frac {g_s}2\, (n_{j}-n_{k}) \, 
\Big) \nonumber 
\eeqa
where $u_i=\log x_i$. This relates the matrix model that arises in $q$-deformed Yang-Mills theory
on $S^{2}$ with $p=1$ (in the second line of (\ref{SWdisccont})) to the matrix model in Chern-Simons theory on $S^{3}$
(in the first line of (\ref{SWdisccont})). It strengthens the analogous relationship
derived in~\cite{Szabo:2010qv} where the integration in
(\ref{SWdisccont}) was interpreted as a Jackson $q$-integral.

\medskip

\subsection{$L(p,1) $ matrix model}\label{se:pinN}~\\[5pt]
We now consider the case of general degree $p\in \IZ_{>0}$. Then the discrete matrix model (\ref{qmat}) for $q$-deformed Yang-Mills theory
on $S^{2}$ is given by
\begin{equation}
\mathcal{Z}_{\mathrm{M}}^{(p) }(q; S^2)=\sum_{u\in 
\mathbb{Z}^N
}\, \exp\Big( -\frac{p\, g_s}2\, \sum_{i=1}^N \, u_{i}^{2}\Big) \ \prod_{j<k}%
\, 4\, \sinh^2 \Big(\,\frac {g_s}2\, (u_{j}-u_{k}) \,\Big) \ .
\label{discrete-p}
\end{equation}%
In the continuous limit $g_s\rightarrow 0$ with the rescaling
$u_i\to g_s \, u_i$, this expression becomes%
\begin{equation}
\lim_{q\to1}\, \mathcal{Z}_{\mathrm{M}}^{(p) }(q; S^2)
=\int_{\IR^N} \ \prod_{i=1}^N \, \dd u_{i} \
\e^{-\frac p{2g_s} \, u_{i}^{2}} \ \prod_{j<k} \, 4\, \sinh^2 \Big(\,
\frac{u_{j}-u_{k}}2\, \Big) \ .  \label{cont}
\end{equation}%
By (\ref{SWdisccont}) the model defined by the partition function (\ref{cont}) is
equivalent to the discrete matrix model%
\begin{equation*}
{\mathcal{Z}}_{\mathrm{SW}}^{(p) }(q;
S^2):=\sum_{v \in 
\mathbb{Z}^N
}\, \exp\Big( -{\frac{g_s}{2p}}\, \sum_{i=1}^N \, v_{i}^{2}\Big) \ \prod_{j<k}\,
4\, \sinh^2 \Big( \, {\frac{g_s}{2p}}\, (v_{j}-v_{k})\, \Big) \ .
\end{equation*}%
By making the change of variables $v_{i}=p\, u_{i}$ we obtain%
\begin{equation}
{\mathcal{Z}}_{\mathrm{SW}}^{(p) }(q;
S^2) =\sum_{u\in 
(\mathbb{Z}
/p)^N}\, \exp\Big( -\frac{p\, g_s}2 \, \sum_{i=1}^N \, u_{i}^{2} \Big) \ \prod_{j<k}\, 4\,\sinh^2 
\Big(\, \frac {g_s}2\, (u_{j}-u_{k}) \, \Big) \ .
\label{Ztot}
\end{equation}%
This shows that, when $p> 1$, the $q$-deformed gauge theory is actually a subsector of the
discrete version of the Stieltjes-Wigert ensemble. This
observation suggests a decomposition into sectors
${\mathcal{Z}}_{\mathrm{SW}}^{(p) }[n](q;
S^2)$ labelled by elements ${n}_i=0,1,\dots,p-1$ for $i=1,\dots,N$ of the cyclic
group $\IZ_p$ in each of which the summation
variables are restricted to $u_{i}\in 
\mathbb{Z}
+\frac{{n}_i}p$. Then
\begin{equation*}
{\mathcal{Z}}_{\mathrm{SW}}^{(p) }(q;
S^2) =\sum_{n\in 
\mathbb{Z}_p^N
}\, {\mathcal{Z}}_{\mathrm{SW}}^{(p) } [n](q;
S^2) \ .
\end{equation*}%
The different torsion sectors ${n}_i$ are interlaced with
one another through the hyperbolic sine function in (\ref{Ztot}); the model (\ref%
{discrete-p}) is recovered as the trivial sector ${n}_i=0$ for $i=1,\dots,N$.
Therefore, to study the discrepancy between the continuous matrix model (\ref%
{cont}) and the discrete matrix model (\ref{discrete-p}), we will write the latter as
a projection from the full ensemble $$\mathcal{Z}_{\mathrm{M}}^{(p) }(q; S^2)= {\mathcal{Z}}_{\mathrm{SW}}^{(p) }[
0 ](q;
S^2) \ . $$

For this, we assume that $q=\zeta_k$ is a $k$-th root of unity, so
that $q^{k}=1$ (this is, in particular, the case relevant for the
connection with Chern-Simons gauge theory). Consider the polynomial%
\begin{equation}
\Gamma _{k,p}(z)=\frac{z^{k\, p}-1}{z^{k}-1}= \sum_{m=0}^{p-1}\,
z^{k\, m} \ .  \label{pol}
\end{equation}%
This is an equally spaced polynomial which arises in the study of
finite fields~\cite{ESP} (here $\IZ_p$); for $k=1$ it is known as
the ``all in one polynomial''. One then
has
\begin{equation*}
\Gamma _{k,p}(q^{u})=p\ \delta _{n,0} \qquad \mbox{for} \quad
u\in\IZ+\mbox{$\frac np$} \ .
\end{equation*}%
Thus by introducing the multivariable polynomial extension of
(\ref{pol}) given by
\begin{equation*}
\Gamma _{k,p}(z_{1},\dots ,z_{N})=\frac{1}{p^{N}} \ \prod_{i=1}^{N}\,
\Gamma _{k,p}(z_{i}) \ ,
\end{equation*}%
we obtain%
\begin{equation*}
\mathcal{Z}_{\mathrm{SW}}^{(p) }[
0 ](q=\zeta_k ;
S^2) =\big\langle \Gamma
_{k,p}(q^{u_{1}},\dots,q^{u_{N}})\big\rangle _{\rm SW}
\end{equation*}%
where the average is taken in the full discrete Stieltjes-Wigert
matrix model (\ref{Ztot}%
). Since $\Gamma_{k,p}(z_{1},\dots ,z_{N}) $ is a polynomial, only
integer moments are involved in computing this correlator and
the discrete/continuous equivalence, due to the indeterminate moment problem,
applies to it as well. Hence we can write the average in the continuous
matrix model to get
\begin{equation*}
\mathcal{Z}_{\mathrm{SW}}^{(p) }[
0 ](q=\zeta_k;
S^2) = \int_{\IR^N} \ \prod_{i=1}^N \, \dd u_{i} \ \e^{-%
{\frac{p}{2g_s}}\, u_{i}^{2}} \ \frac{\e^{k\, p\, u_{i}}-1}{\e%
^{k \, u_{i}}-1} \ \prod_{j<k}\, 4\, \sinh^2
\Big(\,\frac{u_{j}-u_{k}}2\, \Big) \ ,
\end{equation*}%
or alternatively by setting $u_i=\log x_i$ we may write
\beq
\mathcal{Z}_{\mathrm{SW}}^{(p) }[
0 ](q=\zeta_k ;
S^2) = \int_{\IR_{>0}^N } \ \prod_{i=1}^N \,
\mathrm{d}x_{i} \ \omega
_{\mathrm{SW}}\big(x_{i}\,;\,\mbox{$\sqrt{\frac p{2g_s}}$} \, \big) \, \Gamma_{k,p}(x_i) \
\prod_{j<k}\, \left(
x_{j}-x_{k}\right) ^{2} \ . \nonumber
\eeq
For $p=1$ these expressions respectively recover the continuous Stieltjes-Wigert matrix model
representations (\ref{SWdisccont}) and (\ref{SWCSS3}).

\medskip

\subsection{Unitary matrix models}\label{se:unitarymm}~\\[5pt]
For later reference, we briefly discuss the unitary matrix model
that describes the $U(N)$ Chern-Simons gauge theory on
$S^3$~\cite{Okuda:2004mb}. It is given by the partition function
\beq
Z_N(q) = \int_{[0,2\pi)^N} \ \prod_{i=1}^N \, \frac{\dd\phi_i}{2\pi} \
\Theta\big(\e^{\ii\phi_i}\,;\, q\big) \ \prod_{j<k}\,
\big|\e^{\ii\phi_j}-\e^{\ii\phi_k}\big|^2 \ ,
\label{UCS}\eeq
where the weight function of the matrix model is the Jacobi elliptic function
\beq
\Theta(z;q):=\sum_{n=-\infty}^\infty\, q^{n^2/2} \, z^n \
. \label{theta3zq}
\eeq
This theta-function can be written in a product form by using the
Jacobi triple product identity
\beq
\Theta(z;q) = \big(q\,;\,q\big)_\infty\, \big(\sqrt q\, z\,;\,
q\big)_\infty\, \big(\sqrt q\, z^{-1}\,;\, q\big)_\infty \ .
\label{triple}\eeq
By the Heine-Szeg\H{o} identity (see Appendix~B), one can write the
unitary matrix integral (\ref{UCS}) as a Toeplitz determinant and
derive the exact analytical expression~\cite{Szabo:2010sd}
\beq
Z_N(q)= \prod_{k=1}^{N-1}\, \big(1-q^{k}\big)^{N-k}
\label{UCSexact}\eeq
for the partition function (\ref{SWdisccont}) of the
Stieltjes-Wigert matrix model.

The $U(N)$ Chern-Simons theory has a dual formulation as a
$U(\infty)$ matrix model with the product form (\ref{triple}) of the theta-function
truncated at $N$: It can be written as
\beq
Z_N(q)= (q;q)_\infty^N\ \int_{[0,2\pi)^\infty} \ \prod_{i=1}^\infty \, \frac{\dd\phi_i}{2\pi} \
\frac{\Theta_N\big(\e^{\ii\phi_i}\,;\, q\big)}{(q;q)_N} \ \prod_{j<k}\,
\big|\e^{\ii\phi_j}-\e^{\ii\phi_k}\big|^2 \ ,
\label{UCStrunc}\eeq
where the infinite-dimensional integration is defined as the
$N\to\infty$ limit of the finite $N$ eigenvalue model (see Appendix~B
for a discussion of the related convergence issues); the weight function of the matrix model is now a truncated
theta-function which satisfies a finite version of the Jacobi triple
product identity (see e.g.~\cite{Foata})
\begin{eqnarray}
&& \Theta _{N}(z;q)
=(q;q)_N\, \sum_{n=-N}^{N}\ \left[\begin{matrix}2N \\ n+N \end{matrix} \right]_{q}\, q^{n^{2}/2}\, z^{n}
= \big(q\,;\,q\big)_N\, \big(\sqrt q\, z\,;q\big)_N\, \big(\sqrt q\,
z^{-1}\,;\, q\big)_N  \label{trunc}
\end{eqnarray}
and
$$
\left[\begin{matrix}n \\ k\end{matrix}\right]_q:= \frac{(q;q)_n}{(q;q)_k\, (q;q)_{n-k}}
$$
is the $q$-binomial coefficient defined for $n,k\in\IZ_{\geq0}$ with
$n\geq k$; the weights $(q;q)_n$ here have an algebraic
interpretation as the polynomials which compute the number of flags in
an $n$-dimensional vector space over a field with $q$ elements. To prove the formula (\ref{UCStrunc}), we use the
Cauchy-Binet formula for the normalization of the Schur measure~\cite{Macdonald}
\beq
\sum_\lambda\, s_\lambda(x)\, s_\lambda(y)=\prod_{i,j\geq1} \,
\frac1{1-x_i\, y_j} \ ,
\label{SC}\eeq
together with the Gessel
identity (see e.g.~\cite{Szabo:2010sd}) which writes the left-hand
side of (\ref{SC}) as a Toeplitz
determinant. This gives
\begin{equation*}
\prod\limits_{i,j=1}^{N}\,\frac{1}{1-x_{i}\,y_{j}}= \int_{[0,2\pi)^\infty
} \ \prod_{i=1}^\infty \,\frac{\mathrm{d}\phi _{i}}{2\pi } \ 
\prod_{j=1}^{N}\, \big( 1+x_{j}\, \e^{\ii\phi_i}\big)\, \big(
1+y_{j}\, \e^{-\ii\phi_i} \big)
~\prod\limits_{k<l}\,\big\vert\e^{\ii\phi_k}-\e^{\ii\phi_l}\big\vert^{2}
\ .
\end{equation*}%
At the principal specialization $x_{i}=y_{i}=q^{i-\frac12}$ for $%
i=1,\dots,N$, this identity shows that (\ref{UCStrunc}) is equal to
(\ref{UCSexact}).

This result can also be checked by explicit computation, using the Selberg
integral: The $U(k)$ version of the matrix integral (\ref{UCStrunc})
can be computed as~\cite{Mat} 
\begin{eqnarray}
&& \int_{[0,2\pi)^k} \ \prod_{i=1}^k \, \frac{\dd\phi_i}{2\pi} \
\frac{\Theta_n\big(\e^{\ii\phi_i}\,;\, q\big)}{(q;q)_n} \ \prod_{j<k}\,
\big|\e^{\ii\phi_j}-\e^{\ii\phi_k}\big|^2 \nonumber \\
&& \qquad \qquad \qquad \ = \ \prod_{i=0}^{k-1}\, \frac{(q;q)_{i+2n}\, (q;q)_i}{(q;q)_{i+n}^2}
 \ = \ \prod_{j=0}^{n-1}\, \frac{(q;q)_{j}\,
  (q;q)_{k+j+n}}{(q;q)_{j+n}\, (q;q)_{k+j}} \ . \label{Selberg}
\end{eqnarray}%
The two equivalent expressions in (\ref{Selberg}) make manifest the
duality described above: Using the property
$(q;q)_n=(1-q^{n-1})\, (q;q)_{n-1}$, with either the limit
$k\to\infty$, $n=N$ or the dual limit $k=N$, $n\to\infty$ we arrive at
the partition function for $U(N)$ Chern-Simons theory on $S^3$.

The principal specialization of the Cauchy-Binet formula (\ref{SC})
also demonstrates that, up to area-dependent renormalization, the
topological limit $p=0$ of the partition function (\ref{qYM1}) for
$q$-deformed Yang-Mills theory on $S^2$ is related to the unitary
matrix models discussed here via
\beq
\cz^{(0)}(q;S^2)=\sum_\lambda\,
(\dim_q\lambda)^2=\frac{Z_N(q)}{(q;q)_\infty^N} \ .
\label{BFS2N}\eeq

Notice that while the consideration of the q-Pochammer symbol piece in (\ref{triple}) in the corresponding 
unitary matrix model leads to the Chern-Simons partition function, its absence in the matrix model gives 
rise to the MacMahon function (see \S4.4 for more details). This was discussed in \cite{Szabo:2010sd} and it 
corresponds to the relationship between the partition functions of Chern-Simons theory on $S^3$ and 
Donaldson-Thomas theory on $\IC^3$. This type of relationship has later on emerged again in the study of 
five-dimensional maximally supersymmetric Yang-Mills theory on $S^5$ \cite{Kim:2012ava,Kim:2012qf}, where it was found that 
the perturbative partition function of the Yang-Mills theory is given by the partition function of Chern-Simons 
theory on $S^3$.

\section{Classification\label{se:Class}}

In this section we analyse the effect of a full $q$-deformation of the
Migdal partition function (\ref{HK}), which is a simple and natural
modification from the perspective of quantum group theory. By ``full $q$-deformation'' we mean a $q$-deformation of the
dimensions, of the Casimir operator, and of the exponential function,
the three main ingredients that figure into the combinatorial
formula (\ref{HK}). This problem is partly inspired by the form of
Klim\v{c}\'{\i}k's partition function (\ref{klim}), which involves
$q$-deformed Casimir eigenvalues $[C_2(\lambda)]_q$ and is obtained
from the partition function (\ref{qYM1}) by a full $q$-deformation
of the heat kernel action. Another source of inspiration comes from
the observation of~\cite{deHRT} that the generic usage of quantum
dimensions in the combinatorial quantization of two-dimensional
Yang-Mills theory necessitates a modification of the usual Migdal
gluing rules.

We shall now argue that (\ref{qYM1}) is essentially equivalent to a full $q$%
-deformation, and that the analogous combinatorial expansion involving a quantum dimension together with $%
q $-deformed Casimir invariants, as in (\ref{klim}), yields ordinary two-dimensional
Yang-Mills theory. We will also classify the qualitative effects of partial $q$%
-deformations of the partition function (\ref{HK}). For example,
we shall find that incorporating a $q$-exponential function instead of an ordinary
Boltzmann weight in the
heat kernel expansion has a very similar effect to using quantum dimensions instead of
ordinary dimensions; we suggest a geometric interpretation of the
gauge theory involving $q$-exponentials on a quantum torus.

In \S\ref{se:crystal} we give a concrete application of these equivalences
to the problem of crystal melting with an
external potential~\cite{Nakatsu:2007dk}, whose partition function is
the generating function for correlators in a five-dimensional $\cn=1$
supersymmetric gauge theory; it can be interpreted in terms of $q$-deformed two-dimensional
BF-theory and certain supersymmetric extensions. We will show that the
potentials considered in this model can be regarded as $q$-deformed Casimir operators, and then apply the
analysis of \S\ref{se:dimCasdef} which shows that the heat kernel
expansion involving quantum dimensions together with
a $q$-deformed Casimir operator is essentially equivalent to ordinary
generalized two-dimensional Yang-Mills
theory~\cite{Douglas:1994pq,Ganor:1994bq} which is defined by the
partition function
\beq
\mathcal{Z}_{\rm M}^{\rm gen}(g_s,t; \Sigma _{h}) = \sum_{\lambda }\,\left(
\dim \lambda \right) ^{2-2h}\,\exp \Big(-\frac {g_s}2\, \sum_{k=1}^\infty\,
t_k\, C_{k}(\lambda )\Big) \ ,
\label{genYMHK}\eeq
where $t=(t_1,t_2,\dots)$ is a set of coupling constants and
$$
C_k(\lambda)= \sum_{i=1}^N \, (\lambda_i-i+1)^k \ \prod_{j\neq i}\, \Big(1-\frac1{\lambda_i-\lambda_j+j-i}\, \Big)
$$ is the $k$-th Casimir operator eigenvalue in the
representation $\lambda$. The addition of higher Casimir operators in
(\ref{genYMHK}) corresponds to the deformation of the BF-theory type
action (\ref{YMpartfnBFaction}) by adding the operators $\sum_k\,
t_k\, \Tr\big( \phi^k \big)$ in the ring of invariant polynomials $\mathfrak{S}(\frg^*)^G$
on the Lie algebra $\frg$; we suggest a geometric interpretation of
the gauge theory involving $q$-Casimirs on a quantum sphere. We
further demonstrate that this gives a new way of relating these gauge theories to unitary matrix models
which are large $N$ limits of the unitary matrix models for $U(N)$
Chern-Simons gauge theory that we considered in~\S\ref{se:unitarymm}.

\medskip

\subsection{Klim\v{c}\'{\i}k deformations}\label{se:dimCasdef}~\\[5pt]
While it is possible to work directly with the heat kernel expansions,
in this section we shall find it more convenient to use the partition
functions of the associated discrete matrix models. The combinatorial
series with both quantum dimensions and a $q$-deformed
Casimir invariant leads to a matrix model of the form%
\begin{equation}
\mathcal{Z}_{\rm K}(g_s,q; \Sigma _{h}) =\sum_{n\in 
\mathbb{Z}^N
} \ \prod\limits_{i<j}\, \left( q^{n_{i}}-q^{n_{j}}\right) ^{2-2h}\, \exp
\Big( -\frac{p\, g_s}2 \, \sum_{i=1}^{N}\, \left[ n_{i}\right]
_{q}^{2}\Big) \ ,
\label{mm}
\end{equation}%
which is the type of matrix model that follows from (\ref{klim}). In analogy with the solution of the
Chern-Simons matrix model~\cite{Marino:2002fk} in terms of a
Stieltjes-Wigert matrix model~\cite{Tierz}, we consider the change of
variables%
\begin{equation}
n_{i}=\frac{\log(m_{i}+1)}{\log q} \label{map}
\end{equation}%
for $i=1,\dots,N$ which maps the $q$-deformed Vandermonde determinant into the standard one%
\begin{equation*}
\prod\limits_{i<j}\, \left( q^{n_{i}}-q^{n_{j}}\right)
^{2-2h}=\prod\limits_{i<j}\, \left( m_{i}-m_{j}\right) ^{2-2h} \ ,
\end{equation*}%
and also the $q$-deformed Gaussian potential becomes the standard one%
\begin{equation*}
\sum_{i=1}^{N} \, \left[ n_{i}\right] _{q}^{2}=\frac{1}{\left( 1-q\right) ^{2}}%
\, \sum_{i=1}^{N}\, m_{i}^{2}
\end{equation*}%
where here we have used the asymmetric $q$-number $\left[ n\right]
_{q}=( 1-q^{n}) /( 1-q )$. Then the partition
function (\ref{mm}) becomes
\begin{equation}
\mathcal{Z}_{\rm K}(\tilde g_s; \Sigma _{h})= \sum_{m\in 
\mathbb{Z}_{\geq0}^N} \ \prod\limits_{i<j}\, \left( m_{i}-m_{j}\right)
^{2-2h} \, \exp\Big(- \frac{p\, \tilde g_s}2 \, \sum_{i=1}^N\, m_{i}^{2}\Big) \ , \label{Gauss+}
\end{equation}%
which is essentially the discrete Gaussian matrix model that
describes ordinary Yang-Mills theory on the Riemann surface $\Sigma_h
$ with a renormalized coupling constant
\beq
\tilde g_s=\frac{g_s}{(1-q)^2} \ . \nonumber
\eeq

Recall from \S\ref{se:contmm} that without $q$-deformation of the
Casimir eigenvalues, the relevant matrix model at genus zero is
a discrete Hermitian Stieltjes-Wigert ensemble, which is equivalent to the continuous Stieltjes-Wigert matrix model that describes
Chern-Simons gauge theory on $S^{3}$. In addition, due to the change of variables (\ref%
{map}), the range of the eigenvalues in the resulting
Gaussian matrix model (\ref{Gauss+}) is restricted to the positive
weight lattice $\IZ_{\geq0}^N$, similarly to the domain of the weight
function for the
Stieltjes-Wigert ensemble (\ref{SW}). However, the summand of the matrix
model is invariant under Weyl reflections of $m_i$,
so that the
summations can be extended from the Weyl alcove to $m\in \IZ^N$; this
situation also arises when the unitary gauge group is replaced by orthogonal or symplectic
groups, and the same extension to
the whole weight lattice is carried out in that case in~\cite{Crescimanno:1996hx}.
The important feature here is that
the resulting potential is not of
the Stieltjes-Wigert type $\log ^{2}x$, which as we have
seen corresponds to $q$%
-deformed Yang-Mills theory. The latter model is uniquely characterised by
the self-similarity property of its discrete orthogonality measure in
(\ref{wdtransl}), known as the $q$-Pearson equation~\cite{deHaro:2005rz}.

The $q$-deformed
Vandermonde determinant in (\ref{mm}) is directly related to the more common form
used in (\ref{discrete-p}), which is also the form that follows from
(\ref{qmat}), via
\begin{equation*}
\prod\limits_{i<j}\, \left( q^{n_{i}}-q^{n_{j}}\right)
^{2-2h}=\prod\limits_{i=1}^{N}\, q^{(N-1)\, (1-h)\, n_{i}} \
\prod\limits_{i<j}\, \bigg(
2\, \sinh \Big(\, \frac {g_s}2\, (n_{i}-n_{j})\, \Big) \bigg)^{2-2h} \ .
\end{equation*}%
Hence depending on the particular $q$-deformation employed in the dimensions and in
the Casimir invariants, the compensation may not be exact, in which case the
resulting discrete matrix model will
not be exactly Gaussian. However, it will in any case always be well described by a
polynomial potential; since any polynomial in the representation
weights $\lambda_i$ can be written as a linear combination of Casimir
invariants $C_k(\lambda)$, the matrix model will correspond to a deformation via
the addition of
higher Casimir operators in ordinary (generalized) two-dimensional Yang-Mills
theory (\ref{genYMHK}).

We conclude that $q$-deformation of both the dimensions and the
Casimir operators together compensate
each other, and the resulting heat kernel expansion is equivalent to
its undeformed version, i.e. to that of ordinary (generalized) two-dimensional
Yang-Mills theory.

This calculation also illustrates the effect of having only the Casimir
eigenvalues $q$-deformed in (\ref{mm}). With $q=\e^{-g_s}$ it leads to a matrix model with an exponential potential%
\begin{equation}
\mathcal{Z}_{\mathrm{Cas}}(q; \Sigma _{h}) =\sum_{n \in 
\mathbb{Z}^N
} \ \prod\limits_{i<j}\, \left( n_{i}-n_{j}\right) ^{2-2h}\, \exp
\Big( -\frac{p\,\tilde g_s}2 \, \sum_{i=1}^{N}\,\big( 1-\e^{-g_s \, n_i} \big)
^{2}\Big) \ .
\label{Casmat}\end{equation}

\medskip

\subsection{Quantum torus deformations}\label{se:expCasdef}~\\[5pt]
We will now demonstrate that a similar cancellation occurs when we use a $q$%
-deformation of the exponential function together with a $q$-deformed
Casimir operator in the heat kernel expansion. The relevant matrix
model is defined by the partition function
\beq
\mathcal{Z}_{\rm Exp}(g_s,q; \Sigma _{h}) =\sum_{n\in 
\mathbb{Z}^N
} \ \prod\limits_{i<j}\, \left( {n_{i}}-{n_{j}}\right) ^{2-2h}\, \Exp_q
\Big( -\frac{p\, g_s}2 \, \sum_{i=1}^{N}\, \left[ n_{i}\right]
_{q}^{2}\Big) \ ,
\label{Expmat}\eeq
where the $q$-deformed exponential function is defined by
\beq
\Exp_{q}(z) :={\left( -z;q\right) _{\infty }} =
\sum_{n=0}^\infty\, \frac{q^{n\,(n-1)/2}\, z^n}{(q;q)_n} \nonumber
\eeq
for $z\in\IC$.
It has an asymptotic expansion given by~\cite{asym}
\begin{eqnarray*}
\Exp_{q}(x) &=& \frac{1}{\big(
-q\, x^{-1} \,;\, x\big) _{\infty }}\, \exp \Big( \mbox{$\frac{1}{2}$}\,
\log x-\frac{1}{\log q}\, \big( 
\mbox{$\frac{\pi ^{2}}{6}$}+\log^{2}x\big) -\mbox{$\frac1{12}$}\, {\log q}\Big) \\
&& \qquad \qquad \qquad \qquad \times\, \exp \bigg(\,
\sum_{k=1}^{\infty }\,\frac{\cos \big( 2\pi\, k \log
x/\log q \big)}{k \sinh \left( \pi ^{2}\, k/\log q\right) }\, \bigg)
\end{eqnarray*}%
for $x\in\IR_{>0}$, from which it follows that the $q$-exponential series has an asymptotic behaviour $\Exp_{q}(x)\sim \e^{-\log ^{2}x}$ for $x\to\infty$. An analogous result holds for the dual $q$%
-exponential
\beq
\exp_{q}(z):=\frac1{\left( z;q\right) _{\infty }} = \sum_{n=0}^\infty\,
\frac{z^n}{(q;q)_n} = \Exp_{q}(-z)^{-1}
\label{dualqexp}\eeq
defined for $|z|<1$; for $x$ varying inside compact subsets of $\IR$,
both $\Exp_q\big((1-q)\, x\big)$ and $\exp_q\big((1-q)\, x\big)$
converge uniformly to $\e^x$ as $q\to1$. Hence we find that using a $q$-exponential function for the
Boltzmann weight of the theory leads rather directly, without any changes
of variables, to a matrix model of the Stieltjes-Wigert type.

This result
agrees with the fact that the log-normal distribution (\ref{SW}) is
equivalent to the weight function~\cite{Szabo:2010qv}
\beq
w_{\Exp}(x;q) = \Exp_{q^{-1}}(q\, x)\,
\exp_{q}\big(-(q\,x)^{-1}\big) \ . \nonumber
\eeq
It is also
analogous to the equivalent formulation of the Chern-Simons matrix
model as a unitary matrix integral (\ref{UCS}) whose weight function is the
Jacobi elliptic function (\ref{theta3zq}).
This theta-function is itself a $q$-exponential series, and by the
Jacobi triple product identity (\ref{triple}) it can be written in terms of the
$q$-exponentials (\ref{dualqexp}) as~\cite{asym}%
\begin{eqnarray}
\frac{\Theta( x;q) }{\left( q;q\right) _{\infty }} &=&\exp_{q}\big(-%
\sqrt{q}\, x\big) \, \exp_{q}\big(-\sqrt{q}\, x^{-1} \big) \notag
 \\[4pt]
&=&\exp \bigg( -\frac{1}{\log q}\, \big( \mbox{$\frac{\pi ^{2}}{12}$} +\log^{2}x\big) +%
\mbox{$\frac1{12}$}\, {\log q}+\sum_{k=1}^{\infty }\, (-1)^{k}\, \frac{\cos
\big( 2\pi\, k \log x/\log q \big)}{k\sinh \left( \pi ^{2}\, k/\log
    q\right) }\, \bigg)
\label{theta} \end{eqnarray}%
for $x\in\IR_{>0}$. The infinite series in (\ref{theta}) is a
$q$-periodic function which has no effect on the Stieltjes-Wigert matrix model, again due to the moment
problem~\cite{deHaro:2005rz}.

On the other hand, the $q$-deformation of the Casimir operator in
(\ref{Expmat}) essentially
consists in using the $q$-numbers $\left[
n_i \right] _{q}$, and hence in the weights $\Exp_q\big([n_i]_q^2 \big)$ the $q$%
-deformation of the exponential function undoes the $q$-deformation of
its argument, leaving an ordinary Boltzmann weight. Let us
look more closely at an explicit example of such an exact cancellation. Consider the $q$%
-deformation of the Gaussian distribution given by the orthogonality
measure for the continuous $q$-Hermite polynomials. An
explicit form is given by~\cite{q-H} 
\begin{equation*}
f_{q}( x) =\frac{2\, q^{1/16}}{\sqrt{\pi \log q^{-1}}}\, \exp\Big(\,
\frac{4}{\log q}\, \log^{2}\big( x+\sqrt{x^{2}+1}\, \big) \, \Big) =C( q) 
\,\exp\Big(\, \frac{4}{\log q}\,{\rm arcsinh}^2 \, x \, \Big)
\end{equation*}%
for $x\in 
\mathbb{R}
$. If we now substitute the $q$-variable $\left[ x\right] _{q}=\sinh (
g_s \,x /2 ) $ (where as in \cite{Aganagic:2004js} we use here the symmetric $q$%
-number without its proper normalization) we get
\begin{equation}
f_{q}\big( \left[ x\right] _{q}\big) =C( q) \, \e^{ (\log q) \, x^{2}} \ , \label{can}
\end{equation}%
and therefore the $q$-deformation of the Gaussian distribution cancels out the $%
q$-deformation of its argument.

Let us now consider the effect of using a $q$-deformed
exponential alone instead of a quantum dimension in (\ref{qYM1}). This
does not exactly lead to the same
situation as above. We already know that (\ref{qmat}) leads to a Stieltjes-Wigert
type ensemble, i.e. a Hermitian Stieltjes-Wigert ensemble for genus $%
h=0$. On the other hand, using a $q$-exponential of the theta-function type (%
\ref{theta}) in the heat kernel expansion (\ref{HK}) instead of the
usual Boltzmann weight leads to a discrete matrix model of the form
\begin{equation}
\mathcal{Z}_\Theta(g_s; \Sigma _{h}) =\sum_{m\in 
\mathbb{Z}
_{>0}^N} \ \prod\limits_{i<j}\, \left( m_{i}-m_{j}\right) ^{2-2h}\, \exp\bigg(-
\frac{p\, g_s}2 \, \log^{2}\Big(\, \sum_{i=1}^{N}\, m_{i}^{2}\, \Big)
\bigg) \ ,
\label{qexpmat}\end{equation}%
which is not of the random matrix theory type because it involves a
potential of the form $\log^{2}( m_{1}^{2}+\cdots
  +m_{N}^{2})$ instead of the Stieltjes-Wigert potential
$\log^2(m_1)+\cdots + \log^2(m_N)$. However, the confining properties of
the two potentials are qualitatively the same, and in fact the
$q$-exponential matrix model (\ref{qexpmat}) can be bounded from below
by the Stieltjes-Wigert matrix model (\ref{qmat}) using the inequality
\beq
\log^2\Big(\, \sum_{i=1}^N\, m_i^2\,\Big)\geq \frac4N\, \sum_{i=1}^N\,
\log^2(m_i)  \ . \nonumber
\eeq
This inequality follows directly from Jensen's inequality.

A geometric interpretation of the deformation of the gauge theory by
$q$-exponentials could be realised in the following way. The
transcendental function
$$
\Psi_q(z)=\exp_q\big(-\sqrt q \, z\,\big)
$$
is called the quantum dilogarithm function (see e.g.~\cite{ANKirillov}); it is related to the
classical Euler dilogarithm ${\rm Li}_2(x)= \sum_{n\in\IZ_{>0}} \,
  \frac{x^n}{n^2}$ for $|x|<1$ by the asymptotic expansion
$$
\Psi_q(x) = \exp\Big(-\frac{{\rm Li}_2(-x)}{\log q}\, \Big) \,
\big(1+{\mathcal{O}}(\log q)\big) \qquad \mbox{for} \quad q\to 1^- \ .
$$
Consider the quantum torus algebra $\Omega^0(T_q^2)$, which is the associative noncommutative algebra
over $\IC$ generated by two operators $\hat u$ and $\hat v$ which
satisfy the Weyl algebra
$$
\hat u\, \hat v = q \ \hat v\, \hat u \ .
$$
It can be represented by $q$-difference operators on $\Omega^0(\IC)$ by taking $\hat u=z$ to be multiplication by $z\in\IC$ and $\hat
v= \exp\big(-\log(q)\, z\, \frac\dd{\dd z}\big)$.
Then the quantum dilogarithm function with operator arguments
satisfies the relations~\cite{Faddeev:1993pe}
$$
\Psi_q(\hat u+\hat v)=\Psi_q(\hat v)\, \Psi_q(\hat u) \qquad \mbox{and}
\qquad \Psi_q\big(\hat v+\sqrt q\, \hat v\,\hat u+\hat u\big) =
\Psi_q(\hat u)\, \Psi_q(\hat v) \ .
$$
These relations suggest that the gauge theory with $q$-deformed Boltzmann
weights may be systematically treated as a gauge theory on the quantum
torus $T_q^2$; the role of this quantum torus algebra will be elucidated
within a $q$-deformed Hamiltonian framework in
\S\ref{se:Hilbertref}. The path integral for gauge theory on the
noncommutative torus (for which $q\in S^1$) is defined and studied
in~\cite{paniak1}; the analog of the Migdal expansion in this context is developed in~\cite{Paniak:2003gn} and related
to generalized Yang-Mills theory
on the torus $\Sigma_1=T^2= S^1\times S^1$ with infinitely many higher
Casimir operators.

\medskip

\subsection{Other $q$-deformations}\label{se:otherdef}~\\[5pt]
Altogether there
are six possible quantum deformations of the heat kernel
expansion (\ref{HK}) of two-dimensional Yang-Mills theory, in addition
to the standard one (\ref{qYM1}), incorporating at least one $%
q $-deformation of either the dimensions, the Boltzmann weight or the
Casimir operator.
The different choices can be succinctly summarized in the following table:%

\begin{equation*}
\begin{tabular}{|c||c|c|c|}
\hline
Theory & dimensions & exponential & Casimir \\ \hline \hline
$q$-Yang-Mills* & deformed & deformed & deformed \\ \hline
Yang-Mills & deformed & standard & deformed \\ \hline
Double $q$-deformation & deformed & deformed & standard \\ \hline
$q$-Yang-Mills* & standard & deformed & standard \\ \hline
Yang-Mills* & standard & deformed & deformed \\ \hline
Exponential potential & standard & standard & deformed \\ \hline
\end{tabular}%
\end{equation*}%

\medskip

As we have seen, the replacement of the standard Boltzmann weight with a $q$-exponential does not lead to a random
matrix theory form, although the confining potential is qualitatively
similar  --- we
emphasize this with an asterix label on the respective gauge theory in
this table; in \S\ref{se:expCasdef} we proposed an interpretation of the matrix
model (\ref{qexpmat}) at genus one in terms of a noncommutative gauge theory on a
quantum deformation of the torus $\Sigma_1=T^2= S^1\times S^1$.

We have already thoroughly discussed the cases leading to two-dimensional Yang-Mills
theory (\ref{HK}) (and (\ref{genYMHK})) and its $q$-deformation
(\ref{qYM1}); in \S\ref{se:Cat} we will provide a more precise explanation of why the
standard $q$-deformed gauge theory (\ref{qYM1}) is singled out.

There are two
cases in this table that are
not known to have an interpretation in gauge theory. The first case corresponds to
a combinatorial expansion with quantum dimensions together with a
$q$-deformed Boltzmann weight: This yields a rather
complicated
double $q$-deformation involving intricate combinations of $\log^2$ functions, which appears difficult to treat
analytically. The second case is easier to describe: It corresponds to
a pure Casimir $q$-deformation and leads to the matrix model
(\ref{Casmat}); in
\S\ref{se:Sq2} we give a gauge theory interpretation of the matrix
model (\ref{Casmat}) at genus zero in terms of noncommutative gauge
theory on a quantum deformation of the sphere $\Sigma_0=S^2$. In
\S\ref{se:Hilbertref} we shall see how the heat kernel action with
$q$-deformed Casimir operator eigenvalues arises in a Hamiltonian
formalism with quantum group gauge symmetry.

\medskip

\subsection{Five-dimensional gauge theory}\label{se:crystal}~\\[5pt]
The melting crystal partition function with an external potential is given by~\cite{Nakatsu:2007dk}
\begin{equation}
\cz_{p}(q,t) = \sum_{\lambda }\, s_{\lambda }( q^{\rho })
^{2}\, \exp \big(
\Phi(q,t;\lambda ,p)\big) \ , \label{1}
\end{equation}%
where the sum is over arbitrary sequences of partitions
$\lambda=(\lambda_1,\lambda_2,\dots)$, and $s_{\lambda }(
  q^{\rho }) = \dim_q\lambda $
denotes the Schur function $s_\lambda(x)$,
involving an infinite number of variables, with the principal
specialization $x_i=q^{i-\frac12}$, $i\geq1$. The potential is a function on charged partitions
$(\lambda,p)$, $p\in\IZ+\frac12$,  depending on a further set of coupling constants
$t=(t_1,t_2,\dots)$ and is given by
\begin{eqnarray*}
\Phi (q,t;\lambda ,p) = \sum_{k=1}^\infty\, t_{k}\, \Phi
_{k}(q;\lambda ,p)
\end{eqnarray*}
with
\begin{eqnarray}
\Phi _{k}(q;\lambda ,p) = \sum_{i=1}^{\infty }\, q^{k\, (p+\lambda
_{i}-i+1)}-\sum_{i=1}^{\infty }\, q^{k\, (p-i+1)}+ q^{k} \,
\Gamma_{k,p}(q) \ ,
\label{qpol} \end{eqnarray}%
where $\Gamma_{k,p}(z)$ is the equally spaced polynomial (\ref{pol}). Note that
(\ref{qpol}) is a polynomial in $q$. The important feature of these potentials is that they are essentially $q$%
-deformed Casimir operator eigenvalues, as we now explain. 

The partition function (\ref{1}) has an interpretation in
a five-dimensional $\cn=1$ supersymmetric gauge theory
compactified on a circle of radius $R=g_s$, where
$q=\e^{-g_s}$~\cite{Maeda:2004iq}; in this case the potential
(\ref{qpol}) corresponds to the Wilson line operator
$\Tr\big(\cp\exp\ii\oint_{S^1}\, A\big)^k$ on a loop with winding
number $k$ around the compactification circle $S^1$. It possesses an
affine Lie algebra symmetry based on the quantum torus algebra of
\S\ref{se:expCasdef} which in this setting is interpreted as providing
a realisation of the trigonometric basis for $\frsl(\infty)$~\cite{Nakatsu:2007dk}. Its reduction to
four dimensions gives the Nekrasov partition
function~\cite{Nekrasov:2002qd} for $\cn=2$ noncommutative $U(1)$
gauge theory deformed by higher Casimir operators
\beq
\cz_p^{4D}(t)=\sum_\lambda\, (\dim\lambda)^2\, \exp\Big(\,
\sum_{k=1}^\infty\, \frac{t_k}{k+1} \, \Ch_{k+1}(p,\lambda) \, \Big) \nonumber
\ ,
\eeq
where the Chern polynomials $\Ch_k(p,\lambda)=\Tr \big( \varphi_p^k \big)$ can be
expressed in terms of Casimir operators of $U(\infty)$ with order $k-1$ and
lower; here $\varphi_p$ is a complex scalar field in the $\cn=2$
vector multiplet with vacuum expectation value
$\langle\Tr\varphi_p\rangle=p$ which arises from dimensional reduction of
the five-dimensional gauge field $A$. They can be computed from
the generating function~\cite{Marshakov:2006ii}
\beq
\label{chgenfn}
\sum_{i=1}^\infty\, \big(\e^{u\, (p+\lambda_i-i+1)}- \e^{u\,
  (p+\lambda_i-i)} \big) = \sum_{k=0}^\infty\, \Ch_k(p,\lambda) \
\frac{u^k}{k!}
\eeq
which is the Chern character of the universal sheaf at a fixed point
of the instanton moduli space, parametrized by the partition
$\lambda$, of the four-dimensional supersymmetric gauge theory.
For example, for the first two relevant polynomials we find
explicitly
\begin{eqnarray}
\Ch_{2}(p,\lambda) = p^{2}+2\, C_1(\lambda)  \qquad
\mbox{and} \qquad 
\Ch_{3}(p,\lambda) = p^{3}+6\, p\, C_1(\lambda)
+3\, C_2(\lambda) \ , \notag
\end{eqnarray}%
where $C_1(\lambda)=|\lambda|$ is the linear Casimir operator and the
shifted symmetric polynomial of second order
\beq
C_{2}(\lambda )=\frac{1}{2}\, \sum_{i=1}^\infty\, \Big( \big( \lambda _{i}-i+\mbox{$\frac{1}{2}$}%
\big) ^{2}-\big( -i+\mbox{$\frac{1}{2}$}\big) ^{2}\Big) \label{f2}
\eeq
is the quadratic Casimir operator. In general, one can also write
the Chern characters as
\beqa
\Ch_k(p,\lambda) &=& p^k+\sum_{i=1}^\infty\, \big((p+\lambda_i-i+1)^k-
(p-i+1)^k\big)  \nonumber \\ && \qquad -\, \sum_{i=1}^\infty\, \big(
(p+\lambda_i-i)^k - (p-i)^k \big)
\label{Chkpgen}\eeqa
for any $k\in\IZ_{\geq0}$.

From (\ref{chgenfn}) and (\ref{Chkpgen}) it follows that the external potential
perturbation (\ref{qpol}) of the melting crystal model may be regarded as a $q$-deformation of the
higher Casimir $C_{k-1}(\lambda)$, or more precisely of the Chern
character $\Ch_k(p,\lambda)$. In particular, the partition function
(\ref{1}) may be regarded as a deformation of the Klim\v{c}\'{\i}k
partition function (\ref{klim}) by higher order
$q$-Casimirs. As it involves quantum dimensions
together with $q$-deformed
Casimir operators, we may treat it along the lines of
\S\ref{se:dimCasdef}. By the same arguments that led to (\ref{Gauss+}) from (\ref{mm}), we
conclude that the melting crystal model (\ref{1}) is equivalent to $U(\infty )$ generalized two-dimensional Yang-Mills theory
on the sphere $S^{2}$; the infinite rank of the gauge group owes to its origin
in a noncommutative gauge theory in four dimensions (see e.g.~\cite{LSzZ,SzaboPR}).

When $t=0$, the expansion
(\ref{1}) is also the Donaldson-Thomas partition function of an
$\cn=2$ noncommutative $U(1)$ gauge theory on
$\IC^3$~\cite{Iqbal:2003ds,Cirafici:2008sn}. This cohomological gauge
theory is equivalent to the
$q$-deformed $U(\infty)$ BF-theory on $S^2$ which can be obtained as
the $N\to\infty$ limit of (\ref{BFS2N}), and hence it has a
representation as a $U(\infty)$ unitary matrix model with partition
function
\beq
\cz^{6D}(q) := \cz_p(q,0) = \int_{[0,2\pi)^\infty} \
\prod_{i=1}^\infty \, \frac{\dd\phi_i}{2\pi} \
\frac{\Theta\big(\e^{\ii\phi_i}\,;\, q\big)}{(q;q)_\infty} \ \prod_{j<k}\,
\big|\e^{\ii\phi_j}-\e^{\ii\phi_k}\big|^2 \ .
\label{DT}\eeq
Gauge theories on more general local toric Calabi-Yau threefolds with
no compact divisors lead to
more complicated matrix models characterized by weight
functions which are combinations of
theta-functions, as studied in~\cite{Ooguri:2010yk,Szabo:2010sd,Sulkowski:2010eg}
(see~\cite{Yamazaki:2011wy,Sulkowski:2011qs} for reviews). In these
works it is shown how $U(\infty )$ matrix models with combinations of theta-functions
as their weight functions possess partition functions which are corresponding
combinations of MacMahon
functions. For later reference, here we will redo these computations
using the strong Szeg\H{o} limit theorem for Toeplitz determinants (see Appendix~B).

Let
\beq
f(z;q)=\frac{\Theta(z;q)}{(q;q)_\infty}= \prod_{n=1}^\infty\,
\big(1+q^{n-1/2}\, z\big)\, \big(1+q^{n-1/2}\, z^{-1}\big)
\label{double}\eeq
be the weight function of the
matrix integral (\ref{DT}), where we have used the Jacobi triple
product identity (\ref{triple}). To apply the Szeg\H{o} theorem, we need to
determine the Fourier coefficients $[\log f]_k$, $k\in\IZ$ of
\begin{eqnarray*}
\log f(z;q) &=& \sum_{n=1}^{\infty }\, \Big( \log \big( 1+q^{n-1/2}\, z\big) +\log \big(
1+q^{n-1/2}\, z^{-1}\big) \Big) \\[4pt] &=&\sum_{n=1}^{\infty } \
\sum_{k\neq 0}\, \frac{\left( -1\right) ^{k+1}}{k}\, q^{\left(n-1/2\right)\,
k}\, z^{k} \ = \ \sum_{k\neq 0}\, \frac{\left( -1\right) ^{k+1}\,
q^{k/2}}{k\, \big(1-q^{k}\big)} \ z^{k} \ .
\end{eqnarray*}%
It follows that
\begin{equation}
\left[ \log f\right] _{k}=\left[ \log f\right] _{-k}=\frac{\left(
-1\right) ^{k+1}\, q^{k/2}}{k\, \big(1-q^{k}\big)} \ ,  \label{Fourier}
\end{equation}%
and by the Szeg\H{o} theorem one has
\begin{eqnarray}
\log \cz^{6D}(q) &=&\sum_{k=1}^{\infty }\, k\, \left[ \log f
\right] _{k}\, \left[ \log f\right] _{-k}  \label{E} \\[4pt]
&=& \sum_{k=1}^{\infty }\, \frac{q^{k}}{k\, \big(1-q^{k}\big)^{2}} \nonumber \\[4pt] 
&=& \sum_{k=1}^{\infty
} \ \sum_{n=1}^{\infty }\, n\, \frac{q^{k\,n}}{k} \ = \
-\sum_{n=1}^{\infty } \, n\, \log \big( 1-q^{n}\big) \ . \nonumber
\end{eqnarray}%
Thus the partition function (\ref{DT}) evaluates to 
$$
\cz^{6D}(q)= M(q) \ ,
$$
where $M(q)$ is the MacMahon function
\beq
M(q) = \prod_{n=1}^\infty\, \frac1{\big(1-q^n\big)^n} \ .
\label{eq:macmah}\eeq
This expression agrees with the $N\to\infty$ limit of (\ref{BFS2N}).

We can generalise this calculation to the family of $U(\infty)$ matrix
models
$$
\cz_L^{6D}(\alpha;q,Q):= \int_{[0,2\pi)^\infty} \ \prod_{i=1}^\infty \, \frac{\dd\phi_i}{2\pi} \
F_L\big(\e^{\ii\phi_i}\,;\, \alpha\,;\,q,Q\big) \ \prod_{j<k}\,
\big|\e^{\ii\phi_j}-\e^{\ii\phi_k}\big|^2
$$
with weight functions
\beq
F_L(z;\alpha;q,Q):= \prod_{a=1}^L \, \frac{\Theta(Q_a\, z;q)^{\alpha_a}}{(q;q)^{\alpha_a}_\infty}
\label{FKweight}\eeq
where $\alpha_a\in\IC$ for $a=1,\dots,L$. With this product of
theta-functions as a symbol, the Fourier coefficients
(\ref{Fourier}) generalize to
\beq
\left[ \log F_L\right] _{k}=\frac{\left( -1\right) ^{k+1}\, q^{k/2}}{%
k\, \big(1-q^{k}\big)}\, \sum_{a=1}^{L}\, \alpha _{a}\, Q_{a}^{k} \
,
\label{FKFourier}\eeq
and hence the partition function is given by
$$
\log\cz_L^{6D}(\alpha;q,Q)=\sum_{k=1}^{\infty }\, \frac{q^{k}}{%
k\, \big(1-q^{k}\big)^{2}}\, \ \sum_{a,b=1}^{L}\, \alpha _{a}\,
\alpha_b\, 
Q_{a}^{k}\, Q_b^{-k} \ .
$$
Proceeding analogously as before, it is then straightforward to obtain
that this expression evaluates to a product of generalised MacMahon functions
\begin{equation}
\cz_L^{6D}(\alpha;q,Q) =\prod_{a=1}^{L}\, M(q)
^{\alpha _{a}^{2}} \ \prod_{b\neq c}\, M\big(Q_{b}\, Q_{c}^{-1} ,
q\big)^{\alpha _{b}\, \alpha _{c}} \ ,  \label{E-alpha}
\end{equation}%
where%
\begin{equation*}
M( Q,q) =\prod_{n=1}^{\infty }\, \frac1{\big(1-Q\, q^n\big)
  ^{n}} \ .
\end{equation*}
Note the symmetry of the expression (\ref{E-alpha}) under the
interchange $Q_b\leftrightarrow Q_c$ for any $b,c=1,\dots,L$, despite
the absence of this symmetry in the original weight function (\ref{FKweight}).

For example, when $L=1$ and $2\alpha_1^2=\chi$ is the topological Euler characteristic of
a Calabi-Yau threefold, the partition function (\ref{E-alpha})
computes the constant map contributions to the generating function for
Gromov-Witten invariants~\cite{Faber,Behrend}. The corresponding $U(\infty)$
matrix models
\beq
\cz_1^{6D}\big(\pm\, \sqrt{\mbox{$\frac\chi2$}}\,;\,q\big) =M(q)^{\chi/2} =
\int_{[0,2\pi)^\infty} \ \prod_{i=1}^\infty \, \frac{\dd\phi_i}{2\pi} \ \Big(\,
\frac{\Theta\big(\e^{\ii\phi_i}\,;\, q\big)}{(q;q)_\infty} \,
\Big)^{\pm\, \sqrt{\mbox{$\frac\chi 2$}}}
 \ \prod_{j<k}\,
\big|\e^{\ii\phi_j}-\e^{\ii\phi_k}\big|^2
\label{Mqchi}\eeq
then give an explicit realization of the proposal
of~\cite{Bouchard:2011fm,Bouchard:2011ya} to study topological
recursion in an associated matrix model.

Since the theta-function (\ref{theta3zq}) is holomorphic, the
Szeg\H{o} theorem suffices to determine the asymptotics of the
$U(\infty)$ gauge theory partition function (\ref{DT}) (see
Appendix~B). In fact, in our context the Szeg\H{o} theorem gives exact
results. This follows from the fact that for the partition function (\ref{DT}) the statement of
the theorem is equivalent to the Cauchy-Binet formula (\ref{SC})
written in Miwa variables
\beq
\sum_\lambda\, s_\lambda(x)\, s_\lambda(y) = \exp\Big(\,
\sum_{k\geq1}\, k\, s_k\, t_k \, \Big) \ ,
\label{Sc}\eeq
where
$$
s_k=\frac1k\, \sum_{i\geq1}\, x_i^k \qquad \mbox{and} \qquad
t_k=\frac1k\, \sum_{i\geq1}\, y_i^k
$$
are power sums of the sets of variables $x$ and $y$. In the present
case the Miwa variables $s$ and $t$ are precisely the Fourier
coefficients (\ref{Fourier}) of
the potential of the matrix model (\ref{DT}), and in fact the equality
(\ref{DT})
arises from the normalization of the Schur measure (\ref{SC}) via
Gessel's identity~\cite{Szabo:2010sd}. 

To generalise this argument to more complex
combinations of theta-functions, we consider the analog of the
expression (\ref{SC}) involving supersymmetric Schur polynomials
$\HS_\lambda(x|z)$~\cite{BR} (also known as
Schur-Littlewood or hook-Schur polynomials \cite{BG}). They are
defined by
$$
\HS_\lambda(x|z) = \sum_{\mu,\nu}\, N_{\mu\nu}{}^\lambda\, s_\mu(x)\,
s_{\nu'}(z)
$$
where $N_{\mu\nu}{}^\lambda\in\IZ_{\geq0}$ are the
Littlewood-Richardson coefficients defined by expressing the ring
structure on the space of symmetric polynomials in the basis of Schur
functions as
\beq
s_\mu(x)\, s_\nu(x) = \sum_{\lambda}\, N_{\mu\nu}{}^\lambda \
s_\lambda(x) \ ,
\label{LRcoeffs}\eeq
and $\nu'$ denotes the conjugate partition to $\nu$.
The analogous
Cauchy-Binet identity is~\cite{PG,BG} 
\begin{eqnarray}
\sum_{\lambda }\,\HS_{\lambda }(x|z)\,\HS_{\lambda }(y|w)
= \prod_{i,j\geq1}\ \frac{(1+x_i\, w_j)\, (1+y_i\, z_j)}{(1-x_i\,
  y_j)\, (1-z_i\, w_j)} \ ,
\label{ScSUSY}\end{eqnarray}%
which we note is symmetric under interchange $(x,y)\leftrightarrow
(z,w)$. There is also an extension of the Gessel identity which leads to a unitary
matrix model description~\cite{PG}%
\begin{eqnarray}
\sum_{\lambda}\,\HS_{\lambda }(x|z)\,\HS_{\lambda }(y|w) &=&
\int_{[0,2\pi)^\infty } \ \prod_{i=1}^\infty \, \frac{\mathrm{d}\phi_i}{2\pi 
}~\prod_{j\geq1}\, \frac{\big(1+x_j\, \e^{\ii\phi_i}\big)\, \big(1+y_j\,
  \e^{-\ii\phi_i}\big)}{\big(1-z_j\, \e^{\ii\phi_i}\big)\, \big(1-w_j \,
  \e^{-\ii\phi_i}\big)} \nonumber \\ && \qquad \qquad \qquad \qquad \times \ 
\prod\limits_{k<l}\,\big\vert \e^{\ii\phi_k}-\e^{\ii\phi_l}\big\vert^{2}
  \ .
\label{HS}\end{eqnarray}%
After specialisation of the variables
$x_i=Q_1\, q^{i-\frac12}$, $y_i=Q_1^{-1}\, q^{i-\frac12}$, $z_i=-Q_2\,
q^{i-\frac12}$ and $w_i=-Q_2^{-1}\, q^{i-\frac12}$, this result shows that the formula (\ref{E-alpha}) with $K=2$ and
$\alpha_1=\pm\, 1=-\alpha_2$ is exact, with
\beqa
\cz_2^{6D}(\pm\,1,\mp\,1;q,Q_1,Q_2) & = & \frac{M(q)^2}{M\big(Q_1\,
  Q_2^{-1}\,,\, q\big)\, M\big(Q_1^{-1}\, Q_2\,,\, q\big)}  \nonumber
\\[4pt] &=& \int_{[0,2\pi)^\infty } \ \prod_{i=1}^\infty \, \frac{\mathrm{d}\phi_i}{2\pi 
}~ \Big(\, \frac{\Theta\big(Q_1\,
  \e^{\ii\phi_i}\,;\,q\big)}{\Theta\big(Q_2\,
  \e^{\ii\phi_i}\,;\, q\big)} \, \Big)^{\pm\, 1} \ \prod\limits_{j<k}\,\big\vert \e^{\ii\phi_j}-\e^{\ii\phi_k}\big\vert^{2}
  \ . \nonumber
\eeqa
When $Q_1=1$, $Q_2=Q$ this is the partition function of the $\cn=2$
gauge theory on the noncommutative
conifold~\cite{Ooguri:2010yk,Szabo:2010sd}; in the context of
five-dimensional gauge theories, the parameter $Q=R^2\, \Lambda^2$
where $\Lambda$ is the dynamical scale which is identified as the
coupling of the four-dimensional $U(1)$ gauge theory obtained in the
reduction limit.
Moreover, by the generalised Cauchy-Binet formula (\ref{ScSUSY}) we
may identify this partition function as that of a \emph{supersymmetric}
extension of the topological $q$-deformed two-dimensional gauge theory
(\ref{BFS2N}) given by
$$
\cz_2^{6D}(\pm\,1,\mp\,1;q,Q_1,Q_2) = \sum_\lambda\,
\HS_\lambda\big(Q_1\, q^\rho\, \big|\, -Q_2\, q^\rho\big) \,
\HS_\lambda\big(Q_1^{-1}\, q^\rho\, \big|\, -Q_2^{-1}\, q^\rho\big) \ .
$$

In the general case, since the weight function (\ref{FKweight}) is of
Szeg\H{o} class and the Toeplitz determinant is of infinite dimension,
the formula (\ref{E-alpha}) is exact. That the Szeg\H{o} class
condition is satisfied can be easily checked in a number of ways; in
our case it suffices to notice the exponential decay of the Fourier
coefficients (\ref{FKFourier}), see Appendix~B. In fact, a closely
related exact formula exists even in the finite-dimensional case,
which is valid for weight functions which are arbitrary rational
functions (Appendix~B). In our case we consider theta-functions in
(\ref{FKweight}) which involve infinite products, but by truncating
the products at some finite order $m$, making the corresponding weight
$F_L^{(m)}$ a rational function, we immediately find a direct
relationship between the two cases: The effect of this truncation on
the Fourier coefficients in (\ref{FKFourier}) is simply
$$
\big[\log F_L^{(m)}\big]_k = \big[\log F_L\big]_k\, \big(1-q^{k\, m}
\big) \ ,
$$
which follows from the truncation of the corresponding geometric
series. The case where each product in (\ref{FKweight}) has a
different truncation follows manifestly in the same manner. The
behaviour is then of the same type, with the exponential decay of the
Fourier coefficients maintained and with a simplification of the final
result in the non-rational case $m\to\infty$; this is due to the
exponential decay of the coefficients $q^{n-1/2}$ in the products
(\ref{double}). Since this simplification is exact for the
infinite-dimensional Toeplitz determinant, it would be interesting to
study the non-rational case also in the finite-dimensional setting of
Appendix~B.

\medskip

\subsection{Torus bundles on the quantum sphere}\label{se:Sq2}~\\[5pt]
In Klim\v{c}\'{\i}k's partition function (\ref{klim}), the
$q$-deformed Casimir eigenvalues arise as a result of the group-valued
fields which appear in the action of the gauged WZW theory based on
the Drinfel'd double $D(G)=G^{\IC}$. We shall now propose a natural
geometrical interpretation of this $q$-deformation at genus zero based on
noncommutative gauge theory.

Let us first recall the reduction onto torus
bundles on $S^2$ when we evaluate the path integral using the diagonalization
techniques from \S\ref{se:2DYMgen}. It involves a splitting of the
original (trivial) vector bundle associated with the
given $G$-bundle into line bundles $\bigoplus_i\, {\mathcal
  L}_{n_i}$ parametrized by sequences of integers $n_1,\dots,n_N$. 
The corresponding gauge potential is $A=\sum_i\,a_{n_i}$, where
$a_{n_i}$ is the monopole potential of first Chern class $n_i\in\IZ$ and the
$i$-th block is an abelian connection on the bundle ${\mathcal L}_{n_i}={\mathcal L}^{\otimes
  n_i}$. Here ${\mathcal L}\to {S}^2$ is
the standard ${SU}(2)$-equivariant monopole line bundle of
degree one which is classified by the Hopf fibration ${S}^3\to
{S}^2$ and whose isomorphism class is the generator of $H^1({S}^2,{U}(1))\cong
H^2({S}^2,\IZ)\cong\IZ$. Since $f_n:=\dd a_n=2\pi\, n \,\dd\mu$,
the curvature of this connection is given
by
\beq
F_A=\dd A=
\sum_{i=1}^N \,2\pi\,n_i\, \dd\mu \nonumber
\eeq
where $\dd\mu$ is the unit area form on ${S}^2$.

We shall now demonstrate how this construction can be extended to the
standard Podle\`s quantum sphere $S_q^2$~\cite{Po87}, via a $q$-deformation of
the Hopf fibration which is the well-known quantum principal
$U(1)$-bundle over $S_q^2$ whose total space $S_q^3$ is the manifold of the
quantum group $SU_q(2)$~\cite{brzmajid1}. With $q\in(0,1)$, the
algebra $\Omega^0(SU_q(2))$ is the $*$-algebra generated by elements
$a,c$ with relations defined by requiring that the matrix
$$
U=\begin{pmatrix} a& -q\, c^* \\ c& a^* \end{pmatrix}
$$
is unitary: $U\, U^*=U^*\, U=1$. It has the usual Hopf algebra
structure defined by the coproduct $\Delta(U)=U\otimes U$, the
antipode $S(U)=U^*$, and the counit $\varepsilon(U)=1$. The quantum
universal enveloping algebra $\cu_q(\frsu(2))$, with its standard Hopf
$*$-algebra structure described in Appendix~A, naturally acts on this
algebra: There is a bilinear dual pairing $\langle-,-\rangle_q$
between $\cu_q(\frsu(2))$ and $\Omega^0(SU_q(2))$ which defines
canonical left and right $\cu_q(\frsu(2))$-module structures on $\Omega^0(SU_q(2))$
such that
$$
\langle g,g'\triangleright f\rangle_q:=\langle g\,g',f\rangle_q \qquad
\mbox{and} \qquad \langle g,f\triangleleft g'\,\rangle_q:= \langle
g'\, g,f\rangle_q
$$
for all $g,g'\in\cu_q(\frsu(2))$ and $f\in\Omega^0(SU_q(2))$.

Consider now the action of the abelian circle group $U(1)\cong
S^1=\{z\in\IC\ | \ |z| =1\}$ on the algebra $\Omega^0(SU_q(2))$ given by
  the automorphism
$$
\alpha_z(a)= z\, a \qquad \mbox{and} \qquad \alpha_z(c) = z\, c
$$
extended as a $*$-algebra map. The algebra $\Omega^0(S_q^2)$ of the
standard Podle\`s sphere $S_q^2$~\cite{Po87} is the corresponding
fixed-point subalgebra
$$
\Omega^0(S_q^2) = \Omega^0\big(SU_q(2)\big)^{U(1)} = \big\{
f\in\Omega^0\big(SU_q(2)\big) \ \big|\ \alpha_z(f)=f\big\} \ .
$$
The algebra inclusion $\Omega^0(S_q^2)\hookrightarrow
\Omega^0(SU_q(2))$ is a quantum principal bundle which can be endowed
with compatible differential calculi~\cite{brzmajid1}.

There are natural finitely-generated projective
$\Omega^0(S_q^2)$-bimodules of rank one associated to irreducible
one-dimensional representations of $U(1)$ of weight $n\in\IZ$ by
$$
\cl_n=\big\{f\in\Omega^0\big(SU_q(2)\big) \ \big|\ \alpha_z(f) =
(z^*)^n\, f\big\} \ ,
$$
which we regard as sections of $SU_q(2)$-equivariant line bundles over
the quantum sphere $S_q^2$ with monopole charges $n\in\IZ$. The left
action of the group-like element $K$ on $\Omega^0(SU_q(2))$ gives a
dual presentation of these line bundles as
$$
\cl_n=\big\{ f\in\Omega^0\big(SU_q(2)\big) \ \big| \ K\triangleright
f= q^{n/2}\, f\big\} \ ;
$$
this presentation can be understood via the identification $K=q^{H/2}$
from Appendix~A, with $H$ the generator of $U(1)\subset SU_q(2)$.

For the canonical left-covariant two-dimensional calculus on
$\Omega^0(S_q^2)$, there are natural gauge potentials
$a_n\in\Hom_{\Omega^0(S_q^2)}\big(\cl_n,\cl_n \otimes_{\Omega^0(S_q^2)}
\Omega^1(SU_q(2))\big)$
with curvatures~\cite[\S2.3]{LandiSzabo}
$$
f_n= \dd a_n= 2\pi\, q^{\frac12\,(n+1)}\, [n]_q\ \beta
$$
in $\Hom_{\Omega^0(S_q^2)}\big(\cl_n,\cl_n \otimes_{\Omega^0(S_q^2)}
\Omega^2(S_q^2)\big)$, where $\beta$ is the natural generator for the free $\Omega^0(S_q^2)$-bimodule
$\Omega^2(S_q^2)$. Inspired by the reduction of the usual
Yang-Mills gauge theory on $S^2$ to an abelian gauge theory based on
torus bundles, we wish to integrate these gauge field curvatures over
the quantum sphere $S_q^2$. This requires the introduction of a
``twisted'' integral which is a linear functional
$\int_{S_q^2}:\Omega^2(S_q^2)\to \IC$ defined by restriction of the
Haar state on the algebra $\Omega^0(SU_q(2))$ (see
e.g.~\cite[\S2.7]{LandiSzabo}); one has
\beq
\int_{S_q^2}\, (f\cdot f'\,)\ \beta = \int_{S_q^2}\,
\big((f'\triangleleft K^2)\cdot f\big)\ \beta
\label{Twistedtrace}\eeq
for $f,f'\in\Omega^0(S_q^2)$. This is the unique functional which is
invariant under the (quantum adjoint) action of
$\cu_q(\frsu(2))$. Using the normalization $\int_{S_q^2}\, \beta=1$
for the Haar state on $\Omega^0(S_q^2)$, one finds that the integral of
the curvature $f_n$ of the canonical gauge field $a_n$ on $S_q^2$ is
given by
\beq
\frac1{2\pi} \, \int_{S_q^2}\, \Tr_q(f_n) = q^{1/2}\, [n]_q \ ,
\label{intTrqfn}\eeq
where $\Tr_q(M)= \Tr(q^{\langle\rho,H\rangle} \, M) \in\Omega^0(S_q^2)$, for an $\Omega^0(S_q^2)$-valued
matrix $M$ of dimension $|n|+1$, is the quantum trace with the
``twisted'' cyclicity $\Tr_q(M_1\, M_2) = \Tr_q\big((M_2\triangleleft
K^2)\, M_1\big)$, see Appendix~A. The $q$-integer (\ref{intTrqfn}) has a natural
geometric interpretation: It is the $q$-index of the standard Dirac
operator in the Hopf algebraic $SU_q(2)$-equivariant K-theory of
$S_q^2$, i.e. the difference between the quantum dimensions of its
kernel and cokernel computed using the quantum trace $\Tr_q$
above. For further details, see e.g.~\cite[\S2.7]{LandiSzabo}.

It follows that abelianised gauge theory based on the corresponding
torus bundles $\bigoplus_i\, \cl_{n_i}$ over $S_q^2$ can reproduce
Klim\v{c}\'{\i}k's $q$-deformed heat kernel expansion (\ref{klim}) via
a
putative extension of the diagonalisation technique to this quantum
homogeneous space. The problem now is that it is not clear how to even
{define} the classical gauge theory on the quantum sphere,
because it is not obvious in what sense gauge theories on $S_q^2$ are
actually gauge-invariant for $q\neq1$: The integral $\int_{S_q^2}$ is a
quantum trace and the twisted cyclicity property (\ref{Twistedtrace})
breaks the usual gauge symmetry. This problem is discussed
in~\cite{grosse}, where path integral quantization for field theories
on the quantum two-sphere is considered. It is shown there that the
group of gauge transformations $\cg$ is generated by a real sector of a
quotient of a Hopf algebra, and the space of gauge fields $\ca$ is a
subspace of one-forms valued in a Hopf module algebra of differential
forms on $S_q^2$; however, the role of these gauge symmetries
in the quantum field theory is not clear. On the other hand, it is shown
in~\cite{LandiSzabo} that gauge transformations act trivially on the
line bundles $\cl_n$. Hence by \emph{defining} the quantum gauge
theory within the approach of abelianization sketched here, we arrive
at a concrete gauge-invariant definition of Yang-Mills theory on
$S_q^2$ which could cure the problems in the quantization of gauge
fields on $S_q^2$ observed in~\cite{grosse} (see
also~\cite{brzmajid1}). 

\section{Categorification\label{se:Cat}}

In \S\ref{se:Class} we argued that the particular $q$-deformation
(\ref{qYM1}) is, at least qualitatively, essentially the only
non-trivial quantum
deformation of the standard Yang-Mills partition function (\ref{HK}). 
The purpose of this section is to provide a more precise and intrinsic characterization
of this argument which will also demonstrate the appearence of a quantum group 
gauge symmetry. For this, we will use the formalism of two-dimensional
topological quantum field theory to construct Yang-Mills amplitudes;
for ordinary Yang-Mills theory this construction is described
in~\cite{Witten:1991we,review}, and in~\cite{BP,Aganagic:2004js,Szabo:2009vw} for
its $q$-deformation. In
this setting the amplitudes give a representation of a certain
geometric category $\scrs$ in the linear category $\scrr=\Vect$ of complex
vector spaces,
whose gluing properties are concisely formulated as a functor of tensor
categories. In the following we will instead take $\scrr=\Rep(\cu_q(\frg))$ to be the
category of representations of the quantum group
$\cu_q(\frg)$; our description parallels the argument of~\cite{deHRT}
that quantum characters of $\cu_q(\frg)$ are required to capture the
general solution of $q$-deformed Yang-Mills theory defined via gluing rules. Then $\scrr$ has the structure of a semisimple ribbon
category endowed with certain additional canonical
objects and morphisms. Most notably it contains a ``ribbon element'',
and altogether these structures categorify the basic
building blocks of Yang-Mills amplitudes. The corresponding numerical
invariants are ``generalized characters''. By integrating to the standard
characters, corresponding to morphisms with target the trivial object $V=\IC$
of $\Rep(\cu_q(\frg))$, we find that the standard $q$-propagator (with undeformed
Casimir elements) is a fundamental object. More general characters
will be considered in \S\ref{se:Ref}.

\medskip

\subsection{Semisimple ribbon categories}\label{se:MCT}~\\[5pt]
We begin by briefly reviewing the relevant category theory that we
need to abstract and axiomatize the gluing data of two-dimensional
gauge theory into a certain abelian tensor
category. For further details, see e.g.~\cite{BKLect,FRSI}.

The categories $\scrc$ that we are interested in are known as \emph{ribbon
categories}, together with some additional structure; they are defined as follows. First
of all, $\scrc$ is a monoidal (or tensor) category. This means that
$\scrc$ is equipped with a covariant exterior product bifunctor
$\otimes:\scrc\times\scrc\to\scrc$ and a unit object $\Idd\in\Ob(\scrc)$
together with three natural functorial isomorphisms 
\beq
(X\otimes Y)\otimes Z= X\otimes(Y\otimes Z) \qquad \mbox{and}
\qquad \Idd\otimes X= X= X\otimes\Idd
\label{isobasic}\eeq
for all objects $X,Y,Z\in\Ob(\scrc)$, called the associativity and
unity relations. Throughout we exploit Mac~Lane's coherence theorem
(see e.g.~\cite{BKLect})
which states that any monoidal category $\scrc$ is equivalent
to a strict monoidal category in which these isomorphisms become
equalities; this equivalence typically also preserves additional
structures on the categories~\cite{NgSch}, and hence we take all
isomorphisms to be equalities in what follows. The isomorphisms (\ref{isobasic}) satisfy the pentagon relations, which state that
the five ways of bracketing the exterior products of four objects commute, and also the
triangle relations which state that the 
associativity constraint with $Y=\Idd$ is compatible with the unity
relations. Given morphisms $f\in\Hom_\scrc(X,Y)$ and
$g\in\Hom_\scrc(Z,W)$, their exterior product is $f\otimes
g\in\Hom_\scrc(X\otimes Z,Y\otimes W)$. 

We call $\scrc$ braided when there are natural bifunctor isomorphisms
$$
B_{X,Y}\in\Hom_{\scrc}(X\otimes Y,Y\otimes X)
$$ for any
$X,Y\in\Ob(\scrc)$, called commutativity relations. The braiding
$B_{X,Y}$ satisfies the hexagon relations which give two conditions,
one expressing $B_{X\otimes Y,Z}$ in terms of associativity relations
$\id_X\otimes B_{Y,Z}$ and $B_{Z,X}\otimes\id_Y$, and a similar one
for $B_{X,Y\otimes Z}$. In this paper we will only work with
categories $\scrc$ which are
$\IC$-linear and abelian, which means that the morphism spaces
$\Hom_\scrc(X,Y)$ are vector spaces over $\IC$ and there is an
associative $\IC$-bilinear composition product
$\Hom_\scrc(X,Y)\times\Hom_\scrc(Y,Z)\to\Hom_\scrc(X,Z)$, denoted
$(f,g)\mapsto g\circ f$. For any exact sequence $X\to Y\to Z$ of 
objects of $\scrc$ and any morphism $f\in\Hom_\scrc(Y,Z)$, there is
a kernel morphism denoted by ${\rm Ker}(f)\in\Hom_\scrc(X,Y)$ with
$f\circ{\rm Ker}(f)=0$, and similarly there are cokernel morphisms. There
are also direct sums $X_1\oplus\cdots\oplus X_n$ for any finite collection of objects
$X_i\in\Ob(\scrc)$. Basic examples are the category $\Vect$ of vector
spaces over $\IC$ and linear maps, and more generally the representation category $\Rep(\ca)$ of modules
over suitable associative $\IC$-algebras $\ca$ and intertwining operators, both with the usual tensor
product.

For an object $X\in\Ob(\scrc)$, a right dual to $X$ is an object
$X^\vee$ with two morphisms
$$
e_X\,:\,X^\vee\otimes X~\longrightarrow~\Idd \qquad \mbox{and} \qquad
i_X\,:\,\Idd~\longrightarrow~X\otimes X^\vee
$$
which obey the composition laws
\beq
(\id_X\otimes e_X)\circ(i_X\otimes\id_X)=\id_X \qquad \mbox{and}
 \qquad 
(e_X\otimes\id_{X^\vee})\circ(\id_{X^\vee}\otimes i_X)=\id_{X^\vee} \
.
\label{dualitycomps}\eeq
Similarly, one defines left duals ${}^\vee X\in\Ob(\scrc)$. Dual
objects are canonically defined (when they exist), and duality
canonically extends to a contravariant functor $(-)^\vee: \scrc\to\scrc^{\rm
  op}$. (Here $\scrc^{\rm op}$ is the opposite or dual category to
$\scrc$ with $\Ob(\scrc^{\rm op})=\Ob(\scrc)$ and
$\Hom_{\scrc^{\rm op}}(X,Y)=\Hom_{\scrc}(Y,X)$ for all
$X,Y\in\Ob(\scrc)$.) If $X,Y,Z\in\Ob(\scrc)$ and $Y$ has dual
$Y^\vee$, then there exist canonical isomorphisms~\cite{BKLect}
\beq
\Hom_\scrc(X\otimes Y,Z)=\Hom_\scrc(X,Z\otimes Y^\vee\,) \qquad
\mbox{and} \qquad
\Hom_\scrc(X,Y\otimes Z)=\Hom_\scrc(Y^\vee\otimes X,Z) \ .
\label{Homdualisos}\eeq
We henceforth work only with objects that have dual objects.

An object $U\in\Ob(\scrc)$ is simple if any injection of objects
$V\hookrightarrow U$ is either $0$ or an isomorphism. The category
$\scrc$ is semisimple if any object $X$ is isomorphic to a direct sum 
$$
X=\bigoplus_{i\in I}\,n_i\,U_i \ ,
$$
where $U_i$ are simple objects, $I$ is the set of isomorphism classes
of simple objects in $\scrc$, and $n_i\in\IZ_{\geq0}$ with $n_i\neq0$ for
only finitely many $i\in I$. We assume
that the set $I$ includes the tensor unit as $\Idd=U_0$. One has $\Hom_\scrc(U_i,U_j)=0$ for $i\neq j$,
whereas 
$\End_{\scrc}(U_i):=\Hom_\scrc(U_i,U_i)=\IC~\id_{U_i}$. Using
(\ref{Homdualisos}) we can define
fusion coefficients $N_{ij}{}^k\in\IZ_{\geq0}$ by
$$
U_i\otimes U_j=\bigoplus_{k\in I}\, N_{ij}{}^k\ U_k \qquad
\mbox{with} \quad N_{ij}{}^k=\dim\, \Hom_{\scrc}(U_i\otimes U_j\otimes
U_k^\vee,\Idd) \ .
$$
If the tensor product multiplicity $N_{ij}{}^k\neq0$, then we may
choose a basis $(f_l)_{l=1,\dots,N_{ij}{}^k}$ of
$\Hom_\scrc(U_i\otimes U_j,U_k)$ and a dual basis
$(f_l^\vee)_{l=1,\dots,N_{ij}{}^k}$ of $\Hom_\scrc(U_k,U_i\otimes
U_j)$ such that the composition product $f_l\circ f_m^\vee=0$ for
$l\neq m$ and $f_l\circ f_l^\vee$ is proportional to $\id_{U_k}$ for
each $l,m=1,\dots,N_{ij}{}^k$.

We also assume the existence
of a twist $\theta$, defined to be a system of functorial isomorphisms
$\theta_X\in\End_\scrc(X)$ for all $X\in\Ob(\scrc)$ satisfying the
compatibility condition
\beq
\theta_{X\otimes Y}=B_{Y,X}\circ(\theta_Y\otimes\theta_X)\circ
B_{X,Y} \ , 
\label{thetatensor}\eeq
together with a compatible duality; in particular, this additional
structure ensures that every left dual of an object is also a right
dual and vice versa. On simple objects we
define scalars $\theta_i\in\IC^*$ by
$$
\theta_{U_i}=\theta_{U_i^\vee}= \theta_i\ \id_{U_i} \ ,
$$
with $\theta_0=1$. 

For any object $X\in\Ob(\scrc)$, and for any
endomorphism $f\in\End_\scrc(X)$, we define its categorical or Markov trace
${\sf Tr}_X\, (f) \in\End_{\IC}(\Idd)\cong\IC$ by
\beq
{\sf Tr}_X\, (f) =  e_{X^\vee}\circ\big((\psi_X^{-1}\circ\theta_X) \otimes
\id_{X^\vee}\big)\circ(f\otimes\id_{X^\vee})\circ i_X
\label{TrXf}\eeq
where $$
\psi_X=(\id_X\otimes
e_{X^\vee})\circ\big(\id_{X}\otimes B_{X^{\vee\vee}, X^\vee}^{-1}\big)
\circ (i_X\otimes\id_{X^{\vee\vee}})
$$
is a functorial isomorphism in
$\Hom_{\scrc}(X^{\vee\vee},X)$. Taking $f=\id_X$ defines the categorical
dimension of the object $X$ as
\beq
{\sf dim} (X):={\sf Tr}_X(\id_X)= e_{X^\vee}\circ\big((\psi_X^{-1}\circ\theta_X) \otimes
\id_{X^\vee}\big)\circ i_X \ .
\label{quantDim}\eeq
The
braiding is taken to be maximally nondegenerate with respect to the
semisimple structure, in the sense that 
\beq
S_{ij}:={\sf Tr}_{U_i\otimes U_j} \big( B_{U_j,U_i}\circ
B_{U_i,U_j} \big)
\label{Sijdef}\eeq
for $i,j\in I$ is a symmetric invertible matrix. Comparing
(\ref{Sijdef}), (\ref{TrXf}) and
(\ref{quantDim}) with (\ref{thetatensor}) we have
$$
S_{i0}= {\sf dim} (U_i) \qquad \mbox{and} \qquad S_{ij}= \frac1{\theta_i\,
  \theta_j}\, \sum_{k\in I}\, N_{ij}{}^k\, \theta_k\, {\sf dim} (U_k) \ .
$$
If $\scrc$ contains only a finite number of isomorphism classes of
simple objects, i.e. $I$ is a finite set, then this structure makes it
into a \emph{modular tensor category}.

We choose square roots $\nu_i\in\IC^*$ such that $\nu_i^2=\Dim(U_i)$ for $i\in I$. Then
the trivalent basis morphisms introduced above may be normalized so that
$$
f_l\circ f_m^\vee = \delta_{l,m} \ \frac{\nu_i\,\nu_j}{\nu_k} \ \id_{U_k} \ ,
$$
and hence they satisfy
$$
\id_{U_i\otimes U_j}=\sum_{k\in I\, :\, N_{ij}{}^k\neq0} \
\frac{\nu_k}{\nu_i\, \nu_j} \ f_l^\vee\circ f_l \qquad \mbox{and} \qquad
{\sf Tr}_{U_k}\big(f_l\circ f_l^\vee\,\big) = \nu_i\, \nu_j\, \nu_k
$$
for each $l=1,\dots,N_{ij}{}^k$.
This choice of basis will be exploited in our gluing constructions
later on. 

As we explain in Appendix~C, for our purposes we may assume that, as
an abelian category, $\scrc$
arises in the form of representation categories $\scrr=\Rep(\ca)$ of associative
algebras; however, in general there is no natural monoidal
structure. What is needed here is a strengthening of this result
proven by Ostrik~\cite[\S4.1]{Ostrik}: If $\scrc$ is a semisimple
rigid monoidal category enriched over $\Vect$ with finitely many
simple objects, then $\scrc\cong\Rep(\ca)$ where $\ca$ is a (weak)
Hopf algebra. Then the objects $\Ob(\scrr)$ are the representations of the
algebra $\ca$, while the morphisms
$\Hom_\scrr(V,W)$ are intertwiners $f:V\to W$ between $\ca$-modules
$V,W$, i.e. equivariant maps $f(a\triangleright v)=a\triangleright
f(v)$ for $a\in\ca$, $v\in V$. The simple objects $U_i$, $i\in I$ are the irreducible representations of $\ca$, the
monoidal structure $\otimes$ is the tensor product of
$\ca$-modules, and the tensor unit $\Idd$ is the trivial
representation $U_0 \cong\IC$. The
category $\scrr$ is then evidently rigid, i.e. left and right
duals exist for every object $V$, which in this case coincide with the
usual vector space dual $V^\vee={}^\vee V=V^*$ regarded as a conjugate
representation with $V^{**}\cong V$. The linear map
$e_V:V^*\otimes V\to\IC$ is the evaluation $e_V(\varphi\otimes_\ca
v)=\varphi(v)$, while $i_V:\IC\to V\otimes V^*\cong\End_\ca(V)$ is the
coevaluation given by
 $i_V(1)=\id_V$. For a generic noncommutative Hopf
 algebra $\ca$, the tensor products $V\otimes W$ and $W\otimes V$ of
 $\ca$-modules need
 not be isomorphic. However, if $\ca$ is a quasitriangular Hopf algebra, then there
 is also a natural non-trivial isomorphism $V\otimes W\to W\otimes V$ of
 $\ca$-modules, which defines a braiding on the category $\scrr$. These
 facts will all be exploited in our ensuing constructions.

The corresponding Grothendieck group $K_0(\scrr)$ is generated by the isomorphism classes $[U_i]$
of simple objects modulo the subgroup generated by the elements
$[U]+[W]-[V]$ whenever $0\to U\to V\to W\to 0$ is an exact sequence in
the category $\scrr$; in
particular, with this definition we have $[U\oplus V]=[U]+[V]$. In
most instances the Grothendieck group also has a product structure
defined by
$[U\otimes V]=[U]\cdot[V]$. Then $K_0(\scrr)$ is
the representation ring with structure constants $N_{ij}{}^k$ and unit
element $[\Idd]$, which is
isomorphic to the commutative ring of characters on $\ca$. The tensor category $\scrr$ may be thought of as a
\emph{categorification} of the character ring $K_0(\scrr)$; inequivalent
categorifications are typically classified by group
cohomology. For example, when $\scrr=\Rep(SU(N)_k)$ is the modular tensor
category of integrable representations of $G=SU(N)$ at level $k\in\IZ$,
there are $N$ monoidal categories with Grothendieck ring
isomorphic to $K_0(\Rep(SU(N)_k))$: the representation category
$\Rep(SU(N)_k)$ itself, and certain twists of $\Rep(SU(N)_k)$ induced
by the natural $\IZ_N$-grading on the fusion algebra of $K_0(\Rep(SU(N)_k))$~\cite{Ostrik}. However, in the
following we will typically deal with cases where this product is not
well-defined on $K_0(\scrr)$.

When $\ca$ is additionally a ribbon Hopf algebra, the tensor product and braiding in the representation
category $\scrr=\Rep(\ca)$ endows $\scrr$ with the structure of
a semisimple ribbon category. The twist coefficients $\theta_V\in\End_\scrr(V)$
provide a ``Casimir operator'' or ``ribbon element'', while the braiding operator (\ref{Sijdef}) gives a representation
of the ``modular $S$-matrix''. Together with the fusion coefficients
$N_{ij}{}^k$ of the representation ring,
these are all the data that we need to fully specify the gauge
theory; the gluing constraints are determined by
the pentagon and hexagon relations. Thus by abstracting the
gluing data into the semisimple ribbon category $\scrr$, we completely
recover the gauge theory.

\medskip

\subsection{Representations of the geometric surface category}\label{se:surfacecat}~\\[5pt]
The geometric surface category $\scrs$ is the tensor category of oriented surfaces with area. Its
objects $\Ob(\scrs)$ are collections of disjoint oriented circles $S^1$, its
morphisms $\Hom_\scrs$ are oriented 2-bordisms given by surfaces with
area between the circles, i.e. a morphism from an object $B_1$ to an
object $B_2$ is an oriented surface $C$ with area and boundary $\partial
C=B_1\sqcup(-B_2)$, and the monoidal
structure on the category is given by disjoint union of
circles. Composition is defined by concatenation of bordisms such that the
source and target objects have opposite orientation, and areas are additive. The identity
object is the empty one-manifold $\varnothing$; a morphism from
$\varnothing$ to itself is given by a compact oriented Riemann
surface $\Sigma$. The category
$\scrs$ has a duality structure given by the canonical (up to
homotopy) orientation reversing diffeomorphism of $S^1$ and declaring that the cobordisms
in (\ref{dualitycomps}) are equal to the tube bordism defined below.

A representation of the category $\scrs$ is a functor of monoidal
categories with duality
\beq
\scrf_{\ca}\, :\, \scrs \ \longrightarrow \ \Rep(\ca)
\label{functA}\eeq
for a suitable representation category $\scrr= \Rep(\ca)$. We require
$\scrf_{\ca}$ to be invariant with respect to boundary and area
preserving oriented
diffeomorphisms. Because our
categories have duality, it suffices to specify the images of the
basic cobordisms with one, two and three boundary circles
\beqa
\raisebox{-5.5pt}\Disk &\in& \Hom_{\scrs}(S^1,\varnothing) \ , \nonumber \\[4pt]
\raisebox{-6pt}\Tube &\in& \Hom_{\scrs}(S^1,S^1) \ , \nonumber
\\[4pt]
\raisebox{-11pt}\Pants &\in& \Hom_{\scrs}(S^1, S^1\sqcup S^1) \ .
\label{basiccob}\eeqa
Other morphisms (cobordisms) are then obtained by taking compositions
(concatenations) and tensor products (disjoint unions) of the
morphisms (\ref{basiccob}), and using the duality cobordisms
$e_{S^1}$ and $i_{S^1}$.

In order for (\ref{functA}) to be a well-defined functor, we demand
that it preserves the unit objects, so that
$\scrf_\ca(\varnothing)=U_0\cong\IC$, that it takes identity
morphisms to identity morphisms, so that
$\scrf_\ca(\id_B)=\id_{\scrf_\ca(B)}$ for all $B\in\Ob(\scrs)$, and
that it be compatible with the monoidal structures, so that
$\scrf_\ca(B_1\sqcup B_2)=\scrf_\ca(B_1)\otimes \scrf_\ca(B_2)$ for
$B_1,B_2\in\Ob(\scrs)$, which defines the gluing laws. Since
$S^1$ is an injective cogenerator for the category $\scrs$ and the
collection of simple objects $(U_i)_{i\in I}$ is a family of
cogenerators for $\scrr=\Rep(\ca)$ (see Appendix~C), we define
\beqa
\scrf_\ca(S^1)=K=\bigoplus_{i\in I}\, U_i
\label{FAS1KUi}\eeqa
and let
$$
\scrf_\ca(S^1\sqcup\cdots \sqcup S^1)=K\otimes\cdots\otimes K \ .
$$
For a disconnected cobordism $C=C_1\sqcup\cdots\sqcup C_n$, we define
$$
\scrf_\ca(C)=\scrf_\ca(C_1)\otimes\cdots\otimes \scrf_\ca(C_n) \ .
$$
Using duality in $\scrr$, we will identity morphisms in
$\Hom_\scrr(K\otimes\cdots\otimes K,\IC)$ with objects
$K\otimes\cdots\otimes K$. The representation categories we shall consider are not modular as
they contain infinitely many isomorphism classes of simple
objects. Thus one should strictly speaking reformulate all of our constructions in a
suitable ``completion'' of $\Rep(\ca)$, i.e. an ind-category whose objects
are infinite direct sums of objects of $\Rep(\ca)$ and their tensor
products. We shall not delve into such technical issues here; see
Appendix~C for details.

With these identifications made, we define the images of the basic
cobordisms (\ref{basiccob}) by their equivalence classes in a suitable completion
of the Grothendieck group $K_0(\scrr)$ as
\beqa
\bigg[\scrf_\ca\Big(\raisebox{-5.5pt}{\Disk}\, \Big) \bigg] &=& \sum_{i\in
  I}\, \Dim(U_i)\ \theta_i\ [U_i] \ , \label{capA} \\[4pt]
\bigg[\scrf_\ca\Big(\raisebox{-6pt}{\Tube}\, \Big) \bigg] &=& \sum_{i\in
  I}\, \theta_i\ [U_i\otimes U_i^* ] \ , \label{tubeA} \\[4pt]
\bigg[\scrf_\ca
\Big(\raisebox{-11pt}{\Pants}\, \Big) \bigg] &=& \sum_{i\in
  I}\, \frac{\theta_i}{\Dim(U_i)} \ [U_i\otimes U_i^* \otimes
U_i^* ] \ .
\label{pantsA}\eeqa
These definitions make sense in any semisimple ribbon category $\scrr$, see
e.g.~\cite{Kirillovjr1}; in particular, the basic object
(\ref{capA}) is called the
\emph{Gaussian} of the category $\scrr$. These elements are
subjected to the basic gluing axioms of two-dimensional topological
quantum field theory, which in the present case can be stated as
follows:
\begin{enumerate}
\item The commutativity constraint implies that (\ref{pantsA})
  carries an action of the symmetric
group $\mathfrak{S}_3$ which comes from area-preserving diffeomorphisms permuting the three
boundaries of the pair of pants bordism.
\item The associativity constraint implies that
  $$\bigg[\scrf_\ca\Big(\raisebox{-11pt}{\Pantsd} \,\Big)\bigg] =
  \bigg[\scrf_\ca\Big(
  \raisebox{-11pt}{\Pantsu}\, \Big)\bigg] $$ similarly
  carries an action of $\mathfrak{S}_4$.
\item The unit constraint comes from capping any of the three boundary
  circles of the pants bordism and it implies that
  $$\bigg[\scrf_\ca\Big(\raisebox{-11pt}{\DiskPants} \, \Big) \bigg]=
    \bigg[\scrf_\ca\Big(\raisebox{-6pt}{\Tube}\, \Big) \bigg] \ . $$
\end{enumerate}
Here and in the following we suppress the additions of areas in all
gluing laws for notational simplicity.
It follows from these axioms that
$$\bigg[\scrf_\ca\Big(\raisebox{-11pt}{\Pants}\,
\Big)^{\otimes(n-2)} \bigg] $$ carries an
$\mathfrak{S}_n$-action for all $n>0$, where we formally define
$$\bigg[\scrf_\ca\Big(\raisebox{-11pt}{\Pants}\,
\Big)^{\otimes(-1)} \bigg]
:=\bigg[\scrf_\ca\Big(\raisebox{-5.5pt}{\Disk}\, \Big) \bigg] 
\qquad \mbox{and} \qquad
\bigg[\scrf_\ca\Big(\raisebox{-11pt}{\Pants}\,
\Big)^{\otimes(0)} \bigg] :=
\bigg[\scrf_\ca\Big(\raisebox{-6pt}{\Tube}\, \Big) \bigg] \ . $$
There are two applications of
these gluing rules that we are primarily interested in.

First, let us consider the case of a compact oriented Riemann surface
$\Sigma_h$ of genus $h$, which as a morphism $\varnothing\to \varnothing$ of the category $\scrs$
can be decomposed into basic cobordims (\ref{basiccob}) as
follows:
\begin{itemize}
\item For $h=0$ we cut the sphere $\Sigma_0=S^2$ along a circle
$S^1$ into the connected sum of two caps.
\item For $h=1$ we cut the torus
$\Sigma_1=S^1\times S^1$ along two circles into the connected sum
of two tubes.
\item For genus $h>1$ we cut $\Sigma_h$ on $3h-3$ circles into
a connected sum of $2h-2$ pants. 
\end{itemize}
We then apply the functor
(\ref{functA}) using the duality evaluation $e_K:K^*\otimes K\to\IC$
and the fact that $\Hom_\scrr(U_i\otimes
U_j^*,\IC)=\Hom_{\scrr}(U_i,U_j)\cong\delta_{ij}\ \IC$. This results
in the quantity $[\scrf_\ca(\Sigma_h)]= \cz(\ca;\Sigma_h)\ [\Idd]$, where we may generally
write
\beq
\cz(\ca;\Sigma_h)= \sum_{i\in I}\, \Dim(U_i)^{2-2h}\ \theta_i
\label{czcaSigmah}\eeq
for all $h\geq0$.

The second class of surfaces we are interested in involves a chain of
$\ell$ Riemann spheres, which arises when we regard a generic
lens space $L(p,p'\,)=S^3/\Gamma_{p,p'}$ as a Seifert fibration over the two-sphere
(see~\cite{Aganagic:2005wn,Griguolo:2006kp}); here $(p,p'\,)$ are
coprime integers with $p>p' >0$. The base is
described by a projective line $\IP^1$ with an arbitrarily chosen
marked point at which the coordinate neighbourhood is modelled on
$\IC/\IZ_p$, with the cyclic group acting on the local chart
coordinate $z$ as $z\mapsto\e^{2\pi\ii/p}\,z$. We construct a line
V-bundle over this $\IP^1$ orbifold such that the
local trivialization over the orbifold point is modelled by
$\IC^2/\Gamma_{p,p'}$, where $\Gamma_{p,p'}\cong\IZ_p$
acts on the local coordinates $(z,w)$ of the base and fibre as $(z,w)
\mapsto(\e^{2\pi \ii /p}\,z,\e^{2\pi \ii p' /p}\,w)$. This identifies
the lens space $L(p,p'\, )$ as
the total space of the associated unit circle bundle. In the base
model with a chain of $\ell$ spheres, the projective lines intersect
only once
with their nearest neighbours to the right and to the left along the
necklace, and the degrees
$e_a\ge 2$, $a=1,\ldots,\ell$ are obtained by expanding the rational
number $\frac p{p'}>1$ in a simple continued fraction
\beq 
\frac p{p'}=[e_1,\dots,e_\ell]:=
e_1-{1\over\displaystyle e_{2}- {\strut
1\over \displaystyle e_{3}- {\strut 1\over\displaystyle\ddots {}~
e_{\ell-1}-{\strut 1\over e_\ell}}}} 
\nonumber \eeq 
with
$e_1$ the smallest integer $>\frac p{p'}$, and so on. For
example, for $p'=1$ there is only $\ell=1$ sphere with
degree $e_1=p$; the case $p'=p-1$, where the length of the chain is
$\ell=p-1$ and each degree is $e_a=2$, is considered
in~\cite{Aganagic:2005wn}. To glue the neighbouring spheres together,
we braid the composition of two cap bordisms, corresponding to open
disks in the base and fibre directions of the Seifert fibration~\cite{Aganagic:2005wn}, in $\scrs$ together using
the braiding symmetry (\ref{Sijdef}) to define the class
\beq
\bigg[\scrf_\ca\Big(\raisebox{-5.5pt}{\Diskd}\,\Big)^B\, \bigg] = \sum_{i,j\in
  I}\, S_{ij} \ \theta_i\, \theta_j \ [U_i^* \otimes U_j]
\label{sphereB}\eeq
in $K_0(\scrr)$, which carries an action of $\mathfrak{S}_2$. To
account for the non-trivial degrees $e_a$, following~\cite{BP,Aganagic:2005wn,Griguolo:2006kp,Szabo:2009vw} we glue
two caps in this way at the ends of $e_a$ tubes with classes
$$ \bigg[\scrf_\ca\Big(\raisebox{-6pt}{\Tube}\, \Big)^{\otimes e_a}\,
\bigg] \qquad \mbox{for} \quad a=1,\dots, \ell \ . $$
More precisely, in these constructions one should replace the geometric surface
category $\scrs$ with the category $\scrs^{L_1,L_2}$ of 2-cobordisms
$C$ endowed with line bundles $L_1$ and $L_2$ which are trivialized
over the boundary components of $\partial C$, as in~\cite{BP}, but for
brevity we
do not write this explicitly. In
this way we arrive at the partition function
\beqa
\cz(\ca;S^2,p,p'\,) = \sum_{i_1,\dots,i_\ell\in I}\, S_{0i_1}\,
S_{i_1i_2}\cdots S_{i_{\ell-1}i_\ell}\, S_{i_\ell 0}\
\theta_{i_1}^{e_1}\cdots \theta_{i_\ell}^{e_\ell} \ .
\label{czcaS2pp}\eeqa
In particular, for $p'=1$ we have
\beqa
\cz(\ca;S^2,p,1) = \sum_{i\in I}\, \Dim(U_i)^2\ \theta_i^p
\label{czcaS2p1}\eeqa
which coincides with (\ref{czcaSigmah}) at $p=1$ and $h=0$; this
calculation easily generalizes to a degree $p$ circle bundle over an
arbitrary Riemann surface $\Sigma_h$ and amounts to replacing the
categorical dimension factors
$\Dim(U_i)^2$ with $\Dim(U_i)^{2-2h}$ in the formula
(\ref{czcaS2p1}). This formalism could help in explicitly
identifying the suitable two-dimensional gauge theory duals to the
four-dimensional $\cn=2$ gauge theories on $S^1\times L(p,p'\,)$
considered in~\cite{Benini:2011nc,Alday:2013rs}. The
construction presented here can also be extended to generic Seifert fibrations over
Riemann surfaces $\Sigma_h$.

\medskip

\subsection{Constructing $q$-deformed Yang-Mills
  amplitudes}\label{se:qAxioms}~\\[5pt]
We will now specialise the construction of this section to the quantum group
$\ca=\cu_q(\frg)$ associated to the Lie algebra $\frg$ of the gauge
group $G=U(N)$; then $\ca$ has the structure of a quasitriangular
Hopf algebra which is described in Appendix~A. Let us briefly
summarise the structure of the semisimple ribbon category
$\scrr=\Rep(\ca)$ in this instance, restricting to representations
which admit a decomposition into weight spaces; see e.g.~\cite{BKLect} for further details.

As a tensor category, $\scrr$ is equivalent to the category of finite-dimensional representations of $G=U(N)$. In particular, the isomorphism classes of irreducible representations are again
labelled by partitions $\lambda=(\lambda_1,\dots,\lambda_N)$,
$\lambda_i\geq\lambda_{i+1}\geq0$; the unit object is the vacuum
module corresponding to the empty Young diagram with $\lambda=0$. The
dual of an object $V$ is the dual vector space $V^*$ with the left
$\ca$-module structure
$$
(a\triangleright \varphi)(v):=\varphi\big(S(a)\triangleright v\big)
$$
for $a\in\ca$, $\varphi\in V^*$ and $v\in V$, where $S$ is the
antipode of $\cu_q(\frg)$. The braiding on the category $\scrr$ is given by the
functorial isomorphisms
$$
B_{V,W}=P\circ (R\, \triangleright) \, :\, V\otimes W \ \longrightarrow \ W\otimes V \ ,
$$
where $R$ is the universal $R$-matrix for $\cu_q(\frg)$ and $P$ is the trivial
``flip'' braiding $P(v\otimes w)=w\otimes v$ for $v\in V$, $w\in W$;
this yields a non-trivial isomorphism between the tensor product
$\cu_q(\frg)$-modules $V\otimes W$ and $W\otimes V$.

To define the twist, we introduce functorial isomorphisms
$q^{\langle\rho,H\rangle}:V\to V^{**}$ which act as multiplication by
$q^{\langle\rho,\lambda\rangle}$ on the representation $\lambda$, and
set
$$
\theta=q^{\langle\rho,H\rangle}\ u^{-1} \ ,
$$
where $u\in\cu_q(\frg)$ is Drinfel'd's element. The twist $\theta$ is
a central element, and it determines the (universal) Casimir operator
for the quantum group $\cu_q(\frg)$: From the explicit
formulas in Appendix~A, it follows that
$u^{-1}$ acts as multiplication by $q^{\frac12\, \langle\lambda,\lambda\rangle}$
on the irreducible representation $\lambda$ and hence
$$
\theta_\lambda=q^{\frac12\, \langle\lambda,\lambda+2\rho\rangle} =
q^{\frac12\, C_2(\lambda)} \ .
$$
The functorial isomorphisms $\psi_V:V^{**}\to V$ are given by
$\psi_V(x)=u^{-1}\triangleright x$ for $x\in V^{**}$, and hence the
categorical trace of any $\ca$-module endomorphism $f:V\to V$ is the
quantum trace
$$
{\sf Tr}_V(f)=\Tr_V\big(q^{\langle\rho,H\rangle} \,f \big) \ .
$$
In particular, the categorical dimension of a highest weight $\cu_q(\frg)$-module
$U_\lambda$ coincides with the quantum dimension
$$
\Dim(U_\lambda)= s_\lambda\big(q^{\rho}\big)=\dim_q\lambda \ .
$$
The partition function (\ref{czcaSigmah}) is therefore given by
$$
\cz\big(\cu_q(\frg)\,;\,\Sigma_h\big) = \sum_\lambda\,
s_\lambda\big(q^{\rho}\big)^{2-2h}\ q^{\frac12\, C_2(\lambda)} \ ,
$$
which coincides with (\ref{qYM1}) for an $S^1$-fibration over
$\Sigma_h$ of degree $p=1$ when we identify $q=\e^{-g_s}$.

To compute the partition function (\ref{czcaS2pp}), we need the
braiding symmetry (\ref{Sijdef}) which can be expressed in
terms of specializations of Schur functions as
$$
S_{\lambda\mu}= s_\lambda\big(q^{\rho}\big) \
s_\mu\big(q^{\lambda+\rho}\big) = \Big(\, \prod_{k=1}^{N-1} \,
\frac1{\big(q^{-k/2}-q^{k/2} \big)^{N-k} } \, \Big) \ \sum_{w\in \mathfrak{S}_N}\,
(-1)^{|w|}\, q^{\langle w(\lambda+\rho),\mu+\rho\rangle} \ .
$$
This operator is an analytic continuation of the modular $S$-matrix of the $U(N)$ WZW
model in the Verlinde basis~\cite{Griguolo:2006kp}; the second equality here follows from the Weyl
character formula and it makes manifest the symmetry $S_{\lambda\mu}=S_{\mu\lambda}$.
Note that the fusion coefficients $N_{\lambda\mu}{}^\nu\in\IZ_{\geq0}$ of the
category $\scrr$ are the Littlewood-Richardson coefficients for the
fusion of $U(N)$ representations and are represented through Schur
functions by (\ref{LRcoeffs}).
It follows that
\beqa
\cz\big(\cu_q(\frg)\,;\, S^2,p,p'\,\big) &=&
\sum_{\lambda_1,\dots,\lambda_\ell}\, q^{\frac12\, (e_1\, C_2(\lambda_1)+\cdots+
  e_\ell\, C_2(\lambda_\ell))} \ s_{\lambda_1}\big(q^{\rho}\big)^2\,
s_{\lambda_2}\big(q^{\rho}\big) \cdots s_{\lambda_\ell}\big(q^{\rho}\big) \nonumber
\\ && \qquad \qquad \qquad \qquad \qquad \qquad \qquad \times \
s_{\lambda_2}\big(q^{\lambda_1+\rho}\big) \cdots
s_{\lambda_\ell}\big(q^{\lambda_{\ell-1}+ \rho}\big) \ . \nonumber
\eeqa
In particular, at $p'=1$ the partition function (\ref{czcaS2p1})
coincides with the partition function (\ref{qYM1}) at $h=0$ for
$q$-deformed Yang-Mills theory on the sphere.

Note that in our case, where $q$ is not a root of unity, the representation category
$\Rep(\cu_q(\frg))$ is not modular and should be dealt with in the
setting of ind-categories as discussed in
\S\ref{se:surfacecat}. Alternatively, for $q= \zeta_k$ a primitive $k$-th root of unity, one can work with a
``reduced'' version of $\cu_q(\frg)$ defined by imposing the
additional relations $E_i^r=0=F_i^r$ and $K_i^r=1$ on the generators,
where $r=\frac k2$ for $k$ even and $r=k$ when $k$ is odd. This is a
finite-dimensional Hopf algebra whose ribbon category of finite-dimensional integrable
representations has finitely many isomorphism classes of simple
objects; however, this category is not semisimple. See~\cite{Kirillovjr1,BKLect}
for the construction of a modular tensor category of representations
of the quantum group $\cu_q(\frg)$ for $q$ a root of unity.

\medskip

\subsection{Disk amplitudes}\label{se:Disk}~\\[5pt]
From (\ref{capA}) it follows that the Gaussian for the category
$\scrr=\Rep(\cu_q(\frg))$ is given by
$$
\bigg[\scrf_{\cu_q(\frg)}\Big(\raisebox{-5.5pt}{\Disk}\, \Big) \bigg] =
\sum_\lambda\, s_\lambda\big(q^{\rho}\big) \ q^{\frac12\, C_2(\lambda)} \ [U_\lambda] \ .
$$
On the other hand, we can also consider the category of
representations of the quantum group $\ca=\cu_q(\frh)$ corresponding
to the Cartan subalgebra $\frh\subset\frg$, i.e. the Lie algebra of
the maximal abelian subgroup $T= U(1)^N\subset U(N)$. The irreducible
representations $U_n$ are now parametrized by the weight lattice $n\in\IZ^N$
and are all one-dimensional, so that $\Dim(U_n)=1$. The duality
$U_n^*=U_{-n}$ and fusion rules
$$
U_n\otimes U_m=U_{n+m}
$$
furnish the abelian group $\IZ^N$.
From the explicit expression in Appendix~A it follows that the
universal $R$-matrix acts as multiplication by $q^{\frac12\, \langle
  n,m\rangle}$ on $U_n\otimes U_m$, and hence the braiding is given by
$$
B_{U_n,U_m}(v\otimes w)= q^{\frac12\, \langle n,m\rangle} \ w\otimes v \ .
$$
The twist eigenvalue of a simple
object $U_n\in\Ob(\Rep(\cu_q(\frh)))$ is
$$
\theta_n=q^{\frac12\, \langle n,n\rangle} \ ,
$$
while the braiding symmetry is given by
$$
S_{nm}=q^{\langle n,m\rangle} \ .
$$
Whence the Gaussian (\ref{capA}) for this category is given by
$$
\bigg[\scrf_{\cu_q(\frh)}\Big(\raisebox{-5.5pt}{\Disk}\, \Big) \bigg]
=
\sum_{n\in\IZ^N}\, q^{\frac12\, \sum_i\, n_i^2} \ [U_n] \ .
$$

In the approach of Etingof and Kirillov~\cite{EK1}, the Kostant
identity plays a prominent role. It can
be stated as the relationship between Gaussians in the Grothendieck
rings of the two ribbon
categories considered here given by
\beq
\bigg[\scrf_{\cu_q(\frg)}\Big(\raisebox{-5.5pt}{\Disk}\, \Big) \bigg]
=\frac1{Z_N(q)} \ 
\bigg[\scrf_{\cu_q(\frh)}\Big(\raisebox{-5.5pt}{\Disk}\, \Big) \bigg]
\ ,
\label{Kostant}\eeq
where the normalization $Z_N(q)$ is the Chern-Simons partition
function (\ref{UCSexact}) on $S^3$.
The identity (\ref{Kostant}) may be regarded as a categorification of
the Weyl integral formula (\ref{Weylintgroup}) which
relates the Haar measure for integration of $G$-invariant functions on
$G=U(N)$ with the Haar measure for integration of symmetric functions on its
maximal torus $T =U(1)^N$, when we identify classes in the Grothendieck
rings with characters. It is identical to the
result of~\cite{Romo:2011qp} that the character expansion of the
Villain lattice action gives the propagator of $q$-deformed
two-dimensional Yang-Mills theory. The Kostant identity is also derived
in~\cite[App.~C]{Aganagic:2005dh} as an expression for the disk
amplitude in terms of a theta-function on the weight lattice $\IZ^N$, within
the framework of the standard gluing rules for
$q$-deformed two-dimensional Yang-Mills theory. 

Evaluating both sides
of (\ref{Kostant}) at the principal specialization $x=q^{\rho}$ of
the respective characters, we arrive at a simple representation of the
partition function (\ref{qYM1}) of $q$-deformed Yang-Mills theory on
$S^2$ with $p=1$ given by
\beq
\mathcal{Z}_{\mathrm{M}}^{\left(1\right) }(q; S^2)
= \cz\big(\cu_q(\frg)\,;\,S^2\big) =
\frac1{Z_N(q)} \ \prod_{j=1}^N\, \Theta\big(q^{\frac12\, (N+1-2j)} \,;\,
q\big) \ ,
\label{ZMVillain}\eeq
where $\Theta(z;q)$ is the Jacobi theta-function
(\ref{theta3zq}). Using standard modular transformation properties of
$\Theta(z;q)$, one shows that (\ref{ZMVillain})
computes the fractional instanton contributions to the partition
function of $\cn=4$ gauge theory on the surface
$\mathcal{O}(-1)\to\IP^1$ (the blow-up of $\IC^2$ at a point)~\cite{Griguolo:2006kp}.

Nevertheless, the two categorifications of the gauge 
theory are inequivalent: The partition function (\ref{czcaS2pp}) for
the category $\Rep(\cu_q(\frh))$ is given by the theta-function
\beq
\cz\big(\cu_q(\frh)\,;\, S^2,p,p'\, \big) =
\sum_{\mbf{n}\in\IZ^{\ell\,N}} \, q^{\frac12\, \langle \mbf{n} ,\mbf{C} \cdot
    \mbf{n} \rangle} \ ,
\label{ZUqfrh}\eeq
where
$$
\langle \mbf{n} ,\mbf{C} \cdot \mbf{n} \rangle = \sum_{a,b=1}^\ell\,
C_{ab}\, \langle n^a,n^b\rangle \qquad \mbox{for} \quad n^a,n^b \in\IZ^N \ ,
$$
and
$$
C=(C_{ab}) = \begin{pmatrix} e_1 & 1 & 0 & \cdots &0\\
1 & e_2  & 1& \cdots &0\\
0 & 1 & e_3 &\cdots&0\\
\vdots &\vdots & \vdots &\ddots&\vdots\\0&0&0&\cdots& e_\ell
\end{pmatrix}
$$
is the intersection form of the chain of $\ell$ base spheres. For
$p'=p-1$, so that $\ell=p-1$ and $e_a=2$ for $a=1,\dots,\ell$, the matrix $C$ is
essentially the Cartan
matrix of the $A_{p-1}$ Dynkin diagram and (\ref{ZUqfrh}) the
$N$-th power of the theta-function on the weight lattice of the Lie
algebra $\frsl(p)$.

\medskip

\subsection{Combinatorial Hopf algebra structure}\label{se:CombHopf}~\\[5pt]
Our categorical framework points to another way of understanding the
Hopf algebraic structure of two-dimensional Yang-Mills amplitudes, at
least in the infinite rank limit that was considered in
\S\ref{se:crystal}; we briefly describe this structure now, partly to
set the stage for the constructions of \S\ref{se:Defect} and \S\ref{se:Ref}. For this, we first note that it suffices to work
with ordinary representations of the gauge group $G=U(N)$. Let
$\cu(\frg)$ be the ordinary universal enveloping algebra of the Lie
algebra $\frg$, and let $\cu(\frg)[[g_s]]$ be its $g_s$-adelic completion
in the parameter $g_s$, with $q:=\e^{-g_s}$, consisting of formal power
series in $g_s$ with coefficients in $\cu(\frg)$. It is
well-known~\cite[Prop.~3.16]{Drinfeld} that there is an algebra
isomorphism
$$
\varphi\,:\, \cu_q(\frg) \ \longrightarrow \ \cu(\frg)
$$
and an invertible twisting cochain $\cf=1\otimes1+\co(g_s)$ which
relates the underlying Hopf algebras of $\cu_q(\frg)$ and
$\cu(\frg)[[g_s]]$; the cochain
$\cf\in\cu(\frg)[[g_s]]\otimes\cu(\frg)[[g_s]]$ induces a non-trivial
coassociator $\Phi\in
\cu(\frg)[[g_s]]\otimes\cu(\frg)[[g_s]]\otimes\cu(\frg)[[g_s]]$ in this
correspondence. At the level of representation categories, this
induces a functorial equivalence between $\Rep(\cu_q(\frg))$ and
Drinfel'd's category of $\frg$-modules with the usual tensor product but with
non-trivial associativity isomorphisms induced by $\Phi$~\cite{KazLusz}.

We will further describe our Yang-Mills amplitudes in the
representation categories of symmetric groups. For this, we use
Frobenius-Schur duality which establishes a bijection between certain
representations of $\frS_n$ and of $U(N)$ (see e.g.~\cite{review}); this
duality will also play a role in \S\ref{se:Ref} when we study the
corresponding refined gauge theory amplitudes. Let $V_{\rm
  fund}=\IC^N$ be the fundamental representation of $G=U(N)$. Then the
$n$-th symmetric power $V_{\rm fund}^{\odot n}$ for $n\in\IZ_{>0}$
carries, in addition to the $G$-action, an action of the symmetric
group $\frS_n$ by permuting factors. The actions of $G$ and $\frS_n$
commute, so $V_{\rm fund}^{\odot n}$ is a representation of
$G\times\frS_n$ which is completely reducible to the form
$$
V_{\rm fund}^{\odot n} \cong \bigoplus_\lambda\, U_\lambda\otimes
u_\lambda \ ,
$$
where $U_\lambda$ is the irreducible representation of $U(N)$
corresponding to a partition $\lambda=(\lambda_1,\dots,\lambda_N)$
with $\sum_i\, \lambda_i=n$, and $u_\lambda$ is the representation of
$\frS_n$ corresponding to the Weyl character $s_\lambda(x)$; this
yields a \emph{Frobenius-Schur correspondence} between the
representations $U_\lambda$ and $u_\lambda$. A $q$-deformation of this
duality to the quantum group gauge symmetry underlying $q$-deformed
Yang-Mills amplitudes is described in~\cite{deHRT}, wherein the
symmetric group $\frS_n$ is deformed to a Hecke algebra.

This correspondence yields a combinatorial Hopf algebra structure on
$U(\infty)$ Yang-Mills amplitudes in the following way. The ring
$\frSym$ of symmetric functions carries a Hopf algebra structure with
product map $\mu: \frSym\otimes \frSym\to\frSym$ defined by the
Littelwood-Richardson expansion of Schur functions (\ref{LRcoeffs});
its adjoint with respect to the Hall inner product is a coproduct
$\Delta:\frSym\to \frSym\otimes \frSym$ making $\frSym$ into a
coalgebra with
$$
\Delta(s_\lambda) = \sum_{\mu,\nu}\, N_{\mu\nu}{}^\lambda \
s_\mu\otimes s_\nu \ .
$$
The crucial property to check in this claim is coassociativity:
$(\Delta\otimes 1)\, \Delta= (1\otimes\Delta)\, \Delta$. With the
inclusions of symmetric groups $\frS_n\times\frS_m\hookrightarrow
\frS_{n+m}$, this is done by interpreting the product $\mu$ in $\frSym$ as
induction from representations $\frSym\otimes \frSym$ of
$\frS_n\times\frS_m$ to $\frS_{n+m}$, and the coproduct $\Delta$ as
restriction of representations $\frSym$ of $\frS_{n+m}$ to
$\frS_n\times \frS_m$. The coassociativity condition is then
equivalent to the commutativity of the induction and restriction
morphisms of representations, which follows by Mackey theory. The unit
and counit are given respectively by
$$
\eta(1)=s_0 \qquad \mbox{and} \qquad
\varepsilon(s_\lambda)=\delta_{\lambda,0} \ ,
$$
while the antipode
$$
S(s_\lambda)=(-1)^{|\lambda|}\, s_{\lambda'}
$$ 
of the Hopf algebra corresponds to the
well-known involution $\omega$ of the ring of symmetric
functions~\cite{Macdonald}, see e.g.~\cite{Mazza,Fauser}; that the
antipode $S$ here is an involution is a consequence of bicommutativity
of the Hopf algebra structure on $\frSym$.

Geissinger~\cite{Geissinger} shows that the induction and restriction
functors associated to these morphisms induce in this way a bialgebra
structure on the direct sum of Grothendieck groups of representation
categories of $\frS_n$-modules over all $n\geq0$, which corresponds to
the infinite-rank case of the corresponding Yang-Mills
amplitudes. These combinatorial Hopf algebras are primarily associated
with Grothendieck groups, but they can also be reformulated in terms
of the underlying representation categories in a similar vein to the
constructions of this section; via Frobenius-Schur duality, the ring
$\frSym$ is thus interpreted as the self-dual Grothendieck Hopf
algebra on $\bigoplus_{n\geq0}\, \IC[\frS_n]$, where $\IC[\frS_n]$ is
the group ring of the symmetric group $\frS_n$. In particular, Bump
and Gamburd~\cite{BG} show that the coassociativity of this Hopf
algebra is equivalent to the generalized Cauchy-Binet formula
(\ref{ScSUSY}) for the supersymmetric Schur polynomials, and hence the
Hopf algebraic structure is intimately tied to the underlying
supersymmetric structure of $U(\infty)$ Yang-Mills amplitudes
discussed in \S\ref{se:crystal}. Of course, by virtue of the
identification of Schur functions as characters of irreducible
representations of $U(\infty)$, this structure makes the Grothendieck
group $K_0(\Rep(U(\infty)))$ into a graded self-dual, bicommutative Hopf
algebra (see
e.g.~\cite{Fauser}), and ultimately also the representation category
$\Rep(\cu_q(\frg))$ itself in the $N\to\infty$ limit by replacing
$s_\lambda\mapsto U_\lambda$ in the structure maps above. This Hopf
algebra structure is exploited in \S\ref{se:Defect} below.

\medskip

\subsection{Defect operators and module categories}\label{se:Defect}~\\[5pt]
For completeness, and also in preparation for our discussion of
refinement in \S\ref{se:Ref}, we conclude this section by describing
how the computations of correlators in $q$-deformed two-dimensional
Yang-Mills theory fit into our categorical framework. Correlation
functions are associated with insertions of defect operators in
partition functions; generalized defect observables are constructed
via fundamental boundary observables in representations of the
geometric surface category $\scrs$. For each such boundary observable,
we choose a basepoint on each boundary and associated a simple object
$U_i\in\Ob(\scrc)$, $i\in I$, to that boundary. There are two
perspectives one can take in constructing defect observables: one
through the gluing rules of this section, and one through
considerations of the Frobenius algebra (\ref{FAS1KUi}) of isomorphism
classes of simple objects of the category $\scrc$. We will begin here
with the former perspective which can be used to immediately write
down explicit formulas for the correlators, as it is this point of
view that will be taken in \S\ref{se:Ref}. There are three classes of
defect operators which are of relevance for the construction of
generic two-dimensional Yang-Mills amplitudes.

Firstly, there is the extension of the representations considered in
\S\ref{se:surfacecat} to Riemann surfaces with boundaries, which as
morphisms in $\scrs$ live in $\Hom_\scrs(\varnothing, S^1\sqcup\cdots
\sqcup S^1)$. They are easily constructed from the basic amplitudes with
one, two and three boundaries given in (\ref{capA})--(\ref{pantsA})
respectively. For a Riemann surface of genus $h$ having $b$ punctures
with boundary conditions fixed to the holonomy eigenvalues
$u_1,\dots,u_b\in (S^1)^N$ of the gauge connection around the boundary
circles, by evaluation of the corresponding characters at these holonomies we find
that the
general partition function for the usual $q$-deformed gauge theory is given by
$$
\cb_{u_1,\dots,u_b}\big(\cu_q(\frg)\,;\, \Sigma_h,p,1\big) =
\sum_\lambda\, s_\lambda\big(q^\rho\big)^{2-2h-b}\
s_\lambda(u_1)\cdots s_\lambda(u_b) \ q^{\frac p2\,
  C_2(\lambda)} \ .
$$
These amplitudes naturally reflect the monoidal structure of the
representation category $\scrr$, and also the Hopf algebra structure
from \S\ref{se:CombHopf}. These boundary observables are considered
in~\cite{deHRT,Gaiotto:2012xa,ABFH} and related to the four-dimensional $\cn=2$
superconformal index in~\cite{Gadde:2011ik}.

Secondly, we can consider more general closed defect observables which
correspond to closed non-selfintersecting loops on the surface
$\Sigma_h$. They are given by
Wilson loops on $\Sigma_h$ equipped with a
choice of marked point on the loop which corresponds to an insertion of
a defect operator in the partition function; Wilson loop observables
in $q$-deformed Yang-Mills theory are considered
in~\cite{Buffenoir:1994fh,deHRT}. Consider, for example, a
correlation function of a single Wilson loop operator
$\Tr_\lambda\big(\cp\exp\ii\oint_{C}\, A\big)$ in the representation
$\lambda$ on a non-selfintersecting oriented closed curve
$C\subset\Sigma_h$ which divides the Riemann surface $\Sigma_h$ into
inner and outer faces of genera $h_1$ and $h_2$, with $h=h_1+h_2$. This
correlator is determined in terms of the fusion
coefficients $N_{\lambda_1\lambda}{}^{\lambda_2}$ of the
representation category $\scrr$ to be
$$
\cw_\lambda\big(\cu_q(\frg)\,;\, \Sigma_h,p,1\big) =
\sum_{\lambda_1,\lambda_2}\, N_{\lambda_1\lambda}{}^{\lambda_2}\ s_{\lambda_1}\big(q^\rho\big)^{1-2h_1}\,
s_{\lambda_2}\big(q^\rho\big)^{1-2h_2}\
q^{\frac p2\, C_2(\lambda_1)}\,
q^{\frac p2\, C_2(\lambda_2)} \ ,
$$
where the representations $\lambda_1$ and $\lambda_2$ label the inner and outer faces
respectively.

Thirdly, we can consider Wilson loop observables in Chern-Simons
theory on Seifert manifolds which wrap around the $S^1$ fibre; they
are studied in~\cite{Beasley}. In two-dimensional $q$-deformed Yang-Mills
theory they correspond to defect holonomy punctures given by
correlators of the gauge-invariant operators
$\Tr_\lambda\exp(\ii\phi)$ inserted on the base Riemann surface
$\Sigma_h$, which represent the holonomy of the Chern-Simons gauge
connection around the $S^1$ fibre; these defect punctures were
considered in~\cite{Aganagic:2004js,Aganagic:2005dh} and shown
in~\cite{Gaiotto:2012xa,ABFH} to correspond to insertions of supersymmetric surface
operators in the four-dimensional $\cn=2$
superconformal index. The amplitude for $n$ defect punctures in
representations $\lambda_1,\dots,\lambda_n$ on $\Sigma_h$ can also be
written entirely in terms of data in the underlying semisimple ribbon
category $\scrr$, and one has
\beqa
\cp_{\lambda_1,\dots,\lambda_n}\big(\cu_q(\frg)\,;\, \Sigma_h,p,1 \big)
&=& \sum_\lambda\, (\dim_q\lambda)^{2-2h-n}\ S_{\lambda\lambda_1}\cdots S_{\lambda\lambda_n}\ q^{\frac p2\,
  C_2(\lambda)}
\nonumber \\[4pt] &=& \sum_\lambda\,
s_\lambda\big(q^\rho\big)^{2-2h}\
s_{\lambda_1}\big(q^{\lambda+\rho}\big)\cdots
s_{\lambda_n}\big(q^{\lambda+\rho}\big)\ q^{\frac p2\, C_2(\lambda)}
\nonumber \\[4pt] &=& \sum_\lambda\, s_\lambda\big(q^\rho\big)^{2-2h}\
q^{\frac p2\, C_2(\lambda)} \nonumber \\ && \qquad \times \ \sum_{\mu_1,\dots,\mu_{n-1}}\,
N_{\lambda_1\lambda_2}{}^{\mu_1}\, N_{\mu_1\lambda_3}{}^{\mu_2}\cdots
N_{\mu_{n-2}\lambda_n}{}^{\mu_{n-1}}\
s_{\mu_{n-1}}\big(q^{\lambda+\rho}\big)
\label{Pdefect}\eeqa
where we used (\ref{LRcoeffs}) in the last line.

The independence of correlators such as (\ref{Pdefect}) on the
insertion points $x_i\in\Sigma_h$ of the operators
$\Tr_{\lambda_i}\exp(\ii\phi(x_i))$ can be understood by appealing to
an alternative description of these formulas in terms
of a functor from the geometric surface category $\scrs$ to a module
category of Frobenius algebras. Defect punctures correspond to left
$A$-modules and non-selfintersecting Wilson line defects to
$A$-bimodules over a suitable Frobenius algebra object $A$ in the
semisimple ribbon category $\scrc$ (see e.g.~\cite{FRSI}); for the
representation category $\scrc=\scrr=\Rep(\ca)$ this is the Frobenius
algebra $A=K$ of conjugacy classes of $\ca=\cu_q(\frg)$ in
(\ref{FAS1KUi}). In the remainder of this section we briefly explain
how the construction of
correlators of defect operators fits into our categorical framework in
this way.

Again we begin by sketching the relevant structures involved from
category theory. Recall that a Frobenius algebra $A$ is an associative, unital algebra over $\IC$
equipped with a linear function $\varepsilon:A\to\IC$ such that the bilinear
pairing defined by $(a,b):=\varepsilon(a\,b)$ for $a,b\in A$ is
nondegenerate. The basic example is the algebra $A=\IM_n$ of $n\times n$
matrices over $\IC$, with $\varepsilon(a)=\Tr(a)$. More
generally, for any Frobenius algebra $(A,\varepsilon)$ we can enrich
the algebra $\IM_n(A):=\IM_n\otimes A$ with the Frobenius structure
$\varepsilon\circ\Tr:\IM_n(A)\to\IC$. By Wedderburn's theorem, a
finite-dimensional simple algebra is isomorphic to a matrix algebra
over a division ring, and hence every finite-dimensional semisimple algebra admits a
Frobenius structure. Another example is provided by the complex de~Rham cohomology ring
$A=H^*(X)$ of a compact $n$-dimensional complex manifold
$X$. This is an algebra under the wedge product of differential forms
on $X$. Then the integration $\int_X:H^*(X)\to\IC$ over $X$
provides a Frobenius structure $\varepsilon$ on $A$. This example
illustrates a geometric way in which to think of these
algebras. A semisimple Frobenius algebra $A$ is always the algebra of
$\IC$-valued functions on the set $X={\rm Spec}(A)$ of minimal ideals
of $A$, equipped with a ``volume form'' $\varepsilon$ which assigns a
measure $\varepsilon_x$ to each point $x\in X$. The Frobenius
structure $\varepsilon$ thus provides an ``integration'' (or trace)
over the ``space'' $X$. When $A$ is finite-dimensional, $X$ is
just a finite set of points. Frobenius algebras play a well-known role
in two-dimensional topological field theory, see e.g.~\cite{Kock} for
an introduction.

This notion extends to the monoidal categories $\scrc$ we are
interested in. An associative, unital algebra in $\scrc$ is an
object $A\in\Ob(\scrc)$ together with a ``multiplication''
$\mu\in\Hom_\scrc(A\otimes A,A)$ and a ``unit''
$\eta\in\Hom_{\scrc}(\Idd,A)$ satisfying the associativity condition
$$
\mu\circ(\mu\otimes\id_A)=\mu\circ(\id_A\otimes\mu)
$$
and the unit condition
$$
\mu\circ(\eta\otimes\id_A)=\id_A=\mu\circ(\id_A\otimes\eta) \ .
$$
An algebra in $\scrc=\Vect$, with the usual tensor product of vector
spaces, is precisely an associative $\IC$-algebra. An algebra in the
dual category $\scrc^{\rm op}=\Vect^{\rm op}$, with the usual tensor
product of vector spaces, is precisely a coassociative
$\IC$-coalgebra.

Suppose that $A$ is an algebra in $\scrc$. Then $A$ is said to be a
Frobenius algebra in the monoidal category $\scrc$ if there exists morphisms
$\varepsilon\in\Hom_\scrc(A,\Idd)$ and $\Delta\in\Hom_\scrc(A,A\otimes
A)$ such that $(A,\varepsilon,\Delta)$ is a coalgebra and
$$
(\id_A\otimes\mu)\circ(\Delta\otimes\id_A)=\Delta\circ\mu=
(\mu\otimes\id_A)\circ(\id_A\otimes\Delta) \ .
$$
This means that $\Delta:A\to A\otimes A$ is a morphism of
$A$-bimodules. If in addition $\scrc$ is braided, then $A$ is
commutative when
$$
\mu\circ B_{A,A}=\mu
$$
on $A\otimes A\to A$. This is equivalent to the cocommutativity
condition
$$
B_{A,A}\circ\Delta=\Delta
$$
on $A\to A\otimes A$. We say that $A$ is haploid if
$\dim\,\Hom_\scrc(\Idd,A)=1$.

There are two other conditions that we will need below. In the
category $\scrc$, there are two canonical coevaluations
$$
d_A\in\Hom_\scrc(\Idd,A\otimes A^\vee) \qquad \mbox{and} \qquad
\tilde d_A\in\Hom_\scrc(\Idd,A^\vee\otimes A) \ .
$$
Then $A$ is symmetric if the natural isomorphisms of $A$-bimodules 
$$
\Phi_1:=\big((\varepsilon\circ\mu)\otimes\id_{A^\vee}\big)
\circ(\id_A\otimes d_A) \qquad \mbox{and} \qquad
\Phi_2:=\big(\id_{A^\vee}\otimes(\varepsilon\circ\mu)\big)
\circ(\tilde d_A\otimes\id_A)
$$
in $\Hom_{\scrc}(A,A^\vee)$ coincide, $\Phi:=\Phi_1=\Phi_2$. Haploid
algebras are symmetric. The algebra object $A$ is special if $\Delta$
is a right inverse of $\mu$ and
$\varepsilon\circ\eta={\Dim}(A)~\id_{\Idd}$, where ${\Dim}(A)$
is the categorical dimension of $A\in\Ob(\scrc)$. This generalizes the
notion of separable algebras over $\IC$.

For the representation category $\scrc=\scrr=\Rep(\cu_q(\frg))$, we take the
cogenerator $A=K$ from (\ref{FAS1KUi}) and induce a Frobenius
structure on it from the combinatorial Hopf algebra structure of \S\ref{se:CombHopf}. Using
$\Hom_\scrr(U_\lambda,A)\cong\IC$, the multiplication $\mu:A\otimes A\to A$
is provided by the image of the pants bordism (\ref{pantsA}) regarded
as a morphism in the dual category $\scrr^{\rm op}$; explicitly
\beq
\mu(u_\mu\otimes u_\nu)= \sum_\lambda\, N_{\mu\nu}{}^\lambda\
\id_{U_\lambda}(u_\mu\otimes u_\nu)
\label{Amult}\eeq
for all simple subobjects $U_\mu,U_\nu,U_\lambda$ of $A$ and all
vectors $u_\mu\in U_\mu$, $u_\nu\in U_\nu$. Let us explain the meaning of this 
formula. Let $e_\mu\in\Hom_\scrr(U_\mu,A)$ be the basis of canonical
inclusions of simple subobjects, and let
$e^\mu\in\Hom_\scrr(A,U_\mu)$ be the dual basis of projections. Then
the composition $e^{\lambda}\circ\mu\circ(e_\mu\otimes e_\nu)$ is an element of
$\Hom_\scrr(U_\mu\otimes U_\nu,U_{\lambda} )\cong
{N_{\mu\nu}{}^\lambda}\,
\IC \,\id_{U_\lambda}$. The associativity of the product
(\ref{Amult}) follows easily from the associativity
relations for the fusion coefficients $N_{\mu\nu}{}^\lambda$; this multiplication makes $A$
into a (braided) noncommutative algebra.
The comultiplication is given by
\beq
\Delta(u_\lambda)= \sum_{\mu,\nu} \, N_{\mu\nu}{}^\lambda \ \id_{U_\mu\otimes
  U_\nu}(u_\lambda)
\label{Acomult}\eeq
where $u_\lambda\in U_\lambda$. This endows $A$ with the
structure of a noncocommutative coalgebra.
The Frobenius
counit $\varepsilon:A\to\IC$ is provided by the image of
the cap bordism (\ref{capA}); it is dual to the unit $\eta$ of $A$
which is
just the vacuum representation. Nondegeneracy of the inner product $\varepsilon\circ\mu$
follows from the gluing laws, which equate it to the image of the tube
bordism (\ref{tubeA}). The Frobenius algebra object $A$ obtained in this way is
symmetric but not special. Choosing different defects corresponds to selecting $A$-modules
$\tt M$; the original defect operator corresponds to the trivial $A$-module
$\tt A$ acting on itself via its multiplication morphism
$\mu$. We shall now study the Morita equivalence class
of the Frobenius algebra $A$, i.e. the $A$-module structures
provided by all other boundary defect operators.

Let $A$ be an algebra in the semisimple monoidal category $\scrc$. A
(left) $A$-module is a pair $\modM=(M,\varrho)$, where $M\in\Ob(\scrc)$
and $\varrho\in\Hom_{\scrc}(A\otimes M,M)$ with the relations 
$$
\varrho\circ(\mu\otimes\id_M)=\varrho\circ(\id_A\otimes\varrho) \qquad \mbox{and}
\qquad \varrho\circ(\eta\otimes\id_M)=\id_M \ .
$$
The space of morphisms between two left $A$-modules $\modM_1$ and
$\modM_2$ forms a $\IC$-linear subspace of $\Hom_\scrc(M_1,M_2)$
denoted $\Hom_A(\modM_1,\modM_2)$. Let $\Mod_\scrc(A)$ be the category
of left modules over the Frobenius algebra $A$ in the ribbon
category $\scrc$; its objects
are the associated boundary defect fields. When $A$ is a special
Frobenius algebra, then semisimplicity of $\scrc$ implies that the
category $\Mod_\scrc(A)$ is semisimple~\cite[Prop.~5.24]{FS1}. Similarly one defines $A$-bimodules
(see e.g.~\cite[Def.~4.5]{FRSI}), but it suffices to consider left
$A$-modules by~\cite[Rem.~12]{Ostrik}: In any braided tensor category
$\scrc$, an $A$-bimodule can equivalently be regarded as a left $A\otimes
A^{\rm op}$-module. This means that it suffices to focus our attention to
insertions of boundary defects on the Riemann surface $\Sigma_h$.

The category $\Mod_\scrc(A)$ is not a monoidal category, but it
carries the structure of a module category. For this, let
$\modM=(M,\varrho)$ be any left $A$-module with object $M\in\Ob(\scrc)$
and morphism $\varrho\in\Hom_\scrc(A\otimes M,M)$, and let
$X\in\Ob(\scrc)$ be any object of $\scrc$. Then $\modM\otimes
X:=(M\otimes X,\varrho\otimes\id_X)$ has the natural structure of a left
$A$-module. For any $X,Y\in\Ob(\scrc)$, the associativity isomorphism 
$$
M\otimes(X\otimes Y)~\xrightarrow{ \ \approx \ }~(M\otimes X)\otimes Y
$$
yields a morphism of $A$-modules
$$
\modM\otimes(X\otimes Y)~\longrightarrow~(\modM\otimes X)\otimes Y
$$
in $\Hom_A(\modM\otimes(X\otimes Y),(\modM\otimes X)\otimes Y)$. This
endows $\Mod_\scrc(A)$ with the structure of a module category over
$\scrc$, i.e. the ``mixed'' tensor functor 
$$
\otimes\,:\,\Mod_\scrc(A)\times\scrc~\longrightarrow~\Mod_\scrc(A)
$$
is an exact bifunctor with associativity and unit conditions
generalizing the triangle and pentagon axioms. 

For a given defect operator, specified by a fixed non-zero
$A$-module $\modM\in\Ob(\Mod_\scrc(A))$, we demonstrate in
Appendix~C how to construct a canonically defined algebra object
$A_{M}= \,{}^\vee \modM\otimes_A \modM$ of $\scrc$ such that $A_A=A$; given two defect
operators $\modM_1,\modM_2\in\Ob(\Mod_\scrc(A))$, we also prove that the module categories of
$A_{M_1}$ and $A_{M_2}$ are
equivalent. Thus starting from a
single defect operator we get a (symmetric)
Frobenius algebra $A$. Different boundary defects
generically produce distinct Frobenius algebras, but any two such
Frobenius algebras are Morita equivalent. Morita equivalent Frobenius
algebras give rise to equivalent correlation functions in the
two-dimensional gauge theory. For example, this explains the feature
that the correlators of defect holonomy operators are independent of
their insertion points on $\Sigma_h$. In
Appendix~C we give an explicit description of the module category
$\Mod_\scrc(A)$ by constructing a $\IC$-algebra $\ca$ such that
$\Rep(\ca)=\Mod_\scrc(A)$; this functorial equivalence endows the representation
category $\Rep(\ca)$ with the structure of a module category
$$
\otimes\,:\,\Rep(\ca)\times\scrc~\longrightarrow~\Rep(\ca)
$$
over the tensor category $\scrc$, i.e. for $\modV\in\Rep(\ca)$, there are natural
isomorphisms $(\modV\otimes X)\otimes Y\cong\modV\otimes(X\otimes Y)$
for all $X,Y\in\Ob(\scrc)$ and $\modV\otimes U_0\cong\modV$. 

\section{Refinement\label{se:Ref}}

An immediate spinoff from the categorical reformulation of \S\ref{se:Cat} is that one can also
integrate to generalized characters associated to morphisms with target an arbitrary
representation $V$ of $G$. In particular, when $V$ is a symmetric power of
the fundamental representation of $G=U(N)$, we unleash a
\emph{refinement} of the $q$-deformed Yang-Mills amplitudes which
leads to a two-parameter deformation of the heat kernel expansion
(\ref{HK}). This refined $q$-deformed Yang-Mills theory was considered
recently in~\cite{Aganagic:2012si}, and it is an analytic
continuation, in the sense explained in \S\ref{se:qdefgt}, of the refinement of Chern-Simons theory on
Seifert three-manifolds constructed
in~\cite{Aganagic:2011sg,Iqbal:2011kq} which computes the Poincar\'e
polynomials of knot homology (see also~\cite{Fuji:2012pm}). The
topological version of this gauge theory is the refined $q$-deformed BF-theory on the Riemann surface $\Sigma_h$ considered in~\cite{Gadde:2011uv}, or equivalently refined Chern-Simons theory on
$\Sigma_h\times S^1$, which is identified as the two-dimensional
topological field theory computing the topologically twisted
partition function of an $\cn=2$ superconformal field theory on
$S^1\times S^3$, i.e. the $\cn=2$ superconformal index in four
dimensions; the non-topological version identifies the partition
function of an $\cn=2$
supersymmetric non-linear sigma-model on $S^1\times S^3$ with the
propagator of the refined two-dimensional gauge theory~\cite{Tachikawa:2012wi}.

\medskip

\subsection{Constructing refined $q$-deformed Yang-Mills amplitudes}\label{se:refAxioms}~\\[5pt]
To construct refinements of the $q$-deformed two-dimensional gauge
theory, we need to generalize our categorification slightly, as
alluded to in \S\ref{se:Defect}. For this, we
enlarge the morphisms of the source category to include
two-dimensional surfaces with marked points. We then modify the
functor (\ref{functA}) by prescribing additional data at each marked
point given by a fixed finite-dimensional module $V$ over the quantum group
$\cu_q(\frg)$, which may be interpreted as the insertion of a defect
holonomy puncture in the representation $V$ of $G=U(N)$ at each marked point,
i.e. the holonomy of the gauge fields around the marked point is the
representation $V$; the corresponding Yang-Mills amplitude defines a
wavefunction in the Hilbert space associated to the boundary. Thus, for example, the basic class (\ref{capA}) is
correspondingly modified in this case to
$$
\bigg[\scrf_{\cu_q(\frg)} \Big(\raisebox{-5.5pt}{\Diskref}\, \Big) \bigg]
= \sum_{\lambda}\, \Dim(U_\lambda)\ \theta_\lambda \
[U_\lambda\otimes V] \ .
$$
We interpret this refinement as an augmentation of the usual
Grothendieck group to contain ``vector-valued'' characters: Given a non-zero
intertwining operator $\Phi_{\lambda}:U_\lambda\to U_\lambda\otimes V$
for $\cu_q(\frg)$, the
quantity $\Tr_\lambda(\Phi_\lambda \, X)$ for $X\in G$ is a $G$-equivariant
function on the Lie group $G$ with values in
the representation $V$, called a
generalized character in~\cite{Etingof:1994yv,Kirillovjr1}. When $V=\IC$ is the
trivial module we recover the usual characters and the
monoidal functor from \S\ref{se:Cat}. More precisely, since conjugacy
classes in $G$ are the same as orbits of the Weyl group $W=\frS_N$ in the
maximal torus $T=U(1)^N$, the generalized characters are
uniquely defined by their values on $T$ and take
values in the weight~$0$ subspace $V^{(0)}$ in the usual weight
decomposition of the representation $V$ of $G$. The space of
intertwining operators $\Phi_{\lambda}:U_\lambda\to U_\lambda\otimes
V$ is isomorphic to the space $\Hom_\scrr(U_\lambda,U_\lambda\otimes
V)\cong (U_\lambda^*\otimes U_\lambda\otimes V)^{\cu_q(\frg)}$.

We shall take
$V=V_{\rm fund}^{\odot(\beta-1)\, N}$ for fixed $\beta\in\IZ_{>0}$ to be
the $q$-deformation of the $(\beta-1)\, N$-th symmetric power of the fundamental representation
$V_{\rm fund}=\IC^N$ of $G$; it is isomorphic to the space of
homogeneous polynomials in $N$ variables $x_1,\dots,x_N$ of degree $(\beta-1)\,N$. In this case all the weight subspaces of
$V$ are one-dimensional, and in particular
$V^{(0)}\cong\IC$. Since the tensor product multiplicities for
$\cu_q(\frg)$ are the same as those for $\frg$, there is a non-zero $\cu_q(\frg)$-homomorphism
$\Phi_\lambda:U_\lambda\to U_\lambda\otimes V_{\rm fund}^{\odot(\beta-1)\, N}$ if and only if
$\lambda=\mu+(\beta-1)\, \rho$ for a highest weight $\mu$; in this
instance $\Phi_\mu^\bullet:=\Phi_{\mu +(\beta-1)\, \rho}$ is unique up to normalization. 

Etingof and
Kirillov show that in this case the generalized characters
$\Tr_{\lambda+(\beta-1)\,\rho}(\Phi_{\lambda}^\bullet\, X)$
are given by the monic form $M_\lambda(x;q,t)$ of the
Macdonald polynomials in the eigenvalues
$x=(x_1,\dots,x_N)$ of the matrix $X$ at $t=q^\beta$
(see~\cite[Thm.~2.2]{Etingof:1994yv}). Recall~\cite[Chap.~VI]{Macdonald}
that $M_\lambda(x;q,t)$ can be defined algebraically as the unique symmetric polynomials satisfying the following two conditions:
\begin{itemize}
\item[(i)]  Triangular decomposition in dominance order with respect to the basis
of monomial symmetric polynomials:
$$
M_\lambda(x;q,t) = m_\lambda(x)+\sum_{\mu<\lambda}\,
v_{\lambda,\mu}(q,t)\ m_\mu(x) \ ,
$$
where $v_{\lambda,\mu}(q,t)$ are rational functions of $q,t$, the sum runs over $N$-component partitions $\mu$ such that $|\mu|=|\lambda|$
and $\mu_1+\cdots +\mu_i<\lambda_1 +\cdots+\lambda_i$ for all
$i=1,\dots,N$, and  $$m_\lambda(x)=\sum_{w\in\mathfrak{S}_N(\lambda)} \, x_1^{w(1)}\cdots
  x_N^{w(N)}$$
with $\mathfrak{S}_N(\lambda)$ the set of distinct permutations of $\lambda_1,\dots,\lambda_N$.
\item[(ii)]  Orthogonality:
$$
\langle M_\lambda,M_\mu\rangle_{q,t}=0 \qquad \mbox{for} \quad
\lambda\neq \mu \ ,
$$
\end{itemize}
where the inner product here is defined in the basis of power sum symmetric
polynomials $p_\lambda=p_{\lambda_1}\cdots p_{\lambda_N}$, with
$p_n(x):=m_{(n)}(x)=x_1^n+\cdots+ x_N^n$ for $n>0$ and $p_0(x):=1$, as
\beq
\langle p_\lambda,p_\mu\rangle_{q,t}= z_\lambda\ \delta_{\lambda,\mu}\
\prod_{i=1}^{\ell(\lambda)}\, \frac{1-q^{\lambda_i}}{1-t^{\lambda_i}}
\label{innerprodqt}\eeq
with the length $\ell(\lambda)$ the number of non-zero parts of
the partition $\lambda=(1^{m_1} \, 2^{m_2}\cdots)$ and
$$
z_\lambda=\prod_{j\geq1}\, j^{m_j}\, m_j! \ .
$$

The Macdonald polynomials emcompass the various symmetric polynomials
which play a role in this paper:
\begin{itemize}
\item The symmetrized monomials $m_\lambda(x)=M_\lambda(x;q,1)$
are obtained in the limit $t=1$ (independently of $q$). 
\item The Schur polynomials $s_\lambda(x)=M_\lambda(x;q,q)$ are obtained as
the limit $t=q$ of the Macdonald polynomials (also independently of $q$); in this case the inner product
(\ref{innerprodqt}) reduces to the Hall inner product 
$\langle p_\lambda,p_\mu\rangle=z_\lambda\ \delta_{\lambda,\mu}$ such
that $\langle s_\lambda,s_\mu\rangle=\delta_{\lambda,\mu}$. 
\item The Hall-Littlewood polynomials $P_\lambda(x;t)=
  M_\lambda(x;0,t)$ are obtained in the limit $q=0$; they interpolate
  between the Schur polynomials $s_\lambda(x)$ at $t=0$ and the
  monomial symmetric polynomials $m_\lambda(x)$ at $t=1$.
\item The
Jack polynomials $J_\lambda(x;\alpha^{-1})=\lim_{q\to1}\, M_\lambda(x;q,q^\alpha)$ are obtained at $t=q^\alpha$
with $q\to1$ for $\alpha\in\IC$; they are a one-parameter deformation
of the Schur polynomials with $s_\lambda(x)=J_\lambda(x;1)$, and the inner product (\ref{innerprodqt})
in this case is the Jack inner product $\langle
p_\lambda,p_\mu\rangle_\alpha=z_\lambda\, \alpha^{-\ell(\lambda)} \
\delta_{\lambda,\mu}$.
\end{itemize}

The Macdonald inner product (\ref{innerprodqt}) on the space of symmetric
functions can be defined analytically for functions on $T=(S^1)^N$ by the torus scalar product
\beq
\langle f,g\rangle_{q,t}:= \frac1{N!}\, 
\int_{[0,2\pi)^N} \ \prod_{i=1}^N \, \frac{\dd\phi_i}{2\pi} \
\Delta_{q,t}\big(\e^{\ii\phi}\big) \ f\big(\e^{\ii\phi}\big)\,
g\big(\e^{-\ii\phi}\big) \ ,
\label{torusinnerprod}\eeq
where
\beq
\Delta_{q,t}(z) := \prod_{i\neq j}\, \frac{\big(z_i\,z_j^{-1}\,;\,
  q\big)_\infty}{\big(t\, z_i\, z_j^{-1}\,;\, q\big)_\infty}
\label{M-measure}\eeq
is the Macdonald measure for $z=(z_1,\dots,z_N)\in T$ and $0\leq t\leq1$. For
$t=q^\beta$, $\beta\in\IZ_{>0}$, with $z_i=\e^{\ii\phi_i}$ and
$\phi_i\in[0,2\pi)$ for $i=1,\dots,N$, we can write
$$
\Delta_{q,t}\big(\e^{\ii\phi}\big)= \prod_{m=0}^{\beta-1} \
\prod_{i\neq j}\, \Big(1-q^m\, \e^{\ii(\phi_i-\phi_j)}\Big)
$$
which is the $q$-analog of the $\beta$-th power of the Weyl
determinant on $G=U(N)$. Then the norm of the Macdonald polynomials
$M_\lambda(x;q,t)$ is given by
$$
\| M_\lambda\|_{q,t}^2:= \langle M_\lambda,M_\lambda\rangle_{q,t} = \prod_{i<j}\,
\frac{\big(q^{\lambda_i-\lambda_j}\, t^{j-i}\, ;\, q\big)_\infty\,
\big(q^{\lambda_i-\lambda_j+1}\, t^{j-i}\,;\,
q\big)_\infty}{\big(q^{\lambda_i -\lambda_j}\, t^{j-i+1}\,;\,
q\big)_\infty\, \big(q^{\lambda_i-\lambda_j+1}\, t^{j-i-1}\,;\,
q\big)_\infty} \ ,
$$
which for $t=q^\beta$, $\beta\in\IZ_{>0}$ yields Macdonald's inner
product identity
$$
\| M_\lambda\|_{q,t}^2 = \prod_{m=0}^{\beta-1}\
\prod_{i<j}\, \frac{\big[\lambda_i-\lambda_j+\beta\,
  (j-i)+m\big]_q}{\big[\lambda_i-\lambda_j+\beta\,(j-i)-m\big]_q} \ .
$$

To compute the corresponding refinements of the data for the
semisimple ribbon category $\scrr=\Rep(\cu_q(\frg))$, we first consider the
punctured tube amplitude which is the modification of (\ref{tubeA}) in
this case given by
\beq
\bigg[\scrf_{\cu_q(\frg)}\Big(\raisebox{-6pt}{\Tuberef}\, \Big) \bigg]
= \sum_{\lambda}\, \theta_\lambda \ \big[U_\lambda\otimes
U_\lambda^*\otimes V_{\rm fund}^{\odot(\beta-1)\, N}\big] =\sum_\lambda\,
\frac{\theta_\lambda^\bullet}{ \|M_\lambda\|_{q,t}} \ 
[\Phi_\lambda^\bullet] 
\label{puncttube}\eeq
where $\theta_\lambda^\bullet:=\theta_{\lambda+(\beta-1)\, \rho}$,
we used $\bigoplus_\lambda\, \Hom_\scrr\big(U_\lambda\,,\,U_\lambda\otimes
V_{\rm fund}^{\odot(\beta-1)\, N} \big)=\bigoplus_\lambda\,
\IC\,\Phi^\bullet_\lambda/ \|M_\lambda\|_{q,t}$ together with
the canonical duality isomorphisms in the category $\scrr$, and we
accounted for the non-trivial normalisation of the intertwining
operators $\Phi^\bullet_\lambda$~\cite{Kirillovjr1}. It follows
that the twist
$\theta^\bullet$ in this basis is given by~\cite{Kirillovjr1}
\beq
\theta_\lambda^\bullet = q^{\frac12\, \langle\lambda,\lambda\rangle}\
t^{\langle\lambda, \rho\rangle} \ ,
\label{twistref}\eeq
where we used
$$
C_2\big(\lambda+(\beta-1)\,\rho\big)=\langle\lambda+\beta\,\rho,\lambda
+\beta\,\rho\rangle-\langle\rho,\rho\rangle
$$ 
and normalized
$\theta^\bullet$ so that $\theta_0^\bullet=1$. Here and in the
following we should strictly speaking take $t=q^\beta$ with $\beta\in\IZ_{>0}$; however,
in~\cite{Etingof:1993pv,Etingof:1994yv} it is shown how to analytically continue many
of these formulas and results to arbitrary values $\beta\in\IC$, and hence we shall
often write formulas for arbitrary, algebraically independent refinement parameters $t\in\IC$
with $|t|\leq q$ as
well. In this case $V=V_{\rm fund}^{\odot(\beta-1)\, N}$ is regarded
formally as a representation of $\cu_q(\frg)$ with ``highest weight''
$(\beta-1)\, N\, \omega_1$ where $\omega_1$ is the fundamental weight.

Next, computing as we did in (\ref{puncttube}), we consider a sphere
with two marked points which leads to the modification of the
amplitude (\ref{sphereB}) given by
$$
\bigg[\scrf_{\cu_q(\frg)}\Big(\raisebox{-5.5pt}{\Diskdref}\,\Big)^B\,
\bigg] = \sum_{\lambda, \mu}\, S^\bullet_{\lambda\mu} \
\frac{\theta_\lambda^\bullet \,
  \theta_\mu^\bullet}{\|M_\lambda\|_{q,t}\, \|M_\mu\|_{q,t}} \
[\Phi_\lambda^\bullet\,^* \otimes \Phi_\mu^\bullet]
$$
where we have further used the canonical identifications
$\Hom_\scrr(\Idd,U_\lambda\otimes U_\lambda^*)\cong\IC$ in the category
$\scrr$. The braiding symmetry is computed in this
basis in~\cite[Thm.~5.4]{Kirillovjr1} and with suitable normalization
it can be expressed in terms of specializations of the Macdonald functions as
$$
S_{\lambda\mu}^\bullet = M_\lambda\big(t^{\rho}\,;\,q,t\big)\
M_\mu\big(t^{\rho}\,q^{\lambda}\,;\, q,t\big) \ .
$$
Symmetry $S_{\lambda\mu}^\bullet=S_{\mu\lambda}^\bullet$ is now a
consequence of the self-duality identity for Macdonald polynomials
given by~\cite[Chap.~VI]{Macdonald}
$$
\frac{M_\lambda\big(q^{\mu_1}\, t^{N-1} ,q^{\mu_2}\,
t^{N-2},\dots,q^{\mu_N}\, ;\,
q,t\big)}{M_\lambda\big(t^{N-1},t^{N-2},\dots,1\, ;\, q,t\big)}
= \frac{M_\mu\big(q^{\lambda_1}\, t^{N-1},q^{\lambda_2}\,
t^{N-2},\dots,q^{\lambda_N}\, ;\,
q,t\big)}{M_\mu\big(t^{N-1},t^{N-2},\dots,1\, ;\, q,t\big)} \ .
$$
In particular, the categorical dimensions in this basis are given by
Macdonald's special value identity
\beqa
\dim_{q,t} \lambda &:=& S^\bullet_{\lambda0} \nonumber \\[4pt] &=& M_\lambda\big(t^{\rho}\,;\,
q,t\big) \nonumber \\[4pt] &=& t^{|\lambda|/2} \
M_\lambda\big(1,t,\dots,t^{N-1}\,;\, q,t) \ = \ \prod_{i=1}^N\,
t^{(i-\frac12)\, \lambda_i} \ \prod_{j<k}\, \frac{\big(t\, q^{k-j}\, ;\,
  q\big)_{\lambda_j-\lambda_k}}{\big(q^{k-j}\,;\,
  q\big)_{\lambda_j-\lambda_k}} \ . \nonumber
\eeqa
When $t=q^\beta$
with $\beta\in\IZ_{>0}$ we can rewrite this formula as~\cite{Etingof:1994yv}
$$
\dim_{q,t} \lambda = \prod_{i<j} \ \prod_{m=0}^{\beta-1}\,
\frac{\big[\lambda_i-\lambda_j+\beta\, (j-i)+m\big]_q}{\big[\beta\,
  (j-i)+m \big]_q} \ ,
$$
which exhibits the refined categorical dimension as the $q$-analog of
the $\beta$-th power of the quantum dimension $\dim_q\lambda$.
Finally, the structure constants in this basis are no longer
integer-valued but instead are rational functions
$N_{\lambda\mu}^\bullet{}^\nu(q,t)$ of $q,t$ which appear as
generalized Littlewood-Richardson coefficients
$$
M_\lambda(x;q,t)\, M_\mu(x;q,t)= \sum_\nu\,
N_{\lambda\mu}^\bullet{}^\nu(q,t) \ M_\nu(x;q,t) \ .
$$

We thus find that the partition function (\ref{czcaSigmah}) is given
by
\beq
\cz^\bullet\big(\cu_q(\frg)\,;\,\Sigma_h\big) = \sum_\lambda\, \bigg(\, 
\frac{M_\lambda\big(t^{\rho}\,;\, q,t\big)}{\|M_\lambda\|_{q,t}}\, \bigg)^{2-2h}\
q^{\frac12\, \langle\lambda,\lambda\rangle}\, t^{\langle\lambda,\rho\rangle} \
,
\label{cacuqref}\eeq
while (\ref{czcaS2pp}) in this case becomes
\beqa
\cz^\bullet\big(\cu_q(\frg)\,;\, S^2,p,p'\,\big) &=&
\sum_{\lambda_1,\dots,\lambda_\ell}\, \frac{q^{\frac12\, (e_1\, \langle\lambda_1,\lambda_1\rangle+\cdots+
  e_\ell\, \langle\lambda_\ell,\lambda_\ell\rangle)}\, t^{\langle
  e_1\,\lambda_1+\cdots+
  e_\ell\,\lambda_\ell,\rho\rangle}}{\|M_{\lambda_1}\|^2_{q,t}\,
\|M_{\lambda_2}\|_{q,t}\cdots \|M_{\lambda_{\ell-1}}\|_{q,t}\, \|M_{\lambda_\ell}\|^2_{q,t}} \nonumber
\\ && \qquad \qquad \times \ M_{\lambda_1}\big(t^{\rho}\,;\, q,t\big)^2\,
M_{\lambda_2}\big(t^{\rho}\,;\, q,t\big) \cdots
M_{\lambda_\ell}\big(t^{\rho}\,;\, q,t\big) \nonumber
\\ && \qquad \qquad \times \
M_{\lambda_2}\big(t^{\rho}\,q^{\lambda_1}\,;\, q,t\big) \cdots
M_{\lambda_\ell}\big(t^{\rho}\, q^{\lambda_{\ell-1}}\,;\, q,t\big) \ .
\label{czcaS2ppref}\eeqa
It is also straightforward to write down defect observables along the
lines of \S\ref{se:Defect}: In the conventions of \S\ref{se:Defect},
for the correlator of a single Wilson loop in the representation
$\lambda$ we find
\beqa
\cw_\lambda^\bullet\big(\cu_q(\frg)\,;\Sigma_h,p,1 \big) &=&
\sum_{\lambda_1,\lambda_2}\,
N_{\lambda_1\lambda}^\bullet{}^{\lambda_2} \ \bigg(\, 
\frac{M_{\lambda_1}\big(t^{\rho}\,;\, q,t\big)}{\|M_{\lambda_1}
  \|_{q,t}}\, \bigg)^{1-2h_1} \, \bigg(\, 
\frac{M_{\lambda_2}\big(t^{\rho}\,;\, q,t\big)}{\|M_{\lambda_2}
  \|_{q,t}}\, \bigg)^{1-2h_2} \nonumber \\ && \qquad \qquad \qquad \times \
q^{\frac p2\,
  (\langle\lambda_1,\lambda_1\rangle+\langle\lambda_2,\lambda_2
  \rangle)} \, t^{p\, \langle\lambda_1+\lambda_2,\rho\rangle} \ ,
\nonumber
\eeqa
while the general amplitude for a Riemann surface of genus $h$ with
$b$ boundaries fixed at holonomy eigenvalues $u_1,\dots,u_b\in(S^1)^N$ and $n$
defect punctures labelled by irreducible representations
$\lambda_1,\dots,\lambda_n$ is given by
\beqa
\co_{\stackrel{\scriptstyle u_1,\dots,u_b}{\scriptstyle
    \lambda_1,\dots,\lambda_n}}\big(\cu_q(\frg)\,;\, \Sigma_h,p,1 \big)
&=& \sum_\lambda\, \bigg(\, 
\frac{\dim_{q,t}\lambda}{\|M_\lambda\|_{q,t}} \bigg)^{2-2h-b-n}\ q^{\frac p2\,
  \langle\lambda,\lambda\rangle}\, t^{p\, \langle\lambda,\rho\rangle} \
\prod_{i=1}^b\, \frac{M_\lambda(u_i;q,t)}{\|M_\lambda\|_{q,t}}
\nonumber \\ && \qquad \qquad \qquad \qquad \qquad \qquad \times \ \prod_{j=1}^n\,
\frac{S_{\lambda\lambda_j}^\bullet}{\|M_\lambda\|_{q,t}\,
  \|M_{\lambda_j} \|_{q,t}} 
\nonumber \\[4pt] &=& \sum_\lambda\,
\frac{M_{\lambda}\big(t^{\rho}\,;\,
  q,t\big)^{2-2h-b}}{\|M_\lambda\|_{q,t}^{2-2h}} \ q^{\frac p2\,
  \langle\lambda,\lambda\rangle}\, t^{p\, \langle\lambda,\rho\rangle} \ \prod_{i=1}^b\,
  M_\lambda(u_i;q,t) \nonumber \\ && \qquad \qquad \qquad \qquad \qquad \qquad \times \ \prod_{j=1}^n\,
\frac{M_{\lambda_j}\big(t^\rho\, q^\lambda\,;\, q,t\big)}{\|M_{\lambda_j} \|_{q,t}} 
\ . \nonumber
\eeqa

\medskip

\subsection{Refined $L(p,1)$ matrix models}\label{se:refmatrix}~\\[5pt]
Let us consider the partition function (\ref{czcaS2ppref}) for refined
$q$-deformed Yang-Mills theory on $S^2$ at $p'=1$
and with $t=q^\beta$ for $\beta\in\IZ_{>0}$. Setting
$\mu_i=\lambda_i+\beta\,(N-i)$ for $i=1,\dots,N$ yields a $\beta$-deformation
of the discrete Gaussian matrix model (\ref{qmat}) at $h=0$ given by
\beqa
\cz^\bullet_{(p)}(q,t;S^2)& :=& \cz^\bullet\big(\cu_q(\frg)\,;\,
S^2,p,1\big) \nonumber \\[4pt] &=&
\sum_{\mu\in\IZ^N} \ \prod_{i<j} \ \prod_{m=0}^{\beta-1}\,
[\mu_i-\mu_j+m]_q\,[\mu_i-\mu_j-m]_q  \ q^{\frac p2 \,\sum_i\,
  \mu_i^2} \ . \nonumber
\eeqa

For $p=1$, by performing analogous steps to those of \S\ref{se:p=1}
we may rewrite this partition function in the form
\beqa
\cz^\bullet_{(1)}(q,t;S^2) &=& \sum_{u\in\IZ^N}\, q^{\frac12\, \sum_i\, u_i^2} \
\prod_{j\neq k} \ \prod_{m=0}^{\beta-1} \, \Big(q^{\frac12\, (u_j-u_k+m)}-q^{\frac12\,(u_k-u_j-m)}\Big) \nonumber
\\[4pt] 
&=& q^{\frac12\, \beta\, (\beta-1)}\, \sum_{u\in\IZ^N}\, q^{\frac12\, \sum_i\, u_i^2}\,
(\sigma \, q)^{\beta\, \sum_i\,u_i}\ \prod_{j\neq k}\ \prod_{m=0}^{\beta-1}\,
\big(q^{u_j}-q^{-m}\, q^{u_k}\big) \nonumber \\[4pt]
&=& q^{\frac12\, \beta\, (\beta-1)}\, \sigma^{\beta\, N\, (1-\beta\, N)}\,
\int_{\IR_{>0}^N} \ \prod_{i=1}^N \, \dd x_i\ x_i^{\beta-1} \
\sum_{n_i=-\infty}^\infty\, \sigma^{\beta\,n_i}\, q^{\frac12\, n_i^2+n_i}\,
\delta(x_i-\sigma^{\beta} \, q^{n_i}) \nonumber \\ && \qquad \qquad \qquad
\qquad \qquad \qquad \qquad \qquad \qquad \times \
\prod_{m=0}^{\beta-1} \ \prod_{j\neq k}\,
\big(x_j-q^{-m}\, x_k\big) \nonumber \\[4pt]
&=& q^{\frac12\, \beta\, (\beta-1)+\frac12\, N}\, \sigma^{\beta\, N\, (1-\beta\, N)}\,
M(q,\sigma^{\beta})^N \nonumber\\ && \times \ \int_{\IR_{>0}^N} \
\prod_{i=1}^N \, \dd x_i \
w_d(x_i;q,\sigma^\beta) \ \prod_{j<k}\, (x_j-x_k)^2 \
P(x_1,\dots,x_N;q,t) \ , \nonumber
\eeqa
where as before $\sigma=q^{-N}$, the discrete measure $w_d(x;q,\sigma)$ is given by
(\ref{discretemeas}), and we have introduced the polynomial
$$
P(x_1,\dots,x_N;q,t)= \prod_{m=1}^{\beta-1} \ \prod_{i=1}^N\, x_i \
\prod_{j\neq k}\, \big( x_j-q^{-m}\, x_k\big)
$$
with the convention $P(x_1,\dots,x_N;q,q):=1$. The specialization of the family
of measures (\ref{discretemeas}) at $\sigma^\beta=q^{-N\,\beta}$ is again
equivalent to the specialization $\sigma=1$ due to the discrete scaling
symmetry (\ref{wdtransl}). Arguing as in \S\ref{se:contmm}, since the
$\beta$-deformation here amounts to a polynomial average in the matrix
model, only integer moments are involved in its computation, and the
discrete/continuous equivalence of the indeterminate moment problem
for the Stieltjes-Wigert matrix model applies to it as well. With
$q:=\e^{-g_s}$ as previously, we can therefore rewrite the refined
partition function as a correlator in the continuous matrix model
\beqa
\cz^\bullet_{(1)}(q,t;S^2) &=& q^{\frac12\, \beta\,(\beta-1)+\frac12\,
  N}\, \sigma^{\beta\,N\,
  (1-\beta \, N)}\, M(q,\sigma^{\beta})^N \nonumber \\ && \times \ 
\int_{\IR_{>0}^N} \ \prod_{i=1}^N \, \dd x_i \ \omega_{\rm SW}\big(x_i \,;\,
\mbox{$\frac1{\sqrt{2g_s}}$} \big) \ \prod_{j<k}\, (x_j-x_k)^2 \
P(x_1,\dots, x_N;q,t) \nonumber
\eeqa
defined by the Stieltjes-Wigert distribution (\ref{SW}). 
Setting $u_i=\log(x_i)-\frac14 \, g_s\, (\beta-1)$, we may
alternatively rewrite this matrix model average as
\beqa
\cz^\bullet_{(1)}(q,t;S^2) &=& C_{N}(q,t)\ \Big(\, \frac{g_s}{2\pi
}\, \Big) ^{-N/2} \nonumber \\ && \times \ 
\int_{\IR^N } \ \prod_{i=1}^N \,
{\frac{\mbox{d}u_{i}}{2\pi }} \ \e^{-u_{i}^{2}/2g_s} \ 
\prod_{m=0}^{\beta-1} \ \prod_{j\neq k}\, \Big(\e^{\frac12\, (u_j-u_k)}-q^{-m}\,
\e^{\frac12\, (u_k-u_j)}\Big) \label{SWCSS3ref} 
\eeqa
where
$$
C_{N}(q,t)=q^{-N\,\beta\,N^2\,(1-\beta\,N)}\,
  q^{\frac12\, N+\frac18\, (\beta-1)\,(3\beta+1)}\, \big(-q^{\frac32-\beta\, N}\,;q\big)^N_{\infty
}\, \big(-q^{\beta\, N-\frac12}\,;\,q \big)^N_{\infty }\,
\big(q\,;\,q \big)^N_{\infty } \ .
$$
Up to overall normalization, this is just the partition function
$Z_N(q,t):=\cz^\bullet_{(1)}(q,t;S^2)/C_{N}(q,t) $ of
the $\beta$-deformed matrix model for refined Chern-Simons theory on
$S^3$ which was considered in~\cite{Aganagic:2011sg}; the product over
$m$ in (\ref{SWCSS3ref}) yields the $q$-analog of the $\beta$-th power
of the square of the Vandermonde determinant in this case.

For $p>1$, completely analogous calculations to those of
\S\ref{se:pinN} can be similarly carried out. Dropping all normalization
constants and setting $q=\e^{-g_s}$, the partition function
\beqa
\cz^\bullet_{(p)}(q,t;S^2) &=& \sum_{u\in 
\mathbb{Z}^N
}\, \exp\Big( -\frac{p\, g_s}2 \, \sum_{i=1}^N \, u_{i}^{2}\Big) \
\prod_{j\neq k}%
\ \prod_{m=0}^{\beta-1}\, \Big(\e^{\frac {g_s}2\, (u_j-u_k)}-q^{-m}\,
\e^{\frac {g_s}2\, (u_k-u_j)}\Big) \nonumber \\[4pt] 
&=& \cz_{\rm SW}^{(p)}[0](q,t;S^2) \nonumber
\eeqa
is the $\IZ_p$-invariant projection of the partition function of the
corresponding full $\beta$-deformed Stieltjes-Wigert ensemble
\beqa
\cz_{\rm SW}^{(p)}(q,t;S^2) &=& \int_{\IR^N} \ \prod_{i=1}^N \, \dd u_{i} \
\e^{-\frac p{2g_s} \, u_{i}^{2}} \ \prod_{j\neq k} \ \prod_{m=0}^{\beta-1}\, \Big(\e^{
  \frac12 \, (u_j-u_k)}-q^{-m}\,
\e^{\frac12\, (u_k-u_j)}\Big) \nonumber \\[4pt] &=&
\sum_{u\in 
(\mathbb{Z}/p)^N
}\, \exp\Big( -\frac{p\, g_s}2 \, \sum_{i=1}^N \, u_{i}^{2}\Big) \
\prod_{j\neq k}%
\ \prod_{m=0}^{\beta-1}\, \Big(\e^{\frac {g_s}2\, (u_j-u_k)}-q^{-m}\,
\e^{\frac {g_s}2 \, (u_k-u_j)}\Big) \nonumber
\eeqa
considered above. In particular, for $q=\zeta_k$ a primitive $k$-th
root of unity one finds the continuous matrix model representation
\beqa
\cz^\bullet_{(p)}(q=\zeta_k,t;S^2) = 
\int_{\IR^N } \ \prod_{i=1}^N\, 
{\frac{\mbox{d}u_{i}}{2\pi }} \ \e^{-u_{i}^{2}/2g_s} \
\Gamma_{k,p}(\e^{u_i}) \ 
\prod_{m=0}^{\beta-1} \ \prod_{j\neq k}\, \Big(\e^{\frac12\, (u_j-u_k)}-q^{-m}\,
\e^{\frac12\, (u_k-u_j)}\Big) \nonumber
\eeqa
which coincides with (\ref{SWCSS3ref}) for $p=1$.

\medskip

\subsection{Refined unitary matrix model}\label{se:refumm}~\\[5pt]
In~\cite{Aganagic:2011sg} it was shown that the matrix integral
(\ref{SWCSS3ref}) has a representation as a $U(N)$ unitary matrix
model obtained by substituting the usual Haar measure of the matrix
model (\ref{UCS}) with the Macdonald measure
$\Delta_{q,t}(\e^{\ii\phi})$ from (\ref{M-measure}), so that
\beq
Z_N(q,t) = \int_{[0,2\pi)^N} \ \prod_{i=1}^N \, \frac{\dd\phi_i}{2\pi} \
\Theta\big(\e^{\ii\phi_i}\,;\, q\big) \ \Delta_{q,t}\big(\e^{\ii\phi}\big) \ .
\label{subs}\eeq
Below we describe the relevance of such types of matrix integrals to
the refined gauge theories considered
in~\cite{Aganagic:2011sg,Gadde:2011ik,Gadde:2011uv,Iqbal:2011kq}.

The partition function of the unitary matrix model (\ref{subs}) can be
computed analytically from the explicit evaluation~\cite{Mat}
\beq
\int_{[0,2\pi)^k} \ \prod_{i=1}^k \, \frac{\dd\phi_i}{2\pi} \
\frac{\Theta_n\big(\e^{\ii\phi_i}\,;\, q\big)}{(q;q)_n} \
\Delta_{q,t}\big(\e^{\ii\phi}\big) = \prod\limits_{i=0}^{k-1}
\,\frac{\big(t^{i}\, q^{n+1}\,;\,q\big)_n}{\big(t^i\, q\,;\,q\big)_n}  
\label{Mac}\eeq
generalising (\ref{Selberg}). This expression reduces to the two
formulas in (\ref{Selberg}) in the limit $t=q^{\beta }$ with $\beta
=1$. In the limit $k=N$, $n\to\infty$, it evaluates (\ref{subs}) as
\beq
Z_N(q,t)= \prod_{k=1}^{N-1} \,
\frac{(t^k\, q;q)_\infty}{(q;q)_\infty} \ .
\label{CSrefS3}\eeq
When $t=q^\beta$ for $\beta\in\IZ_{>0}$, the expression
(\ref{CSrefS3}) can be written as
$$
Z_N(q,t) = \prod_{k=1}^{N-1} \ \prod_{m=0}^{\beta-1}\,
\big(1-q^{\beta\, k+m}\big)^{N-k} \ ,
$$
which coincides with the exact analytical expression for the partition
function of the $\beta$-deformed Stieltjes-Wigert matrix model for
refined Chern-Simons theory on $S^3$ that we considered in
\S\ref{se:refmatrix}~\cite{Aganagic:2011sg}. However, there are no known
determinantal expressions for Macdonald polynomials which yield a
suitable analog of Gessel's identity for the generalization of the
Cauchy-Binet summation formula (\ref{SC}); already the limiting case
of Jack polynomials involves Toeplitz
hyperdeterminants~\cite{Matsumoto08}. We are not aware of any analog of the matrix model expression (\ref{BFS2N}) for the
topological refined $q$-deformed two-dimensional gauge theory
considered in~\cite{Gadde:2011uv} ($p=0$ in (\ref{czcaS2p1})); we
return to this point below. 

\medskip

\subsection{Refined $q$-deformed BF-theory}\label{se:refUinfinity}~\\[5pt]
We can also consider refined versions of the six-dimensional $\cn=2$
gauge theories that we studied in \S\ref{se:crystal}, whose partition
functions compute refined (or motivic) Donaldson-Thomas invariants~\cite{Cirafici:2011cd}. We start from the
refinement of the $U(\infty)$ matrix model (\ref{DT}) with partition
function
\beq
\cz^{6D}(q,t) := \int_{[0,2\pi)^\infty} \ \prod_{i=1}^\infty \, \frac{\dd\phi_i}{2\pi} \
\frac{\Theta\big(\e^{\ii\phi_i}\,;\, q\big)}{(q;q)_\infty} \ \Delta_{q,t}\big(\e^{\ii\phi}\big) \ .
\label{DTsubs}\eeq
The weight function is given again by (\ref{double}) and there is an
extension of the Szeg\H{o} limit theorem to this generalized case
which is spelled out in Appendix~B. Using the Fourier coefficients
(\ref{Fourier}) and the limit formula (\ref{Szegoextended}), we compute
\begin{eqnarray}
\log\cz^{6D}(q,t) &=&\sum_{k=1}^{\infty }\, k\, \left[
\log f \right] _{k}\, \left[ \log f \right] _{-k} \, \frac{1-q^{k}}{1-t^{k}} \label{Emod} \\[4pt]
&=&\sum_{k=1}^{\infty }\, \frac{q^{k}}{k\,\big(1-q^{k}\big)\,
  \big(1-t^{k}\big)} \nonumber \\[4pt]
&=& \sum_{k=1}^{\infty }\ \sum_{n,m=1}^{\infty }\, \frac{q^{k\,
    n}\, t^{k\, (m-1)}}k \ = \ -\sum_{n,m=1}^{\infty }\, \log
\big(1-q^{n}\, t^{m-1}\big) \ . \nonumber
\end{eqnarray}%
Thus the partition function evaluates explicitly to
\begin{equation*}
\cz^{6D}(q,t) =M(q,t) \ ,
\end{equation*}
where $M(q,t)$ is exactly the refined MacMahon
function~\cite{Iqbal:2007ii}
\begin{equation*}
M(q,t) = \prod\limits_{n,m=1}^{\infty }\ \frac1{1-q^{n}\,
  t^{m-1}} \ ;
\end{equation*}%
this formula also follows from the
$k,n\to\infty$ limit of (\ref{Mac}). 
Whence the change of integration measure from the Haar measure to the
Macdonald measure produces the refined version of the $\cn=2$ gauge theory
partition function, which in this case is the MacMahon function. Employing
the bilinear sum identity
$$
\sum_\lambda\, \frac1{\| M_\lambda \|^2_{q,t}} \
M_\lambda(x;q,t)\, M_\lambda(y;q,t) = \prod_{i,j\geq1}\, \frac{(t\,
  x_i\, y_j;q)_\infty}{(x_i\, y_j;q)_\infty}
$$
for the Macdonald functions, which generalizes the Cauchy-Binet formula
(\ref{SC}), at the specialisation $x=y=t^\rho$ shows that the
partition function (\ref{DTsubs}) can also be expanded in terms of
refined quantum dimensions as
$$
\cz^{6D}(q,t) = \sum_\lambda\, \bigg(\,
\frac{\dim_{q,t}\lambda}{\|M_\lambda\|_{q,t}} \, \bigg)^2 \ .
$$
This is just the $U(\infty)$ version of the topological $(q,t)$-deformed
two-dimensional gauge theory which was derived in~\cite{Gadde:2011uv}
from a four-dimensional superconformal index.

We can also apply the Macdonald measure substitution to define the
family of refined partition functions
$$
\cz_L^{6D}(\alpha;q,t,Q):= \int_{[0,2\pi)^\infty} \ \prod_{i=1}^\infty
\, \frac{\dd\phi_i}{2\pi} \
F_L\big(\e^{\ii\phi_i}\,;\, \alpha\,;\,q,Q\big) \
\Delta_{q,t}\big(\e^{\ii\phi}\big)
$$
with weight functions (\ref{FKweight}). This modifies the result
(\ref{E-alpha}) to
$$
\cz_L^{6D}(\alpha;q,t,Q) =\prod_{a=1}^{L}\, M(q,t)
^{\alpha _{a}^{2}} \ \prod_{b\neq c}\, M\big(Q_{b}\, Q_{c}^{-1} ,
q, t\big)^{\alpha _{b}\, \alpha _{c}} \ ,
$$
where $M( Q,q,t) $ is the refined generalised MacMahon function%
\begin{equation*}
M( Q,q,t) =\prod\limits_{n,m=1}^{\infty }\ \frac1{1-Q\,
  q^{n}\, t^{m-1}} \ .
\end{equation*}

The derivation presented here is an alternative and equivalent method for obtaining
refined partition functions based on refinements of the weight
function of the matrix model~\cite{Sulkowski:2010ux}. Indeed, the form
of the generalized strong Szeg\H{o} theorem (\ref%
{Szegoextended}) suggests the possibility of interpreting the additional
factor $\left( 1-q^{k}\right) \big/ \left( 1-t^{k}\right) $ as a generalization
of the Fourier coefficients $[\log f]_k$, and therefore one can construct the
corresponding $U(\infty)$ matrix model with the usual Haar measure but with a
generalized weight function. These are the types of matrix models that were
considered in \cite{Sulkowski:2010ux}, where a ``refined'' theta-function
\begin{equation}
\widetilde\Theta (z;q,t)= \prod_{n=1}^\infty\, \big(1+q^{n-1/2}\,
z\big)\, \big(1+t^{n-1/2}\, z^{-1}\big) \label{ref-theta}
\end{equation}%
was used as weight function. We can see the relationship explicitly
from the ordinary Szeg\H{o} limit theorem: One has
$$
\big[\log\widetilde\Theta\, \big]_k = \frac{\left( -1\right) ^{k+1}\, q^{k/2}}{%
k\, \big(1-q^{k}\big)} \qquad \mbox{and} \qquad
\big[\log\widetilde\Theta\, \big]_{-k} = \frac{\left( -1\right) ^{k+1}\, t^{k/2}}{%
k\, \big(1-t^{k}\big)}
$$
for $k>0$, and hence
$$
\log\widetilde\cz\,^{6D}(q,t) = \sum_{k=1}^\infty\, k\,
\big[\log\widetilde\Theta\, \big]_k\, \big[\log\widetilde\Theta\, \big]_{-k}
= \sum_{k=1}^{\infty }\, \frac{q^{k/2}\, t^{k/2}}{k\,\big
  (1-q^{k}\big)\, \big(1-t^{k}\big)} \ .
$$
Written in this form,
the refined partition function $\widetilde\cz\,^{6D}(q,t)$ is realised as an elliptic gamma-function~\cite{S}.
By Gessel's identity, the unitary matrix model
with weight function (\ref{ref-theta}) follows from a $U(\infty)$ two-dimensional
gauge theory with partition function
\begin{equation}
\widetilde\cz\,^{6D}(q,t)=\sum_{\lambda }\, s_\lambda\big(q^\rho \big)\, s_\lambda\big(t^\rho\big) \label{ref-schur}
\end{equation}%
involving two deformation parameters $q,t$. Whence the refined six-dimensional
gauge theory is equivalent to the $(q,t)$-deformed topological
Yang-Mills theory on $S^2$ in (\ref{ref-schur}). This result shows
moreover that the selection made in 
\cite{Sulkowski:2010ux} for a refined theta-function (\ref{ref-theta})
is equivalent to that
given by the refined topological vertex introduced in
\cite{Iqbal:2007ii}, i.e. application of Gessel's identity to the expressions in \cite%
{Iqbal:2007ii} directly gives the matrix models of
\cite{Sulkowski:2010ux}. The partition function (\ref{ref-schur}) also
coincides with the perturbative part of Nekrasov's partition function
for five-dimensional gauge theory~\cite{Nek}, with the variables $q,t$
parametrizing the $\Omega$-background.

For the case of $\cn=2$ gauge theory on the noncommutative conifold discussed in~\cite{Sulkowski:2010ux}, this double deformation also gives the correct result if
one generalizes the Schur polynomials to the supersymmetric Schur
polynomials considered in \S\ref{se:crystal}: The weight function
used in the refined matrix model of~\cite{Sulkowski:2010ux} is
reproduced in (\ref{HS}) with the specialisation
$x_i=q^{i-\frac12}$, $y_i=t^{i-\frac12}$, $z_i=-Q\,
q^{i-\frac12}$ and $w_i=-Q^{-1}\, t^{i-\frac12}$, which yields the
refined generating function for the noncommutative conifold
\beqa
\cz_2^{6D}(1,-1;q,t,1,Q) & = & \frac{M(q,t)^2}{M(Q,q,t)\,
  M(Q^{-1}, q,t)}  \nonumber \\[4pt] &=& \int_{[0,2\pi)^\infty } \
\prod_{i=1}^\infty \, \frac{\mathrm{d}\phi_i}{2\pi 
}~\frac{\widetilde\Theta \big(\e^{\ii\phi_i}\,;\, q,t\big)}{\widetilde\Theta \big(Q^{-1}\,
  \e^{\ii\phi_i}\,;\, q,t \big)} \ \prod\limits_{j<k}\,\big\vert \e^{\ii\phi_j}-\e^{\ii\phi_k}\big\vert^{2}
  \ . \nonumber
\eeqa
This expression coincides with the partition function of a
supersymmetric extension of the refined topological gauge theory
(\ref{ref-schur}) given by
$$
\cz_2^{6D}(1,-1;q,t,1,Q) = \sum_\lambda\,
\HS_\lambda\big(q^\rho\, \big|\, -Q\, q^\rho\big) \,
\HS_\lambda\big(t^\rho\, \big|\, -Q^{-1}\, t^\rho\big) \ .
$$
Notice that the expansion based on hook-Schur polynomials also allows
for a richer specialization, with up to four deformation parameters;
below we consider examples of such multiple refinements.

Thus, at this level of refinement, one may equivalently use either
Schur polynomials
or Macdonald polynomials. This observation can also be applied to
provide a new impetus on the partition function (\ref{Mqchi}). It is
possible to obtain a linear relationship between the respective powers
of the theta-function as a weight function of the matrix model and of
the MacMahon partition function, if
one considers $\beta $-ensembles instead of unitary ensembles; these
ensembles naturally arise when one refines to the Jack polynomials
$J_\lambda(x;\alpha^{-1})$ instead of Schur polynomials
$s_\lambda(x)=J_\lambda(x;1)$. They also satisfy a Cauchy identity
\beq
\sum_\lambda\, \frac1{\|J_\lambda\|_\alpha^2} \
J_\lambda(x;\alpha^{-1})\, J_\lambda(y;\alpha^{-1}) =
\prod_{i,j\geq1}\ \frac1{(1-x_i\, y_j)^{\alpha}} \ ,
\label{CauchyJack}\eeq
where the norms $\|J_\lambda\|_\alpha^2$ are rational functions of the
parameter $\alpha$. The same results that lead to unitary matrix
models from the expansion into Schur functions apply to the Jack
functions as well, by replacing Toeplitz
determinants with Toeplitz hyperdeterminants and unitary ensembles
with $\beta $-ensembles (here with $\beta=2\alpha$); this follows from
the Heine-Szeg\H{o}
identity for Toeplitz hyperdeterminants \cite{Matsumoto08}. Hence we may
replace the matrix model representation (\ref{Mqchi}) with
\beqa
\cz_1^{6D}\big(\sqrt{\mbox{$\frac\chi2$}}\,;\,q\big) = \sum_\lambda\, \bigg(\,
\frac{J_\lambda\big(\, q^\rho\,;
  \mbox{$\frac2\chi$}\,\big)}{\|J_\lambda\|_{\chi/2}}\, \bigg)^2
= \int_{[0,2\pi)^\infty} \ \prod_{i=1}^\infty \, \frac{\dd\phi_i}{2\pi} \
\frac{\Theta\big(\e^{\ii\phi_i}\,;\, q\big)}{(q;q)_\infty}
 \ \prod_{j<k}\,
\big|\e^{\ii\phi_j}-\e^{\ii\phi_k}\big|^\chi \ , \nonumber \\
\label{jack-alt}\eeqa
and both matrix models represent the partition function
$M(q)^{\chi/2}$.

These two types of refinements have been previously considered in
the context of the refined topological vertex: Awata and Kanno
introduced a refinement of the topological vertex based on
Macdonald polynomials \cite{Awata:2005fa,Awata:2008ed}, whereas the
refinement of Iqbal, Kozcaz and Vafa \cite{Iqbal:2007ii} is based on
different specializations of the Schur polynomials as in
(\ref{ref-schur}). The Macdonald refinement of two-dimensional
Yang-Mills theory was originally presented in~\cite{Gadde:2011uv}; the
analysis presented above is the first
consideration and comparison of both refinements in the simpler
setting of two-dimensional gauge theory.

\medskip

\subsection{Higher refinement}\label{se:refhigher}~\\[5pt]
With the Macdonald polynomials one can
easily obtain more general refinements by giving a different set of
parameters $(q_a,t_a)$ to each polynomial. We have thus far considered the (related) cases of gauge theories
associated to matrix models
involving a refined theta-function (\ref{ref-theta}) with the ordinary
Haar measure, and an ordinary theta-function (\ref{triple}) with the Macdonald
measure. We will now combine these
two refinements and use distinct refinement parameters. The partition function is
$$
\cz^{6D}(q_1,t_1;q_2,t_2):= \int_{[0,2\pi)^\infty} \
\prod_{i=1}^\infty \, \frac{\dd\phi_i}{2\pi} \
\widetilde\Theta\big(\e^{\ii\phi_i}\,;\, q_1,t_1\big) \
\Delta_{q_2,t_2}\big(\e^{\ii\phi}\big) \ ,
$$
and by the strong Szeg\H{o} limit theorem we have
\begin{eqnarray*}
\log\cz^{6D}(q_1,t_1;q_2,t_2) &=&\sum_{k=1}^{\infty }\, k\, \big[
\log \widetilde\Theta\, \big] _{k}\, \big[ \log \widetilde\Theta\,
\big] _{-k}\ \frac{1-q_{2}^{k}}{%
1-t_{2}^{k}} \\[4pt]
&=& \sum_{k=1}^{\infty }\, \frac{q_{1}^{k/2}\, t_{1}^{k/2}\, \left(
1-q_{2}^{k}\right) }{k\, \big(1-q_{1}^{k}\big)\,
\big(1-t_{1}^{k}\big)\, \big(1-t_{2}^{k}\big)} \\[4pt]
&=&\sum_{k=1}^{\infty }\, \frac{1-q_2^k}{k} \ 
\sum_{n,m,l=1}^{\infty }\, q_{1}^{k\, (n-1/2) } \, t_{1}^{k\,
  (m-1/2)}\, t_{2}^{k\, (l-1)} \\
&=&-\sum_{n,m,l=1}^{\infty }\, \Big( \log
\big(1-q_{1}^{n-1/2}\, t_{1}^{m-1/2}\, t_{2}^{l-1}\big) \\ && \qquad
\qquad \qquad -\, \log
\big(1-q_{2}\, q_{1}^{n-1/2}\, t_{1}^{m-1/2}\, t_{2}^{l-1}
\big) \Big) \ .
\end{eqnarray*}
It follows that
\begin{equation*}
\cz^{6D}(q_1,t_1;q_2,t_2) =\prod\limits_{n,m,l=1}^{\infty }\, \frac{%
1-q_{2}\, q_{1}^{n-1/2}\, t_{1}^{m-1/2}\, t_{2}^{l-1}}{%
1-q_{1}^{n-1/2}\, t_{1}^{m-1/2}\, t_{2}^{l-1}} \ .
\end{equation*}%

Note that $\cz^{6D}(q,t;q,q)=\widetilde\cz\,^{6D}(q,t)$ is the elliptic gamma-function which corresponds to the perturbative part of
the $\cn=1$ gauge theory partition function in five dimensions. On
the other hand, in the Hall-Littlewood
limit $q_2\to0$ of the Macdonald measure \cite{Macdonald}, we get
\begin{equation*}
\cz^{6D}(q_1,t_1;0,t_2)=\prod\limits_{n,m,l=1}^{\infty } \ \frac{1}{%
1-q_{1}^{n-1/2}\, t_{1}^{m-1/2}\, t_{2}^{l-1}} \ .
\end{equation*}%
This is the perturbative part of Nekrasov's partition function for seven-dimensional gauge
theory compactified on a circle~\cite{Nek}. The Hall-Littlewood limit of the
two-dimensional topological field theory underlying the computation of
$\cn=2$ superconformal indices in four dimensions is also studied
in~\cite{Gadde:2011uv}.

Notice also that in the limit of coincident deformation parameters
$(q_1,t_1)=(q,t)=(q_2,t_2)$ we have
\beq
\log \cz^{6D}(q,t;q,t) = \sum_{k=1}^{\infty }\, \frac{q^{k/2}\,
  t^{k/2}}{k\, \big(1-t^{k}\big)^{2}} \ ,
\label{cancel-2}\eeq
which is essentially the partition function corresponding to a single
$t$-deformation equivalent to a $q$-deformation. The additional factor
$q^{k/2}$ appearing in (\ref{cancel-2}) is inconsequential as it can be
removed if one uses, instead of the symmetric product form of the
refined theta-function (\ref{ref-theta}), the asymmetric product form
\begin{equation}
\widetilde{\Theta }\,'(z;q,t)=\prod_{n=1}^{\infty }\, \left(
  1+q^{n-1}\, z\right)\, \left(
1+t^{n}\, z^{-1}\right) \ .  \label{asym}
\end{equation}%
In this case the gauge theory partition function evaluates to the
result (\ref{E}) with deformation parameter $t$ instead of $q$. Hence
this limiting case is equivalent to the matrix model (\ref{DT}) whose
weight function is the usual Jacobi theta-function (\ref{triple}) with
the standard Haar measure; this is just another example of the
cancellation of deformations/refinements that we first encountered in \S\ref{se:Class}.
Both theta-functions (\ref{ref-theta}) and (\ref{asym}) give exactly
the same result in the unrefined limit $q=t$.

\medskip

\subsection{Refined disk
  amplitudes}\label{se:refdisk}~\\[5pt]
Let us now consider the refined version of the Kostant identity
(\ref{Kostant}). The modification of the Gaussian (\ref{capA}) of the category
$\scrr=\Rep(\cu_q(\frg))$ is given by
$$
\bigg[\scrf_{\cu_q(\frg)}\Big(\raisebox{-5.5pt}{\Diskref}\, \Big) \bigg] =
\sum_\lambda\, \frac{M_\lambda\big(t^{\rho}\, ;\, q,t\big)}{\|M_\lambda\|_{q,t}} \
q^{\frac12\, \langle\lambda,\lambda\rangle} \ t^{\langle\lambda , \rho\rangle} \
\big[U_\lambda\otimes V_{\rm fund}^{\odot(\beta-1)\, N} \big] \ .
$$
We can also consider generalized characters as functions on the Cartan
subalgebra $\frh\subset\frg$ with values in the representation
$V_{\rm fund}^{\odot(\beta-1)\, N}$ by restriction. For this, the sole modification of the ribbon category
data that we require is that of the twist, for which we only need to
incorporate the appropriate shift in weights $n\in\IZ^N$. Using
$$
\big\langle n+(\beta-1)\, \rho\,,\,n+(\beta-1)\,\rho \big\rangle= \big\langle
n\,,\,n+2(\beta-1)\,\rho \big\rangle +(\beta-1)^2\,
\big\langle\rho\,,\,\rho \big\rangle \ , 
$$
after suitable normalization we find the refinement of the twist
coefficients
$$
\theta_n^\bullet = q^{\frac12\, \langle n,n\rangle}\ \Big(\, \frac
tq\,\Big)^{\langle n, \rho\rangle} \ .
$$
The corresponding modification of the Gaussian is thus given by
$$
\bigg[\scrf_{\cu_q(\frh)}\Big(\raisebox{-5.5pt}{\Diskref}\, \Big) \bigg]
=
\sum_{n\in\IZ^N}\, q^{\frac12\, \sum_i\, (n_i^2-(N+1-2i)\, n_i)} \
t^{\frac12\, \sum_i\,
  (N+1-2i)\, n_i} \ \big[U_n\otimes V_{\rm fund}^{\odot(\beta-1)\, N} \big] \ ,
$$
where here the representation $V_{\rm fund}^{\odot(\beta-1)\, N}$ is regarded as a $\cu_q(\frh)$-module by
restriction. Hence the modification of the Kostant identity (\ref{Kostant}) is
given by~\cite{Cherednik}
\beq
\bigg[\scrf_{\cu_q(\frg)}\Big(\raisebox{-5.5pt}{\Diskref}\, \Big)
\bigg] = \frac1{Z_N(q,t)} \
\bigg[\scrf_{\cu_q(\frh)}\Big(\raisebox{-5.5pt}{\Diskref}\, \Big)
\bigg] \ ,
\label{Kostantref}\eeq
where the normalization $Z_N(q,t)$ is the refined Chern-Simons partition
function (\ref{CSrefS3}) on $S^3$. As we discuss below, the identity (\ref{Kostantref}) may
be regarded as a
categorification of the generalization of the Weyl integral formula
for $G$-equivariant functions on the Lie group $G$ with values in the
representation $V_{\rm fund}^{\odot(\beta-1)\, N}$,
which was considered in~\cite{EFK}.

Evaluating
the generalized characters on both sides of (\ref{Kostantref}) at the
specialization $x=t^{\rho}$ yields a simple expression for the
partition function (\ref{cacuqref}) of the refined $q$-deformed gauge theory on the
sphere $S^2$ given by
$$
\cz^\bullet\big(\cu_q(\frg)\,;\,S^2\big) = 
\frac1{Z_N(q,t)} \ \prod_{j=1}^N\, \Theta\big(t^{N+1-2j}\,
q^{-\frac12\, (N+1-2j)}\,; \, q\big) \ .
$$
A similar formula is also derived in~\cite{Aganagic:2012si}; by using standard modular properties of
the Jacobi theta-function $\Theta(z;q)$, it is
related there to the Hirzebruch $\chi_y$-genus of the moduli space of
instantons on the toric surface
$\mathcal{O}(-1)\to \IP^1$, which arises in five-dimensional
supersymmetric gauge theory; this can be thought of as a
categorification of the usual Euler characteristic invariants computed
by the $\cn=4$ Vafa-Witten gauge theory.

\medskip

\subsection{Quantum gauge theory perspective}\label{se:Hilbertref}~\\[5pt]
We conclude by describing how to interpret the constructions of $(q,t)$-deformed Yang-Mills
amplitudes of this section from a field theory point of view; this is
also discussed in a similar vein but from different perspectives than ours
in~\cite{Aganagic:2011sg,Aganagic:2012si}. For this, we will describe the physical states of refined $q$-deformed
Yang-Mills theory in more detail through an operator formalism. We
take the classical field theory to be unchanged as
in~\cite{Aganagic:2011sg,Aganagic:2012si}. The Hilbert space of any topological field theory in two dimensions is based on a
circle $S^1$. As in the case of the ordinary (unrefined and
undeformed) two-dimensional Yang-Mills theory~\cite{Witten:1992xu}, canonical quantization of the field theory with action
(\ref{YMpartfnBFaction}) shows that $\phi$ and $A$ are canonically
conjugate variables. If $C=S^1$ is an initial value circle in the Riemann
surface $\Sigma^\bullet$ with a marked point, then the Hilbert space
$\ch_C$ obtained by canonical quantization on $\Sigma^\bullet$
consists of functionals of $A$ in Schr\"odinger polarization, with $A$
taken to be multiplication operators and $\phi$ acting as the functional
derivative $\phi(x)=-\ii \frac\delta{\delta A(x)}$. By gauge invariance, such functionals are
the wavefunctions
$\psi(U)$ depending only on the boundary holonomy
$U=\cp\exp\ii\oint_{C} \, A\in G$ which are valued in the finite-dimensional unitary representation $V=V_{\rm
  fund}^{\odot(\beta-1)\, N}$ of $G=U(N)$. In particular, they are not
conjugation invariant class functions of $U$, but rather define elements
of the vector space $\Omega^0(G,V)^G$ of $G$-equivariant $V$-valued
functions, i.e. $\psi(g\, U\, g^{-1})=g\triangleright \psi(U)$ for all
$g,U\in G$. To define amplitudes of such states, we note that since $V$ is a unitary representation, it has a natural $G$-invariant inner
product $\langle-|-\rangle_V$, i.e. $\langle g\triangleright
v|g\triangleright w\rangle_V=\langle v|w\rangle_V$ for all $g\in G$ and
$v,w\in V$. Hence we can define
an inner product on $\Omega^0(G,V)^G$ by
\beq
\langle\psi|\chi\rangle:= \int_G\, \dd U \ \big\langle\psi(U)\,\big|\,
\chi(U)\big\rangle_V \ .
\label{OmegaGVinnprod}\eeq
This inner product defines the gluing rules for states associated to
surfaces with boundary $C$.
By the isomorphism $G/{\rm Ad(G)}= T/W$, we have
\beq
\Omega^0(G,V)^G = \Omega^0\big(T,V^{(0)}\big)^W \ .
\label{Omega0GVG}\eeq
Since the weight zero subspace $V^{(0)}\cong\IC$ in the case at hand,
this isomorphism identifies equivariant functions on $G$ with symmetric
functions on the maximal torus $T=U(1)^N$. As the scalar function
$U\mapsto\langle\psi(U)|\chi(U)\rangle_V$ on $G$ is conjugation invariant, we
can apply the usual Weyl integral formula (\ref{Weylintgroup}) to
write the inner product (\ref{OmegaGVinnprod}) as
\beq
\langle\psi|\chi\rangle = \frac1{N!}\, \int_{[0,2\pi)^N} \
\prod_{i=1}^N \,
\frac{\dd\phi_i}{2\pi}\ \prod_{j<k}\, 4\sin^2\Big(\,
\frac{\phi_j-\phi_k}2\, \Big) \ \big\langle \psi(\phi)\,\big|\,
\chi(\phi)\big\rangle_V \ .
\label{OmegaGVinndiag}\eeq

Thus the Hilbert space of the quantum gauge theory
$$
\ch_C=L^2(G,V)^G
$$
is the $L^2$-completion of the vector space
$\Omega^0(G,V)^G$ with respect to the norm induced by the inner
product (\ref{OmegaGVinnprod}). A natural basis for $\ch_C$ is
provided by generalised characters, and there is an analog of the Peter-Weyl
theorem for generalised characters which decomposes the Hilbert space
under the action of $G\times G$ into an orthogonal direct sum of
finite-dimensional $G$-modules as~\cite{EFK}
\beq
\ch_C =\bigoplus_{\lambda}\, \big(U_\lambda^*\otimes U_\lambda\otimes
V \big)^G \ ,
\label{PeterWeylgen}\eeq
where the Hilbert space direct sum runs over the orthogonal subspaces
$\Hom_G(U_\lambda,U_\lambda\otimes V)$ of intertwining operators
$\Phi_\lambda$ with respect to the inner
product (\ref{OmegaGVinnprod}). A generalization of the Weyl
orthogonality theorem is proven in~\cite{Etingof:1993pv}, whereby it is shown that the
spaces of intertwining operators $\Hom_\scrr(U_\lambda,U_\lambda\otimes
V)$ for the quantum group $\cu_q(\frg)$ form a mutually orthogonal system with respect to the inner
product (\ref{OmegaGVinndiag}); this suggests an analog of the
decomposition (\ref{PeterWeylgen}) of the Hilbert space of physical
states into generalized characters $(U_\lambda^*\otimes
U_\lambda\otimes V)^{\cu_q(\frg)}$. As discussed previously, when the
highest-weight modules $U_\lambda$ are regarded
as quantum group representations, the basis of generalised characters
for $\ch_C$ coincide with the Macdonald polynomials $M_\lambda(u;q,t)$
in the holonomy eigenvalues $u\in(S^1)^N$, regarded as elements of
$L^2\big(T,V^{(0)}\big)^W$ under the isomorphism (\ref{Omega0GVG}), which are orthogonal with respect to the inner
product (\ref{torusinnerprod}). It follows that any physical state wavefunction
$\psi$ has
an expansion in generalised characters as
$$
\psi(u) = \sum_\lambda\, c_\lambda(q,t) \, M_\lambda(u;q,t)
$$ 
where $c_\lambda(q,t) \in\IC$.

Canonical quantization shows that the Hamiltonian of the gauge theory
with action (\ref{YMpartfnBFaction}) is quadratic and given by
$$
H=-\frac {g_s}2\, \oint_C\, \dd\sigma \ \Tr \phi^2
$$
where $\sigma\in S^1$ is the local coordinate of the initial value
circle $C\subset\Sigma^\bullet$. On the Hilbert space $\ch_C$, this
operator acts via $\Tr\phi^2=-\Tr\big(\frac\delta{\delta A}\big)^2 =
-\Tr\big(U\,\frac\partial{\partial U} \big)^2$,
and hence as usual the Hamiltonian operator is proportional to the
Laplace-Beltrami operator on the group manifold of $G$. It is shown
by~\cite{EFK} that every conjugation invariant scalar differential
operator on $G$ defines an operator on $\Omega^0(G,V)^G$ acting as a
scalar on $\Hom_G(U_\lambda,U_\lambda\otimes V)$ for every
irreducible representation $\lambda$, and using the isomorphism
(\ref{Omega0GVG}) this action can be rewritten in terms of
differential operators on $T$ with coefficients in
$\End_T\big(V^{(0)}\big)$. In particular, the Laplace-Beltrami operator acts diagonally on
the space $L^2(G,V)^G$ in the generalised characters $M_\lambda$
with eigenvalue $C_2(\lambda)$~\cite{EFK}, and hence the Hamiltonian is
diagonalised in this basis as the quadratic Casimir operator
\beq
H\, M_\lambda = \mbox{$\frac {g_s}2$}\, C_2(\lambda)\, M_\lambda \ .
\label{spectrumM}\eeq
Whence the quantum amplitudes, computed as matrix elements of the
operator $\exp(-\tau\, H)$ between external states in $\ch_C$,
involves the standard heat kernel $\e^{-\tau \,g_s \,C_2(\lambda)/2}$ of
the gauge group $G$. With $q=\e^{-g_s}$ and $t=q^\beta$, this heat
kernel coincides (up to area-dependent renormalization ambiguities)
with the twist eigenvalues
(\ref{twistref}) which are used in the building blocks of refined
$q$-deformed Yang-Mills amplitudes. 

The same spectrum (\ref{spectrumM}) and Hilbert space (\ref{PeterWeylgen}) were obtained
in a similar fashion
in ordinary two-dimensional Yang-Mills theory by Gorsky
and Nekrasov in~\cite[\S1.3]{Gorsky:1993dq}; they compute the path
integral for Yang-Mills theory on a cylinder cut by a Wilson line in
the representation $V=V_{\rm fund}^{\odot(\beta-1)\, N}$ along the
temporal direction. They further conjecture that the analogous
calculation in the gauged $G/G$ WZW model should be expressible
through representations of the quantum group $\cu_q(\frg)$, with the
presence of the Wilson line in the representation $V$ yielding a
deformation of the usual gauge group characters, i.e. the Schur polynomials, to Macdonald
polynomials. Given the relation of this two-dimensional topological
gauge theory to Chern-Simons theory on the corresponding trivial
circle bundle, Iqbal and Kozcaz suggest in~\cite{Iqbal:2011kq} that
this deformed WZW theory could be related to refined Chern-Simons
theory; an analogous relationship in the unrefined case is described in~\cite{deHaro}.
The Hamiltonian analysis presented here further agrees
with the results of~\cite{Tachikawa:2012wi}, where the decomposition
(\ref{PeterWeylgen}) and the standard heat kernel (\ref{spectrumM}) on $G$ also
appear in the refined two-dimensional Yang-Mills propagator that
computes the partition function of an $\cn=2$ non-linear sigma-model on
$S^1\times S^3$.

This construction also illustrates how the refined partition functions
${Z}_{\mathrm{YM}}^{(p) }(q,t ; \Sigma^\bullet_{h})$ may be derived
directly in the path integral formalism through the technique of
diagonalisation. For fixed 
$\phi\in\Omega^0(\Sigma_h^\bullet,G)$, it follows from our description of the
physical states of the refined gauge theory that we should extend the functional
(\ref{Fphi}) to a $\cg$-equivariant functional valued in $V=V_{\rm
  fund}^{\odot(\beta-1)\, N}$; in the diagonalisation formula that follows
from (\ref{Omega0GVG}), we should
then employ the Macdonald inner product (\ref{torusinnerprod})
appropriate to generalised characters associated with $\cu_q(\frg)$-modules. The path integral is thus given
by the sum over torus bundles
\beqa
Z^{(p)}_{\rm YM}(q,t;\Sigma^\bullet_h) &=& \frac1{{\rm vol}(\cg)} \ \sum_{n\in\IZ^N} \ \int_{\ca_n} \,
\scrd\mu[A^\frh] \ \int_{\Omega^1(\Sigma^\bullet_h,\cl_n\times_T\frk)} \,
  \scrd\mu[A^\frk] \nonumber\\ && \times \ 
  \int_{\Omega^0(\Sigma^\bullet_h, (\IR/2\pi\, \IZ)^N)} \
  \prod_{i=1}^N \,
  \scrd\mu[\phi_i] \ \big[\Delta_{q,t}(\phi) \big] \
  \exp\big(- S^{\bullet\, (p)}_{\rm BF}[\phi,A^\frh,A^\frk] \big) \ , \nonumber
\eeqa
where
\beqa
S^{\bullet\, (p)}_{\rm BF}[\phi,A^\frh,A^\frk]&=& \frac1{g_s} \, \sum_{i=1}^N \ \int_{\Sigma^\bullet_h} \, \Big(-\ii \phi_i\,
\dd A^\frh_i+\frac p2\, \phi_i^2\, \dd \mu\Big) \nonumber \\ && +\, \sum_{\alpha\in{\rm Ad}(G)}\
\int_{\Sigma^\bullet_h}\, \ \prod_{m=0}^{\beta-1}\, \Big( 1-\e^{-m\, g_s}\, \e^{\ii\langle\alpha,\phi\rangle} \Big) \, A^\frk_\alpha\wedge
A_{-\alpha}^\frk \ . \nonumber
\eeqa
Proceeding as in \S\ref{se:2DYMgen}, this sum yields the appropriate $\beta$-deformation
$$
Z^{(p)}_{\rm YM}(q,t;\Sigma_h^\bullet) = \sum_{\mu\in\IZ^N} \ \prod_{i<j} \ \prod_{m=0}^{\beta-1}\,
[\mu_i-\mu_j+m]^{1-h}_q\,[\mu_i-\mu_j-m]^{1-h}_q \, \exp\Big(-
\frac{p\, g_s}2 \, \sum_{i=1}^N\, \mu_i^2\Big)
$$
of the $q$-deformed discrete Gaussian matrix model
(\ref{qmat}). However, a complete Lagrangian or Hamiltonian
description of the refined two-dimensional Yang-Mills theory, and of the
related refinement of Chern-Simons theory on Seifert three-manifolds,
is currently lacking, and it would be interesting to find a more
precise and physical derivation of these quantum
amplitudes from first principles. In particular, it is not clear at
present how to properly incorporate the quantum group gauge symmetry
based on $\cu_q(\frg)$ into the definition of the quantized gauge theory.

The construction presented here further elucidates the physical meaning of the
class of deformed gauge theories introduced in \S\ref{se:dimCasdef},
and in particular of Klim\v{c}\'{\i}k's partition function
(\ref{klim}). It consists in using a Hamiltonian framework based on
replacing differential operators with
$q$-difference operators on $\Omega^0(T)$, generalizing the quantum
torus deformation of \S\ref{se:expCasdef}, and $G$-modules with
representations of the quantum group $\cu_q(\frg)$. Consider the operators $\hat
u_i=z_i$ acting as multiplication by $z_i\in S^1$ and the
$q$-difference operators $\hat v_i=\exp\big(-\log(q)\, z_i\,
\frac\partial{\partial z_i}\big)$ for $i=1,\dots,N$. They obey the
quantum $N$-torus algebra relations
$$
\hat u_i\, \hat v_j=q\ \delta_{ij}\ \hat v_j\, \hat u_i
$$
for $i,j=1,\dots,N$. It follows by~\cite[Prop.~6.2]{Etingof:1993pv}
that corresponding to the central quantum Casimir element ${\tt
  Cas}_q$ for $\cu_q(\frg)$ (see Appendix~A) there exists a unique $q$-difference
operator $\hat H_{{\tt Cas}_q}(\hat u,\hat v)$ whose spectrum consists
of the twist eigenvalues $\theta_\lambda$ from \S\ref{se:qAxioms},
such that the generalized characters $M_\lambda$ satisfy the
difference equation
$$
\hat H_{{\tt Cas}_q}(\hat u,\hat v)\, M_\lambda = q^{\frac12\,
  C_2(\lambda)} \, M_\lambda \ .
$$
This spectrum generalizes (\ref{spectrumM}) and the corresponding
``heat kernel'' yields $q$-deformed Casimir eigenvalues as in
\S\ref{se:dimCasdef} and (\ref{klim}).

\setcounter{section}{0}
\setcounter{subsection}{0}

\appendix{Quantum groups\label{app:QG}}

The quantum universal
enveloping algebra $\cu_q(\frg)$ for $\frg=\frsl(N)$ is defined as
the unital algebra over $\IC$ generated by elements $E_i,F_i,
K_i^{\pm\, 1}$,
$i=1,\dots, N-1$, with the defining relations
\beqa
K_i\, K_i^{-1}=1=K_i^{-1}\, K_i \quad &\mbox{and}& \quad \big[K_i\,,\, K_j^{\pm\,1}\, \big]= 0 \ ,
\nonumber\\[4pt]
K_i\,E_i=q\ E_i\,K_i \quad &\mbox{and}& \quad K_i\,F_i=q^{-1} \ F_i\,K_i \ ,
\nonumber\\[4pt]
K_i\,E_{i\pm1} =q^{-1/2} \ E_{i\pm1} \,K_i \quad &\mbox{and}& \quad K_i\,F_{i\pm1}=q^{1/2}
\ F_{i\pm1} \,K_i \ ,
\nonumber\\[4pt]
[K_i,E_j]=0 \quad &\mbox{and}& \quad [K_i,F_j]=0 \qquad \mbox{for} \quad
j\neq i,i\pm1 \ , \nonumber\\[4pt]
[E_i,F_j]&=& \delta_{ij} \ \frac{K_i-K_i^{-1}}{q^{1/2}-q^{-1/2}} \ , \nonumber
\eeqa
together with the Serre relations
\beqa
E_i^2\, E_j-\big(q^{1/2}+q^{-1/2}\big)\, E_i\, E_j\, E_i +E_j\, E_i^2=0 \quad
&\mbox{for}& j=i\pm1 \ , \nonumber\\[4pt]
F_i^2\, F_j-\big(q^{1/2}+q^{-1/2}\big)\, F_i\, F_j\, F_i+ F_j\, F_i^2 =0 \quad
&\mbox{for}& j=i\pm1 \ , \nonumber\\[4pt]
[E_i,E_j]=0= [F_i,F_j] \quad &\mbox{for}& j\neq i,i\pm1 \ . \nonumber
\eeqa
A vector space basis for
$\cu_q(\frg)$ is given by the set $\{E_i^{n_i}\, K_i^{m_i}\, F^{l_i}_i
\ | \ n_i,l_i\in\IZ_{\geq0} \ , \ m_i\in\IZ \ , \ i=1,\dots,N-1\}$. We
usually take the deformation parameter $q\in\IR$ to lie in the
interval $0<q<1$ without loss of generality. The quantum Casimir
element for $\cu_q(\frg)$ is given by
$$
{\tt Cas}_q:= \sum_{i=1}^{N-1}\, \Big(\, E_i\, F_i+\frac{q^{-1/2}\,
  K_i+q^{1/2} \,
  K_i^{-1}}{\big(q^{1/2}-q^{-1/2}\big)^2}\, \Big) = \sum_{i=1}^{N-1}\,
\Big(\, F_i\, E_i+\frac{q^{1/2} \, K_i+q^{-1/2}\,
  K_i^{-1}}{\big(q^{1/2}-q^{-1/2}\big)^2}\, \Big) \ .
$$
The element ${\tt Cas}_q$ is central in the algebra $\cu_q(\frg)$.

A Hopf algebra structure on $\cu_q(\frg)$ is provided by the coproduct
$\Delta:\cu_q(\frg)\to\cu_q(\frg)\otimes \cu_q(\frg)$, the counit
$\varepsilon: \cu_q(\frg)\to \IC$ and the antipode $S:\cu_q(\frg)\to
\cu_q(\frg)$ given on generators by
\beqa
\Delta\big(K_i^{\pm\,1}\big)= K_i^{\pm\,1}\otimes K_i^{\pm\,1} \ ,
\qquad \varepsilon\big(K_i^{\pm\,1}\big)=1 \quad &\mbox{and}& \quad
S\big(K_i^{\pm\,1}\big)= K_i^{\mp\,1} \ , \nonumber\\[4pt]
\Delta(E_i)=E_i\otimes K_i+1\otimes E_i \ , \qquad \varepsilon(E_i)=0
\quad &\mbox{and}& \quad S(E_i)=-E_i\, K_i^{-1} \ , \nonumber\\[4pt]
\Delta(F_i)=F_i\otimes1+K_i^{-1}\otimes F_i \ , \qquad
\varepsilon(F_i)=0 \quad &\mbox{and}& \quad S(F_i)=-K_i\, F_i \ ,
\nonumber
\eeqa
and extended as (anti-)algebra homomorphisms. The $*$-structure is the
anti-algebra morphism given on generators by
$$
K_i^*=K_i \ , \qquad E_i^*= F_i \qquad \mbox{and} \qquad F_i^*=E_i \ .
$$

The Hopf algebra $\cu_q(\frg)$ has a quasitriangular structure defined
by a universal $R$-matrix, which is an invertible element $R$ in a certain
completed tensor product algebra $\cu_q(\frg)\,\widehat{\otimes}\,
\cu_q(\frg)$ that intertwines the coproduct $\Delta$ and the opposite
coproduct $\Delta^{\rm op}:=P\circ\Delta$ where $P$ is the flip
isomorphism $P(a\otimes b)=b\otimes a$ for $a,b\in\cu_q(\frg)$. It has the form
$$
R=q^{\frac12\, \sum_i\, H_i\otimes H_i} \ R^\vee \qquad \mbox{with} \quad
  R^\vee\in\cu_q^+(\frg)\,\widehat{\otimes}\, \cu_q^-(\frg) \ ,
$$
where $H_i$, $i=1,\dots,N-1$ is an orthonormal basis for the Cartan
subalgebra $\frh\subset\frg$ with respect to the invariant bilinear
form $\Tr|_\frh$; we formally identify the group-like generators
$K_i=q^{H_i/2}$. Here $\cu_q^+(\frg)$ (resp.~$\cu_q^-(\frg)$) is
the subalgebra of the quantum enveloping algebra $\cu_q(\frg)$ generated by
$K_i^{\pm\,1},E_i$ (resp.~$K_i^{\pm\,1},F_i$). The element
$R^\vee$ satisfies
$$
(\varepsilon\otimes1)\big(R^\vee\,\big)=1\otimes 1=
(1\otimes\varepsilon)\big(R^\vee\, \big) \ .
$$
We use the standard Sweedler
notation
$$
R=R_{(1)}\otimes R_{(2)}
$$
with implicit summation.

From the quasitriangular structure $R$, one
constructs Drinfel'd's element $u$ in a certain completion of
$\cu_q(\frg)$ as
$$
u := S\big(R_{(2)}\big)\, R_{(1)} \ .
$$
It is invertible and has the property that $u\, S(u)$ is a central
element with $S^2=S\circ S$ acting as an inner automorphism
$$
S^2(a)= u\, a\, u^{-1}
$$
for all $a\in\cu_q(\frg)$, and moreover
\beqa
R_{(2)}\, u\, R_{(1)} &=& S\big(R_{(2)}\big)\, u\, S\big(R_{(1)}\big)
\ = \ 1 \ , \nonumber \\[4pt]
\Delta(u) \ = \  (u\otimes u)\, \big(R_{(2)}\, R_{(1)}\otimes R_{(1)}\,
R_{(2)}\big)^{-1} &=& \big(R_{(2)}\, R_{(1)}\otimes R_{(1)}\,
R_{(2)}\big)^{-1} \, (u\otimes u) \ . \nonumber
\eeqa

\appendix{Toeplitz determinants}

Let $f(z)$ be a complex-valued function on $\IC$ with Laurent series
expansion $f(z)=\sum_{k\in\IZ}\, f_k\, z^k$, and let
$T_{N}(f ) =(f_{i-j})_{i,j=1,\dots,N}$ be the associated Toeplitz operator of dimension $N$ and
symbol $f$. By the Heine-Szeg\H{o} identity, the corresponding
Toeplitz determinant is the partition function of a $U(N)$
unitary matrix model
\beq
Z_N[f] := \det T_N(f) = \int_{[0,2\pi)^N} \ \prod_{i=1}^N \, \frac{\dd
  \phi_i}{2\pi}\ f\big(\e^{\ii\phi_i}\big) \ \prod_{j<k}\,
\big|\e^{\ii\phi_j}- \e^{\ii\phi_k}\big|^2 \ .
\label{HSid}\eeq
Let $[\log f]_k$, $k\in\IZ$ denote the coefficients in the Fourier series
expansion on the unit circle $S^1$ of the logarithm of the symbol,
$$
\log f(z)= \sum_{k=-\infty}^\infty\, [\log f]_k \
z^k \ ,
$$
and suppose that they obey the absolute summability conditions
$$
\sum_{k=-\infty}^\infty\, \big\vert \left[ \log f \right]
_{k}\big\vert <\infty \qquad \mbox{and} \qquad
\sum_{k=-\infty}^\infty\, k\, \big\vert \left[ \log f \right]
_{k}\big\vert ^{2}<\infty \ .
$$
Let $G(f)
=\exp ( \left[ \log f \right] _{0}) $ denote the geometric
mean of the symbol $f$. 

Then the strong Szeg\H{o} limit theorem for Toeplitz determinants states~\cite{SS,Simon}
\begin{equation*}
\lim_{N\rightarrow \infty }\, \frac{\det T_{N}( f ) }{G(
f ) ^{N}}=\exp \Big(\, \sum_{k=1}^{\infty }\, k\, \left[ \log f
\right]_{k}\, \left[ \log f \right] _{-k}\, \Big) \ .
\end{equation*}%
By the Heine-Szeg\H{o} identity (\ref{HSid}), the Szeg\H{o} theorem is not only a
statement about Toeplitz determinants but also about unitary matrix models.
In particular, the strong Szeg\H{o} theorem gives
the partition function $Z_\infty[f]$ of a $U(\infty )$ unitary matrix
model, defined as the $N\to\infty$ limit of (\ref{HSid}). The strong
Szeg\H{o} theorem generally involves an exponentially small error
term ${O}\left( \e^{-B\, N}\right) $~\cite{Simon}; if $\log f(z)$ is
real-valued and analytic in a neighbourhood of the unit circle
$S^1\subset\IC$, then the error term is simply ${O}\left( \e^{-B\,
    N}\right) $, i.e. there are no $\frac1N$
corrections~\cite{Simon}. The analyticity condition holds if the
Fourier coefficients $[\log f]_k$ have exponential decay as
$k\to\infty$, which is precisely the case studied in this paper;
although we have considered complex powers $\alpha_a$ in
(\ref{FKweight}), a simple argument shows that the reality condition
can also be relaxed~\cite{Johansson1}. If the function
$f(z)$ is holomorphic, then the Szeg\H{o} theorem is enough to
determine the asymptotics. For the more general case of a symbol with zeroes or poles,
one has to use the more refined Fisher-Hartwig asymptotics.

The Heine-Szeg\H{o} identity for generalized Toeplitz determinants reads as~\cite{Mat}
\beq
Z_N[f;q,t]:= \det \widetilde{T}_N(f;q,t) = \int_{[0,2\pi)^N} \
\prod_{i=1}^N \, \frac{\dd
  \phi_i}{2\pi}\ f\big(\e^{\ii\phi_i}\big) \
\Delta_{q,t}\big(\e^{\ii\phi}\big)
\ ,
\label{HSidgen}\eeq
where
$$
\Delta_{q,t}( z) =\prod\limits_{i<j}\, \bigg\vert\, 
\frac{\big( z_{i}\, z_{j}^{-1}\,;\, q\big) _{\infty }}{\big(
t\, z_{i}\, z_{j}^{-1}\, ;\, q\big) _{\infty }}\, \bigg\vert ^{2}
$$
is the Macdonald measure~\cite{Macdonald}. This class of integrals has
been considered for a long time in the context of
the Selberg integral~\cite{FW,War}.
Under the same conditions on the Fourier
coefficients of the logarithm of the symbol $f$, there is an extension of the Szeg\H{o} theorem to
this generalized case which reads as~\cite{Mat}%
\begin{equation}
\lim_{N\rightarrow \infty }\, \frac{\det \widetilde{T}_{N}(f;
    q,t) }{G( f) ^{N}}=\exp \Big(\,
\sum_{k=1}^{\infty }\, k\,
\left[ \log f\right] _{k}\, \left[ \log f\right] _{-k}\, \frac{1-q^{k}}{%
1-t^{k}}\, \Big) \ .  \label{Szegoextended}
\end{equation}%
This is a statement about the $N\to\infty$ limit of a matrix model in
the refined ensemble rather than the unitary ensemble, obtained as in
(\ref{HSidgen}) by
replacing the usual $U(N)$ Haar measure in the matrix integral with the
Macdonald measure.

Finite-dimensional Toeplitz determinants can be
computed with the method of Day~\cite{Day}; this problem has been revisited
recently in the context of random matrix theory applications to
$L$-functions in number theory. For this, let $R_1,R_2\in\IR$ with
$0\leq R_1<R_2$. Consider complex polynomials
$$
D(z)=\prod_{j=1}^k \, (z-\delta_j) \ , \qquad F(z) = \prod_{j=1}^h\,
\big(1-\rho_j^{-1}\, z\big) \qquad \mbox{and} \qquad
G(z)=\prod_{j=1}^p\, (z-r_j)
$$
where the zeroes $\delta_i$ satisfy $|\delta_i|\leq R_1$ for
$i=1,\dots,k$, the zeroes $\rho_j$ satisfy $|\rho_j|\geq R_2$ for
$j=1,\dots,h$, and the zeroes $r_1,\dots,r_p$ are distinct. Let $f(z)=
G(z)/F(z)\,  D(z)$ on the annulus $\{z\in\IC \ | \ R_1<|z|<R_2\}$. If
$p=k+m$ with $m\geq h$, then
\beq
\det T_N(f)= (-1)^{m\, (N+1)}\, \sum_{\stackrel{\scriptstyle
    I\subset\{1,\dots,k+m\}}{|I|=m}} \ \prod_{i\in I} \
\prod_{s=1}^k \ \prod_{j\in\overline{I}} \
\prod_{t=1}^h \, r_i^{N+1}\, \frac{(r_i-\delta_s) \, (\rho_t-r_j)
}{(r_i-r_j)\, (\rho_t-\delta_s)} \nonumber
\eeq
where $\overline{I}:= \{1,\dots,k+m\}\setminus I$.

\appendix{Embedding theorems for abelian categories}

\subsection{Ind-completions}~\\[5pt]
In category theory the
notions of disjoint unions and direct sums are generalised to
\emph{colimits}, which are diagrams indexed by discrete categories. We
describe here how to construct a category of all countable direct
colimits in an abelian category which contains all necessary features
justifying formal sums over infinite
sets of simple objects; it is obtained
by formally adjoining directed colimits. For further details and
properties of the construction, see~\cite{KashShap}.

We begin by summarizing some of the basic notions that we need,
beginning with that of filtered categories, which generalise the
notion of directed set to category theory. A non-empty category $I$ is
\emph{filtered} when:
\begin{itemize}
\item[(1)] For every pair of objects $i,i'\in\Ob(I)$, there exists an
  object $k\in\Ob(I)$ and morphisms $(f:i\to k)\in\Hom_I(i,k)$ and
  $(f':i'\to k)\in\Hom_I(i',k)$.
\item[(2)] For every pair of morphisms $(u,v:i\to j)\in\Hom_I(i,j)$,
  there exists an object $k\in\Ob(I)$ and a morphism $(w:j\to
  k)\in\Hom_I(j,k)$ such that $w\circ u=w\circ v$.
\end{itemize}
The actual objects and morphisms in $I$ are largely
irrelevant; only the ways in which they are interrelated above
matters. Let $\scrc$ be a small abelian category enriched over $\Vect$, and $I$
a filtered category. A functor $F:I\to \scrc$ is called a ``diagram of
type $I$''; a diagram can be thought of as indexing a collection of
objects and morphisms of $\scrc$ patterned on the directed index
category $I$. Diagrams are also called \emph{direct systems}. A \emph{co-cone} of a diagram $F: I\to\scrc$ is an object
$N\in\Ob(\scrc)$ together with a family of morphisms $(\psi_i:F(i)\to
N) \in\Hom_\scrc(F(i),N)$, $i\in\Ob(I)$, such that
$\psi_j\circ F(f)=\psi_i$ for all morphisms $(f:i\to
j)\in\Hom_I(i,j)$. A \emph{filtered colimit} of a diagram $F:I\to\scrc$ is
co-cone $(L,\varphi)$ of $F$ which is universal: For any other co-cone
$(N,\psi)$, there exists a unique mediating morphism $u\in\Hom_\scrc(L,N)$ such
that the diagrams
$$
\xymatrix{
F(i) \ \ar[rr]^{F(f)} \ar[dr]^{\varphi_i} \ar[ddr]_{\psi_i} & &  \ F(j)
\ar[dl]_{\varphi_j} \ar[ddl]^{\psi_j} \\ & \ L \ \ar@{.>}[d]^u & \\ &
\ N \ &
}
$$
commute for all $i,j\in \Ob(I)$. Colimits are also called ``direct limits'' or
``inductive limits''; if a diagram $F$ has a colimit then it is unique
up to unique isomorphism. Similarly, one defines limits by taking
colimits in the corresponding dual categories.

Let
$\scrc^\vee$ be the cocomplete
 functor category of $\IC$-linear contravariant functors from $\scrc$ to the
category $\Vect$ of complex vector spaces and linear transformations. 
There are two ways in which we can adjoin filtered colimits to the
category $\scrc$: Either formally by regarding objects of the cocompletion
as filtered diagrams in $\scrc$, or concretely as objects in
$\scrc^\vee$ which are expressible as filtered colimits of
representable functors.
The Yoneda embedding $X\mapsto
\Hom_\scrc(-,X)$ for $X\in\Ob(\scrc)$ is a fully faithful exact functor
sending $\scrc\hookrightarrow \scrc^\vee$; we identify $\scrc$ with
this full subcategory of $\scrc^\vee$ in what follows. It has the
properties
\beqa
\Hom_{\scrc^\vee}\big(\Hom_\scrc(-,X)\,,\, F\big)&=& F(X) \ ,
\nonumber\\[4pt] \Hom_{\scrc^\vee}\big(\Hom_\scrc(-,X)\,,\,
\Hom_\scrc(-,Y)\big) &=& \Hom_\scrc(X,Y)
\label{Yonedaprops}\eeqa
for $F\in\Ob(\scrc^\vee\,)$ and $X,Y\in\Ob(\scrc)$.
An ind-object or formal inductive
limit over $\scrc$ is a diagram $\underline{X}:I\to\scrc$, where $I$
is a partially ordered directed set regarded as a small filtered category. We write
$\underline{X}=(X_i)_{i\in \Ob(I)}$ where $X_i:=\underline{X}(i)\in \Ob(\scrc)$. Since
the $\Hom_\scrc$ functor preserves all limits in $\scrc$, it relates
colimits in $\scrc$ to colimits in $\Vect$ and the ind-object
$\underline{X}=(X_i)_{i\in \Ob(I)}$ is uniquely
isomorphic to the object of $\scrc^\vee$ of the form
\beq
Y\ \longmapsto \ \Hom_{\scrc^\vee}\big(\Hom_\scrc(-, Y)\,,\,
\underline{X}\, \big) = \lim_{\overrightarrow{i\in \Ob(I)}}\, \Hom_\scrc(Y,X_i) \
,
\label{indobjectvee}\eeq
which is natural in $Y$ and respects colimiting cones;
here the inductive limit is taken in the category of
vector spaces. This notion of an ind-object is a refinement of the notion of
colimit.

We write $\Ind(\scrc)$ for the category whose objects are ind-objects
over $\scrc$. Two ind-objects $\underline{X}:I\to\scrc$ and
$\underline{Y}:J\to\scrc$ determine a functor
$\Hom_\scrc(X_i,Y_j):I^{\rm op}\times J\to \Vect$; hence by Yoneda's
lemma and the construction of colimits in $\scrc^\vee$, we can use
(\ref{Yonedaprops}) and (\ref{indobjectvee}) to define the morphisms
of $\Ind(\scrc)$ by
$$
\Hom_{\Ind(\scrc)}\big(\, \underline{X}\,,\, \underline{Y}\, \big) =
\lim_{\overrightarrow{i\in \Ob(I)}} \ \lim_{\overleftarrow{j\in \Ob(J)}}\,
\Hom_\scrc(X_i,Y_j) \ ,
$$
where again the limit and colimit are taken in the
category $\Vect$. Then $\Ind(\scrc)$ is a full abelian subcategory of
$\scrc^\vee$, and the embedding $\Ind(\scrc)\hookrightarrow\scrc^\vee$
preserves all  limits. In fact, the functor $\underline{X}\mapsto
\Hom_{\Ind(\scrc)}(-, \underline{X}\,)$ is a fully faithful left-exact
equivalence between $\Ind(\scrc)$ and the abelian category
$\scrc^{\vee,{\rm add}}$ of additive functors $\scrc^{\rm op}\to
\Vect$. The ind-category $\Ind(\scrc)$ is $\IC$-linear and it has a concrete description in many cases; for example, if $\scrc$ is the
category of finite-dimensional vector spaces, then the ind-category
$\Ind(\scrc)=\Vect$ may be identified with the category of all vector spaces
$\Vect$ itself. The category $\Ind(\scrc)$ has the usual nice
properties which can be found in~\cite{KashShap}; in particular, by
(\ref{indobjectvee}) it is
equivalent to the full subcategory of $\scrc^\vee$ consisting of
functors which are filtered colimits of representable functors.

The Yoneda embedding identifies $\scrc$ with a full subcategory of its
ind-category $\Ind(\scrc)$. Then the objects
$\Ob(\scrc)$ are identified with the constant ind-objects: The map
\beq
\scry\,:\, \scrc \ \longrightarrow \ \Ind(\scrc) \ , \quad \scry(X)= (X_i)_{i\in
  I_\emptyset} \qquad \mbox{with} \quad I_\emptyset=\{\emptyset\} \ ,
\ X_\emptyset=X \ ,
\label{indembedding}\eeq
is a natural
$\IC$-linear fully faithful exact functor. It preserves all limits
which exist in $\scrc$. The category $\Ind(\scrc)$ is cocomplete,
i.e. it has all  colimits, and in fact the full embedding
(\ref{indembedding}) makes it into a
\emph{cocompletion} of the category $\scrc$: Every object of
$\Ind(\scrc)$ is a colimit of objects in the image of $\scry$. Note
that the Yoneda embedding $\scrc\hookrightarrow \scrc^\vee$ is also a
cocompletion of $\scrc$. What uniquely characterizes the
ind-completion among all cocompletions of $\scrc$ is
that all objects $X$ in the image of (\ref{indembedding}) are finitely
presentable in $\Ind(\scrc)$, i.e. the functor
$\Hom_\scrc(X,-):\Ind(\scrc)\to\Vect$ preserves partially ordered
directed colimits.

If
$\scrc$ is in addition a monoidal category, then its tensor structure extends to 
$\Ind(\scrc)$ in the following way. Let $\underline{X}=(X_i)_{i\in \Ob(I)}$
and $\underline{Y}=(Y_j)_{j\in \Ob(J)}$ be ind-objects over $\scrc$. Using
the exterior product bifunctor $\otimes:\scrc\times\scrc\to\scrc$ we define
$\otimes:\Ind(\scrc)\times \Ind(\scrc)\to \Ind(\scrc)$ by the formula
$\underline{X}\, \otimes\, \underline{Y}:= (X_i\otimes Y_j)_{i\in
  \Ob(I),j\in\Ob(J)}$. Given another ind-object
$\underline{Z}=(Z_k)_{k\in\Ob(K)}$, there are functorial isomorphisms
$(X_i\otimes Y_j)\otimes Z_k\to X_i\otimes(Y_j\otimes Z_k)$ which
induce an isomorphism between the ind-objects $(\,\underline{X}\otimes
\underline{Y}\, )\otimes \underline{Z}:=
( (X_i\otimes Y_j)\otimes
Z_k)_{i\in\Ob(I),j\in\Ob(J),k\in\Ob(K)}$ and $\underline{X}
\otimes(\,\underline{Y} \otimes
\underline{Z} \,):=(X_i\otimes(Y_j\otimes Z_k) )_{i\in\Ob(I),j\in\Ob(J),
  k\in\Ob(K)}$. One can extend other structures on the category
$\scrc$ to its ind-completion $\Ind(\scrc)$ in similar ways; in
particular, in this manner it is possible to define notions of
ind-algebras, ind-modules, and so on.

\medskip

\subsection{Morita equivalence}\label{app:Morita}~\\[5pt]
Let $\Mod_\scrc(A)$ be the module category over an algebra object $A$
of a semisimple monoidal category $\scrc$. Let $\modM_1,\modM_2\in\Ob(\Mod_\scrc(A))$. Since the functor
$X\mapsto\Hom_A(\modM_1\otimes X,\modM_2)$ is (right) exact, it has a
left adjoint $\inthom_\scrc(M_1,M_2)$ called the internal Hom from
$\modM_1$ to $\modM_2$. This is an ind-object of $\scrc$ representing
this functor, and it is an internal version of the ``space of
morphisms from $\modM_1$ to $\modM_2$''. When both categories $\scrc$
and $\Mod_\scrc(A)$ have finitely-many simple objects, then
$\inthom_\scrc(M_1,M_2)\in\Ob(\scrc)$ is an object of $\scrc$ which is
uniquely defined up to isomorphism by Yoneda's lemma, and so
$\inthom_\scrc(-,-)$ is a bifunctor. Then the internal Hom is defined by
the relation 
\beq
\Hom_\scrc\big(X\,,\,\inthom_\scrc(M_1,M_2)\big):=
\Hom_A(\modM_1\otimes X,\modM_2)
\label{inthomdef}\eeq
for all objects $X\in\Ob(\scrc)$.

By (\ref{inthomdef}) there is an isomorphism
$$
\Hom_\scrc\big(\inthom_\scrc(M_1,M_2)\,,\,\inthom_\scrc(M_1,M_2)\big)=
\Hom_A\big(\modM_1\otimes\inthom_\scrc(M_1,M_2)\,,\,\modM_2\big) \ .
$$
We define a canonical evaluation morphism
$$
{\rm ev}_{M_1,M_2}\,:\,\modM_1\otimes
\inthom_\scrc(M_1,M_2)~\longrightarrow~\modM_2
$$
as the image of $\id_{\inthom_\scrc(M_1,M_2)}$ under this
isomorphism. Now let
$\modM_1,\modM_2,\modM_3\in\Ob(\Mod_\scrc(A))$. Then the sequence of
morphisms
\begin{eqnarray*}
\modM_1\otimes\big(\inthom_\scrc(M_1,M_2)\otimes
\inthom_\scrc(M_2,M_3)\big)\=
\big(\modM_1\otimes\inthom_\scrc(M_1,M_2)\big)\otimes
\inthom_\scrc(M_2,M_3) \\
\qquad \qquad \xrightarrow{{\rm ev}_{M_1,M_2}\otimes
\id_{\inthom_\scrc(M_2,M_3)}}~\modM_2\otimes\inthom_\scrc(M_2,M_3)~
\xrightarrow{{\rm ev}_{M_2,M_3}}~\modM_3
\end{eqnarray*}
defines a canonical composition morphism
\beq
\inthom_\scrc(M_1,M_2)\otimes\inthom_\scrc(M_2,M_3)~
\longrightarrow~\inthom_\scrc(M_1,M_3) \ .
\label{inthomcomp}\eeq
This multiplication is associative and compatible with isomorphisms
involving the internal Hom.

Let us now fix a particular non-zero $A$-module $\modM\in\Ob(\Mod_\scrc(A))$. Then the 
multiplication morphism (\ref{inthomcomp}) defines an algebra
structure on the object
\beq
A_M:=\inthom_\scrc(M,M)
\label{AMinthom}\eeq
of the category $\scrc$. If $\Mod_\scrc(A)$ is a semisimple
indecomposable module category, then $A_M\in\Ob(\scrc)$ is a
semisimple indecomposable algebra.

Define a functor $\scrf:\Mod_\scrc(A)\to\scrc$ by
\beq
\modN~\longmapsto~\inthom_\scrc(M,N) \qquad \mbox{for}
\quad \modN\in\Ob\big(\Mod_\scrc(A) \big) \ .
\label{ModAfunctor}\eeq
The multiplication morphism (\ref{inthomcomp}) defines the structure
of a left $A_M$-module on $\inthom_\scrc(M,N)$, and hence
(\ref{ModAfunctor}) restricts to a functor
$\scrf:\Mod_\scrc(A)\to\Mod_\scrc(A_M)$. Furthermore, for
$X,Y\in\Ob(\scrc)$ one has, by (\ref{inthomdef}) together with the
canonical isomorphisms (\ref{Homdualisos}) and $X=(\,{}^\vee X)^\vee$, 
the isomorphisms 
\begin{eqnarray*}
\Hom_\scrc\big(Y\,,\,\inthom_\scrc(M,N\otimes X)\big)&=&
\Hom_A(\modM\otimes Y,\modN\otimes X) \\[4pt] &=&
\Hom_A\big((\modM\otimes Y)\otimes {\,}^\vee X\,,\,\modN\big) \\[4pt] &=&
\Hom_A\big(\modM\otimes(Y\otimes {\,}^\vee X\,)\,,\, \modN\big) \\[4pt] &=& 
\Hom_\scrc\big(Y\otimes{\,}^\vee X\,,\,\inthom_\scrc(M,N)\big) \\[4pt]
&=& \Hom_\scrc\big(Y\,,\,\inthom_\scrc(M,N)\otimes X\big) \ ,
\end{eqnarray*}
and hence the canonical isomorphism
\beq
\inthom_\scrc(M,N\otimes X)=\inthom_\scrc(M,N)\otimes X
\label{inthomtensor}\eeq
for all $X\in\Ob(\scrc)$. This isomorphism defines a structure of a
module functor on the functor $\scrf:\Mod_\scrc(A)\to\Mod_\scrc(A_M)$
given by (\ref{ModAfunctor}), by compatibility of the multiplication
(\ref{inthomcomp}) with (\ref{inthomtensor}). Using standard
homological algebra, one then proves~\cite{Ostrik} that this functor
is an equivalence of module categories.

It follows that for any two $A$-modules
$\modM_1,\modM_2\in\Ob(\Mod_\scrc(A))$, the module categories of
$A_{M_1}$ and $A_{M_2}$ are equivalent. Hence the algebras $A_{M_1}$
and $A_{M_2}$ in $\scrc$ are \emph{Morita equivalent}. By the explicit
construction (\ref{ModAfunctor}) of the functor $\scrf$ and
(\ref{inthomcomp}), the Morita equivalence bimodules are given
explicitly by the objects $\inthom_\scrc(M_1,M_2)$ and
$\inthom_{\scrc}(M_2,M_1)$ of $\scrc$.

Consider the case where $\modM=\modA:=(A,\mu)$ is the trivial
$A$-bimodule. That the algebra object $A$ is haploid is equivalent to
the statement that $A$ is simple as a bimodule $\modA$ over
itself~\cite{KR1}. Then $A_A=\inthom_\scrc(A,A)$. The unit morphism 
$\eta\in\Hom_\scrc(\Idd,A)$ defines a canonical isomorphism
$\Hom_A(\modA,\modM)=\Hom_\scrc(\Idd,M)$ for any $A$-module
$\modM$. Then for any object $X\in\Ob(\scrc)$ one has the canonical
isomorphism
\beq
\Hom_A(\modA\otimes X,\modM)=\Hom_A(\modA,\modM\otimes X^\vee\,)=
\Hom_\scrc(\Idd,M\otimes X^\vee)=\Hom_\scrc(X,M) \ .
\label{HomAsimpl}\eeq
This isomorphism is generated by the reciprocity map
$f\mapsto\varrho\circ(\id_A\otimes f)$ for $f\in\Hom_\scrc(X,M)$. Using
duality one similarly has a canonical isomorphism
\beq
\Hom_A(\modM,\modA\otimes X)=\Hom_\scrc(M,X)
\label{HomAsimpldual}\eeq
generated by $f\mapsto
(\id_A\otimes(f\circ\varrho))\circ((\Delta\circ\eta)\otimes\id_M)$ for
$f\in\Hom_\scrc(M,X)$. From the definition (\ref{inthomdef}) it then
follows that there is a natural isomorphism
$$
\inthom_\scrc(A,M)=M
$$
and in particular
$$
A_A=\inthom_\scrc(A,A)=A \ ,
$$
as expected.

It follows from these properties of the internal Hom that there is a
natural identification $\inthom_\scrc(M_1,M_2)=\,{}^\vee \modM_1\otimes_A
\modM_2$. In particular, $A_M=\,{}^\vee \modM\otimes_A \modM$, and the Morita
equivalence bimodules above are ${}^\vee \modM_1\otimes_{A_{M_1}}\modM_2\cong
A_{M_2}$ and ${}^\vee\modM_2\otimes_{A_{M_2}}\modM_1\cong
A_{M_1}$. (See~\cite[Def.~5.2]{FS1} for the definition of the tensor
product $\modN\otimes_A\modM$ of a right $A$-module $\modN$ and a left
$A$-module $\modM$.) This is
useful in explicit calculations, and it works whenever both $\scrc$ and
the module category $\Mod_\scrc(A)$ are semisimple with finitely-many
simple objects, so that the
internal Hom is always an object of $\scrc$. In more general cases, one
needs to work with ind-objects of the category $\scrc$ as discussed above. In particular, if $\scrc$ has infinitely many simple
objects, then one can establish a similar Morita equivalence result by
working with ind-algebras and ind-modules~\cite{Ostrik}.

\medskip

\subsection{Freyd-Mitchell embedding theorem}\label{app:GW}~\\[5pt]
It is a classical result in category theory that every small abelian
category is equivalent to a full subcategory of the representation category of
modules over some ring $\ca$. If the category in
question $\scrc$ is enriched over $\Vect$, then $\ca$ can
be taken to be an algebra. In the semisimple case, with
$\scrc$ containing a finite number $N$ of simple objects, we
can easily construct an equivalent category of modules, simply with as
many irreducible representations as the category at hand. By Wedderburn theory, finite-dimensional semisimple
algebras are essentially direct sums of full matrix algebras: The
algebra $\IM_n$ of $n\times n$ complex
matrices contains no non-trivial minimal ideals. We can thus
obtain $N$ irreducible representations by taking $\ca$ to be a direct
sum of full matrix algebras over $\IC$ as
$$
\ca=\bigoplus_{i=0}^{N-1}\,\IM_{n_i} \ .
$$
Each summand has only one irreducible representation, an
$n_i$-dimensional complex vector space $\IC^{n_i}$. This construction is not
canonical since we are free to choose the dimensions $n_i$ as we wish.

The ring in question is formed by taking the endomorphism ring
$\ca=\End_{\scrm}(K)$ of some object $K$ which is
an injective cogenerator in some functor category $\scrm$. This means that every object injects in
$K^{\oplus n}$ for suitable $n$ (depending on the object). In the
semisimple case, with $\scrc$ containing finitely many isomorphism
classes of simple objects, we can actually apply the argument inside the
category $\scrc$ itself, as then the sum $K$ of simple objects is an injective
cogenerator: Every object is a sum of simple objects,
and so will inject into a sufficiently big sum of copies of $K$. In
more general cases, e.g. when $\scrc$ has infinitely many simple
objects, one
needs to work with ind-objects of the category $\scrc$ and pass to an
associated $A_\infty$-category $\scrm$; we shall return to this point
later on. For the moment, we briefly work through some of the
details of the construction of the ring $\ca$; for more details, see~\cite[\S4.4]{Freyd}
and~\cite[\S{IV.4}]{Mitchell}.

We begin with some standard category theory definitions. We call an object
$K\in\Ob(\scrm)$ in an abelian category $\scrm$ injective if the
contravariant functor 
$$
\scrh:=\Hom_{\scrm}(-,K)\,:\,\scrm~\longrightarrow~\Vect
$$ is coexact, i.e. it carries exact sequences into exact sequences
(with the arrows reversed). We say that $K$ is a cogenerator if the
functor $\scrh$ is an embedding, i.e. for $X,Y\in\Ob(\scrm)$, the
map $\Hom_{\scrm}(X,Y)\to\Hom_{\Vect}(\scrh(X),\scrh(Y))$ is
injective. A collection of objects $(K_i)_{i\in I}$ of $\scrm$ is
called a family of cogenerators for $\scrm$ if for every
$X,Y\in\Ob(\scrm)$ and every non-zero morphism
$\alpha\in\Hom_{\scrm}(X,Y)$ there exists $\kappa_i\in\Hom_{\scrm}(Y,K_i)$ such
that $\kappa_i\circ\alpha \neq0$. Note that $K$ is a cogenerator if $(K)$
is a family of cogenerators for $\scrm$. If
$$
K=\bigoplus_{i\in I}\,K_i
$$
and $\Hom_{\scrm}(X,K_i)\neq\emptyset$ for all $i\in I$ and
$X\in\Ob(\scrm)$, then $K$ is a cogenerator for $\scrm$ if and only if
$(K_i)_{i\in I}$ is a family of cogenerators for $\scrm$. Below we
will use the fact that it suffices to construct ind-objects on
cogenerators~\cite{KashShap}. 

Now let $\ca$ be an associative algebra over $\IC$, and let $\Rep(\ca)$
denote the representation category of left $\ca$-modules. Then $\ca$ is an injective
cogenerator for $\Rep(\ca)$: The contravariant functor
$$
\Hom_{\Rep(\ca)}(-,\ca)\,:\,\Rep(\ca)~\longrightarrow~\Vect \ ,
$$
with $\ca$ regarded as the trivial left $\ca$-module, is the ``forgetful''
functor assigning to each $\ca$-module $V$ the underlying complex vector
space (forgetting that $V$ is an $\ca$-module). Thus any category
equivalent to $\Rep(\ca)$ also has an injective cogenerator. We now show
that the converse is also true, i.e. if $\scrm$ is an abelian category
enriched over $\Vect$ which possesses an injective cogenerator $K$,
then $\scrm$ is equivalent to the representation category $\Rep(\ca)$ for
some algebra~$\ca$.

Let
\beq
\ca:=\End_{\scrm}(K)=\Hom_{\scrm}(K,K)
\label{assalgRKdef}\eeq
be the $\IC$-vector space of endomorphisms of $K\in\Ob(\scrm)$. For
every $X\in\Ob(\scrm)$, the $\IC$-vector space $\Hom_{\scrm}(X,K)$ has a
canonical $\ca$-module structure: For $\alpha\in\Hom_{\scrm}(X,K)$ and
$\rho\in \ca$, define $\rho\triangleright\alpha\in\Hom_{\scrm}(X,K)$ to be
the composition $\rho\circ\alpha$. We may thereby define the functor
\beq
\scrt \,:\,\scrm~\longrightarrow~\Rep(\ca) \ , \qquad
\scrt(X)=\Hom_{\scrm}(X,K) \ .
\label{Tsemisimpleequiv}\eeq
Using the fact that $K$ is an injective cogenerator, one shows that $\scrt$
is an equivalence of abelian categories, such that the map
$\Hom_{\scrm}(X,Y)\to\Hom_{\Rep(\ca)}(\scrt(X),\scrt(Y))$ induced by $\scrt$ is an
isomorphism whenever $X$ is finitely-generated. By Yoneda's lemma,
$\scrt$ is a left-exact functor, and the assignment $X\mapsto
\Hom_{\scrm}(X,-)$ yields a duality between $\Rep(\ca)$ and a
subcategory of the category $\scrm^\vee$ of left-exact functors $\scrm\to \Vect$.

If the category $\scrc$ is
semisimple, abelian and finite with simple objects $U_i$, $i\in I$, then we may take $\scrm=\scrc$ and
\beq
K=\bigoplus_{i\in I}\,U_i
\label{modKdef}\eeq
is an injective cogenerator for $\scrc$. Hence the associative
$\IC$-algebra (\ref{assalgRKdef}) and the
equivalence of abelian categories (\ref{Tsemisimpleequiv}) are given
explicitly by
\beq
\ca=\bigoplus_{i,j\in I}\,\Hom_\scrc(U_i,U_j)\cong \IC^{|I|} \qquad
\mbox{and} \qquad \scrt(X)=\bigoplus_{i\in I}\,\Hom_\scrc(X,U_i) 
\label{equivorcat}\eeq
for $X\in\Ob(\scrc)$. As mentioned
before, the algebra $\ca$ is a non-canonical object as it is only
defined up to Morita equivalence.

If the category
$\scrc$ is not semisimple, then one requires
the general construction of the Freyd-Mitchell embedding
theorem~\cite{Freyd,Mitchell}. For this, let $\scrm=\scrc^\vee$ be the category
of left-exact functors $\scrc\to\Vect$, and construct an injective
cogenerator $K$ with endomorphism algebra (\ref{assalgRKdef}). Define
a functor $\scrt:\scrc\to \Rep(\ca)$ by $\scrt(X):=
\Hom_{\scrc^\vee}\big(\Hom_{\scrc}(X,-),K\big)$ for $X\in \Ob(\scrc)$. Then
$\scrt$ is a fully faithful exact functor which yields an
equivalence between $\scrc$ and a subcategory of~$\Rep(\ca)$. If
$\scrc$ contains infinitely many simple objects $U_i$, $i\in I$, then
the cogenerator $K$ may be taken to be the ind-object
$\underline{U}=(U_i)_{i\in I}$; in this instance the functor $\scrt$
coincides with the presentation of $\underline{U}$ given by
(\ref{indobjectvee}).

We now consider the case where $\scrm=\Mod_\scrc(A)$ is the
module category of an algebra object $A\in\Ob(\scrc)$ in a tensor category. For
$U\in\Ob(\scrc)$, the induced $A$-module is the (left) $A$-module
$\modA\otimes U=(A\otimes U,\mu\otimes\id_U)$. This defines
an induction functor $\scrc\to\Mod_\scrc(A)$. There is also a
restriction functor $\Mod_\scrc(A)\to\scrc$ given by the forgetful map
$\modM\mapsto M$ on objects. Induced $A$-modules have the useful
computational properties~\cite[Lem.~5.8]{FS1}
\beq
\modN\otimes_A(\modA\otimes X)\cong \modN\otimes X
\label{IndAmodtimesprop}\eeq
as $A$-modules, for every right $A$-module $\modN$ and every
$X\in\Ob(\scrc)$, and also
\beq
(\modA\otimes X)\otimes_A(\modA\otimes Y)=\modA\otimes(X\otimes Y)
\label{IndAmod2timesprop}\eeq
for all $X,Y\in\Ob(\scrc)$.
When $A$ is a Frobenius algebra, every module
$\modM$ over $A$ is a submodule of an induced
module~\cite[Lem.~5.23]{FS1}, because there is an injection sending
$\modM\mapsto\modA\otimes M$. If the tensor category $\scrc$ is
semisimple, abelian and finite with simple objects $U_i$,
$i\in I$, and the Frobenius algebra $A$ is special, then 
\beq
\modK=\modA\otimes\Big(\,\bigoplus_{i\in I}\,U_i\,\Big)
\label{modKdef}\eeq
is an injective cogenerator for $\Mod_\scrc(A)$. Then the endomorphism
ring $\ca=\End_A(\modK)$ of the $A$-module $\modK$
is the (semisimple) associative $\IC$-algebra we are looking for; the
equivalence of abelian categories given by (\ref{Tsemisimpleequiv}),
$\scrm=\Mod_\scrc(A)=\Rep(\ca)$, can then be used to induce the
structure of a module category over $\scrc$ on $\Rep(\ca)$. Using (\ref{HomAsimpl}) and
(\ref{HomAsimpldual}), we can write the equivalence between abelian
categories explicitly in terms of morphism spaces in the
\emph{original} tensor category $\scrc$ as
\beq
\ca=\bigoplus_{i,j\in I}\,\Hom_\scrc(U_i,A\otimes U_j) \qquad
\mbox{and} \qquad \scrt(\modM)=\bigoplus_{i\in I}\,\Hom_\scrc(M,U_i) 
\label{equivorcat}\eeq
for $\modM\in\Ob(\Mod_\scrc(A))$. If the Frobenius algebra $A$ is not special, then the category
$\Mod_\scrc(A)$ need not be semisimple; in these instances we need
the general construction of the Freyd-Mitchell embedding theorem combined with the ind-object
treatment as above.

\end{document}